\documentclass[12pt]{amsart}
\usepackage{large_games}
\usepackage{placeins}

\makeatletter
\newcommand{\startappendix}{
	\appendix
	\phantomsection\addcontentsline{toc}{part}{Appendix}
	\let\@makechapterhead\@makeschapterhead
}

\newcommand{\typ}{\tau}

\title{}
\begin{document}
	\vspace*{5ex minus 1ex}
	\begin{center}
		\Large \textsc{Econometric Inference on a Large Bayesian Game with Heterogeneous Beliefs}
		\bigskip
	\end{center}

	\date{\today}

	\vspace*{3ex minus 1ex}
	\begin{center}
		\textsc{Denis Kojevnikov} and \textsc{Kyungchul Song} \\
		\textit{Tilburg University and University of British Columbia}\\
		\bigskip\bigskip
	\end{center}

	\thanks{
		Corresponding address: Kyungchul Song, Vancouver School of Economics, University of British Columbia, Vancouver, BC, Canada. Email address: kysong@mail.ubc.ca.
	}

\begin{bibunit}[elsart-harv]
\begin{abstract}
 	Econometric models of strategic interactions among people or firms have received a great deal of attention in the literature. Less attention has been paid to the role of the underlying assumptions about the way agents form beliefs about other agents. We focus on a single large Bayesian game with idiosyncratic strategic neighborhoods and develop an approach of empirical modeling that relaxes the assumption of rational expectations and allows the players to form beliefs differently. By drawing on the main intuition of \cite{Kalai:04}, we introduce the notion of hindsight regret, which measures each player's ex-post value of other players' type information, and obtain the belief-free bound for the hindsight regret. Using this bound, we derive testable implications and develop a bootstrap inference procedure for the structural parameters. Our inference method is uniformly valid regardless of the size of strategic neighborhoods and tends to exhibit high power when the neighborhoods are large. We demonstrate the finite sample performance of the method through Monte Carlo simulations.
\medskip

\medskip

\end{abstract}

\maketitle

\section{Introduction}

Many economic outcomes arise as a consequence of agents' decisions under the influence of others' choices. Endogeneity and simultaneity of such influence pose a challenge for an empirical researcher. In response to this challenge, a strand of empirical methods has employed game-theoretic models to capture strategic interactions among agents. (See \cite{Bresnahan/Reiss:91}, \cite{Tamer:03}, \cite{Ciliberto/Tamer:09}, \cite{Aradillas-Lopez:10}, \cite{Beresteanu/Molchanov/Molinari:11}, \cite{Aradillas-Lopez/Tamer:08}, and \cite{dePaula/Tang:12}, among many others.) However, these models often adopt a framework of many independent copies of the same game to facilitate identification and inference. Furthermore, they usually rely on a strong assumption on the way the agents form beliefs, namely, a \textit{common prior assumption} where the payoff types of the players are drawn from a common distribution, and the distribution belongs to common knowledge among all the players.

These two features of empirical modeling of games do not fit very well with many empirical settings in practice. In a typical empirical setting of interactions, each player tends to have a different set of other players whose actions jointly affect the player's payoff. In this case, the common prior assumption, apart from its restrictiveness in belief formation, does not help us in empirical modeling because the observed actions from an equilibrium have different distributions across the players, and it is not possible to aggregate the observed actions to recover the beliefs from data.\footnote{\cite{Manski:04} proposed using data on subjective probabilities in choice studies. See \cite{Dominitz/Manski:97} for a study on subjective income expectations and \cite{Li/Lee:09} for an investigation of rational expectations assumptions in social interactions using subjective expectations data. However, in strategic environments with many players, it is often far from trivial to obtain data on players' expectations about the other players' types prior to the play.}

In this paper, we focus on a large incomplete information game where the researcher observes actions that arise from a pure strategy Bayesian equilibrium and develop a new approach to empirical modeling that does not rely on the two commonly used features.

First, our approach adopts a large game perspective where each player faces a different set of other players whose actions affect his payoff - we call this set the player's \textit{strategic neighborhood} - and chooses an action from a finite set. As in \cite{Ciliberto/Tamer:09}, we pursue an inference procedure that does not require any restrictions on the equilibrium selection rules and, thus, seek to find a set of testable implications from the model to construct a confidence region for the payoff parameters. We develop a bootstrap inference method that is uniformly valid regardless of the sizes of the strategic neighborhoods and exhibits high power when the neighborhood size is large. Therefore, our approach is most useful for settings with large-scale interactions where the strategic neighborhoods are large. However, our approach is not useful for games with a small number of players or network formation games and matching games where the action space becomes larger as the number of players increases.

Second, our approach departs from the common prior assumption by allowing each player to form beliefs differently from the other players. The role of beliefs in generating predictions from a game has long been a fundamental issue in game theory. Despite its crucial role in modeling strategic interactions, it seems to have received relatively scant attention in the literature of econometrics.\footnote{One notable exception is \cite{Aradillas-Lopez/Tamer:08} which we will discuss in detail when we review the literature.}

The distinctive feature of our empirical model is that while the equilibrium is driven by the heterogeneous subjective beliefs of the players, the validity of statistical inference is measured in terms of Nature's objective probability. In this paper, we allow each player to form beliefs differently using different priors. Furthermore, the subjective beliefs do not need to coincide with Nature's objective probability. In this sense, our model departs from the commonly used framework of rational expectations. Using this model, we propose an inference procedure that is robust to how individual players form their beliefs about other players' types. If the predictions from game models should be robust to players' belief formation processes, as emphasized by \cite{Wilson:87} and \cite{Bergemann/Morris:05}, the same applies all the more to econometric inference on such models.

Instead of attempting to recover subjective beliefs from data, we develop a \textit{hindsight regret approach} drawing on the insights of \cite{Kalai:04} and \cite{Deb/Kalai:15}. The \textit{hindsight regret} of a player measures the \textit{ex post} payoff loss due to his inability to observe the other players' types. More specifically, the hindsight regret quantifies the additional compensation needed to preserve each player's incentive compatibility constraint in equilibrium even after all the players' types are revealed.

Using the hindsight regret, we derive moment inequalities in a spirit similar to \cite{Ciliberto/Tamer:09}. The tightness of the moment inequalities depends on how strongly any two players are strategically interdependent. In a social interaction model, when the strategic neighborhood is large, and each player's payoff is affected by the action of another player in inverse proportion to the group sizes, the inequalities can be fairly tight. On the other hand, inference tends to be overly conservative in the situation with small private information games as in \cite{Aradillas-Lopez:10} and \cite{dePaula/Tang:12}.

For inference, we propose a bootstrap-based approach and establish its uniform asymptotic validity as the number of players increases to infinity. The asymptotic validity is uniform over Nature's probabilities for drawing the players' types. Our approach for inference is inspired by the work of \cite{Andrews:05}, who investigated the inference problem in the presence of common shocks in short panel data (see \cite{Kuersteiner/Prucha:13} for related research on dynamic panel models). However, we cannot use the random norming as he did to pivotize the test statistic because the restrictions here are inequalities rather than equalities. Instead, we use a bootstrap procedure that is inspired by the Bonferroni approach of \cite{Romano/Shaikh/Wolf:14}.

Game-theoretic models have been frequently used in the literature of empirical research and econometrics. See \cite{Bresnahan/Reiss:91}, \cite{Tamer:03}, \cite{Krauth:06}, \cite{Ciliberto/Tamer:09}, \cite{Aradillas-Lopez:10}, \cite{Bajari/Hong:10}, \cite{Beresteanu/Molchanov/Molinari:11}, \cite{Aradillas-Lopez/Tamer:08}, \cite{dePaula/Tang:12}, and \cite{Aradillas-Lopez/Gandhi:16} among others. See \cite{dePaula:13} for further references and discussions.

This paper's framework is also related to various models of social interactions. As a seminal paper in the structural modeling and estimation of social interactions, \cite{Brock/Durlauf:01a} developed discrete choice models of social interactions. Their discrete-choice approach influenced many subsequent studies such as \cite{Krauth:06}, \cite{Ioannides/Zabel:08}, and \cite{Li/Lee:09} to name but a few. See the monograph by \cite{Ioannides:13:FNN} for methodological progress in the literature on social interactions. More recently, \cite{Blume/Brock/Durlauf/Jayaraman:15:JPE} considered a Bayesian game of social interactions on a network.

Our paper is closely related to \cite{Aradillas-Lopez/Tamer:08}, who considered game models and studied the identifying power of the solution concept as it is weakened from Nash equilibria to level $k$-rationalizability. Part of their results concern an incomplete information game, which, like our paper, permits the players' beliefs to be heterogeneous and incorrect. The main difference between their work and ours is that their work focuses on the identifying power of the solution concept as we depart from Nash equilibria, whereas our paper stays with pure strategy Bayesian equilibria. Hence, robustness to beliefs in our paper is narrower than that in their context of rationalizability. On the other hand, our primary focus is on producing a framework of empirical modeling and developing uniformly valid asymptotic inference which can be applied to a large game setting.

A recent stream of literature considers a setting in which the researcher observes one large game. For example, \cite{Xu:18} studied a single large Bayesian game similar to ours, focusing on a setting that yields uniqueness of the equilibrium and point-identification of the parameters. On the other hand, \cite{Bisin/Moro/Topa:11} admit multiple equilibria, but their equilibrium concept requires asymptotic stability of aggregate quantities (as the number of players increases). \cite{Menzel:12} developed an asymptotic inference for large complete information games where type-action profiles are (conditionally) exchangeable sequences.\footnote{The fundamental difference between \cite{Menzel:12} and this work lies in modeling the probability of observations. Menzel employs a complete information game model where the randomness of the observed outcomes is mainly due to the sampling variations. Thus, random sampling schemes and variants justify his exchangeability conditions. On the other hand, we consider an incomplete information game, where the randomness of observed outcomes stems from the inherent heterogeneity across players due to Nature's drawing of types.} More recently, \cite{Canen/Schwartz/Song:20} considered a large local interactions model with quadratic utilities and developed an inference procedure when the players observe their neighbors' types that are unobserved by the researcher.

The main departure of this paper from this literature is that it relaxes the assumption of rational expectations in a single large game setup and develops a uniformly valid bootstrap inference procedure on the parameter of interest.

The game theory literature has explored various solution concepts that relax the common prior assumption and/or the assumption that the information structure of the players is known to the researcher. (See \cite{Morris:95:EP} for a critical discussion of the common prior assumptions and references.) More recent literature develops solution concepts that are informationally robust. We cannot give an adequate review of this large, growing literature in this limited space. We would instead refer the reader to \cite{Bergemann/Morris:17:GEB} for a review of this literature and references. Our paper focuses on relaxing the common prior assumption and does not consider the robustness to the information structures. We adopt a solution concept of the pure strategy version of Bayesian equilibrium (see \cite{Maschler/Solan/Zamir:13:GameTheory}), which does not rely on the common prior assumption.

As mentioned before, our paper's approach is inspired by the main ideas in \cite{Kalai:04} and \cite{Deb/Kalai:15}. They considered the problem of characterizing a class of Bayesian games whose set of equilibria exhibit approximate \textit{ex post} stability. More specifically, they defined a strategy profile to be $(\epsilon,\rho)$-hindsight-stable (or $(\epsilon,\rho)$-\textit{ex post} Nash), if the players have the incentive to deviate from the strategy profile only with a low probability, not greater than $\rho$, after the types of their opponents are revealed, as long as their payoffs are compensated by at least an additional amount of $\epsilon$. This additional amount $\epsilon$ is what we call the hindsight regret here. Despite this close connection, there is one important difference between their papers and ours. \cite{Kalai:04} and \cite{Deb/Kalai:15} study the \textit{ex post} robustness of equilibria in large Bayesian games with a common prior assumption, whereas our paper uses the \textit{ex post} robustness property to derive testable implications from a large Bayesian game which does not satisfy a common prior assumption. The hindsight regret $\epsilon$ in \cite{Kalai:04} and \cite{Deb/Kalai:15} is a margin of error permitted for \textit{ex post} robustness; here, it plays a more substantial role, as it determines the strength of the testable implications for econometric inference.

Recently, \cite{Aguirregabiria/Magesan:20:ReStud} studied the identification of dynamic game models where the beliefs are allowed not to be in equilibrium. They treated the beliefs as nuisance parameters and showed that when certain exclusion restrictions on the payoff functions are satisfied, there exist testable restrictions for testing the null hypothesis that the beliefs are in equilibrium. Furthermore, they provided conditions under which the payoffs and the beliefs are fully identified. There are two major differences between their and our settings. First, our setting focuses on a static game, while their games are dynamic. Second, unlike their setting, we consider a situation where the econometrician observes a single large game in which players form heterogeneous beliefs. In this situation, we cannot hope for consistently estimating the beliefs, even if we identify them using the joint distribution of observed variables from the entire game.

This paper is organized as follows. The next section introduces a large Bayesian game and a belief-free version of hindsight regrets. The section turns to econometric inference, deriving testable implications, and presents a general inference method and its uniform asymptotic validity. For simplicity of exposition, most of the results in the paper are obtained assuming a binary action space. Their extension to the case of a general finite action set is provided in the Appendix. Section 3 presents a finite sample study of the inference using Monte Carlo simulations. In Section 4, we conclude. The Supplemental Note contains mathematical proofs and additional simulation results.

\section{A Large Bayesian Game with Heterogeneous Beliefs}

\subsection{The Setup}
\label{subsec:setup}

In this section, we introduce a Bayesian game. A finite set $N$ of players simultaneously choose a binary action from $\{0,1\}$ during the play of the game. (In the Appendix, we extend our proposal to the large Bayesian games with a general finite action set.) We let $n = |N|$ be the number of the players in the game throughout the paper.

We assume that there is a measurable space $(\Omega,\H)$ such that once Nature draws $\omega\in \Omega$, each player $i \in N$ is given the payoff state $\typ_i(\omega)$ as a realized random vector $\tau_i$. Given an action profile $y = (y_j)_{j \in N}$, each player $i \in N$ receives payoff
\begin{align*}
	u_i(y_i,y_{-i} ;\typ_i(\omega)),
\end{align*}
by choosing $y_i\in \{0,1\}$, when the player faces the other players in $N \setminus\{i\}$ who choose action profile $y_{-i} \eqdef (y_j)_{j \in N\setminus \{i\}}$. For each player $i \in N$, we assume that there exists a subset $N(i) \subset N \setminus \{i\}$ such that $u_i(y_i,y_{-i} ;\typ_i(\omega))$ depends on $y_{-i}$ only through $y_{N(i)} = (y_j)_{j \in N(i)}$. We call the set $N(i)$ the \textit{strategic neighborhood} of player $i$, which refers to the set of players whose actions potentially affect the payoff of player $i$. We let $n(i) = |N(i)|$ denote the size of the strategic neighborhood of player $i$. We assume that the payoff state $\typ_i$ is specified as
\begin{equation}
	\label{eq:Typ_i}
	\typ_i = (X_i,\eta_i),
\end{equation}
where $\eta_i$ is a random vector that represents unobserved heterogeneity, i.e., an idiosyncratic payoff component unobserved by the researcher, and $X_i$ the vector of observable covariates of player $i$.

The information for each player $i$ is given by
\begin{align}
	\label{eq:info}
	\mathcal{I}_i \eqdef \sigma\left(\eta_i, \X \right), \quad \X \eqdef (X_i)_{i \in N},
\end{align}
i.e., the $\sigma$-field generated by $(\eta_i, \X)$. A \textit{pure strategy} $Y_i:\Omega \to \{0,1\}$ of player $i\in N$ is an $\IF_i$-measurable function from the state space to the action set, and a \textit{pure strategy profile} $\Y\eqdef (Y_j)_{j \in N}$ is the vector of individual pure strategies. The measurability with respect to $\IF_i$ reflects the fact that each player needs to form a strategy using only information $\IF_i$.

In evaluating uncertainty, there are two types of probability measures on the measurable space $(\Omega, \H)$. First, Nature uses the \textit{objective probability} $\PM$ to determine the probability of any event involving $(\eta_i,X_i)_{i \in N}$. It is the objective probability $\PM$ that the researcher uses to express the validity of his inference method (such as the coverage probability of a confidence interval or the size and power of a test). On the other hand, each player $i$ uses the \textit{subjective probability} $\QM_i$ to evaluate his expected payoff.\footnote{
	One might consider modeling the subjective beliefs using Choquet capacities. (See, e.g., \cite{Epstein/Seo:15} for the De Finetti type results for exchangeable capacities.) The main difficulty ixn pursuing this direction in our context is to introduce McDiarmid's inequality under conditional independence restrictions.  While we believe that this extension might be feasible, it requires introducing a substantial amount of additional mathematical notions and establishing some of the basic results in this paper. Hence, we relegate this extension to future research.
}
The subjective probability determines the shape of the equilibrium strategies that we introduce below. As pointed out by \cite{Aumann:76}, when $\PM$ belongs to common knowledge, we have $\QM_i=\PM$ for all $i\in N$ so that the distinction between the objective and subjective probabilities is not necessary. However, we do not assume that $\PM$ belongs to common knowledge in our paper. 

Let us introduce the solution concept of the game that we use in this paper.

\begin{definition}
\label{def:BNE}
A strategy profile $\Y$ is a \textit{pure strategy Bayesian equilibrium} if for each player $i\in N$ and any pure strategy $Y_i'$,
\begin{equation}
\label{eq:icc}
	\E_{\QM_i}[u_i(Y_i,Y_{-i};\typ_i)\mid \IF_i]\ge \E_{\QM_i}[u_i(Y_i',Y_{-i};\typ_i)\mid \IF_i], \quad(\QM_i\text{-a.s.}),
\end{equation}
where $\E_{\QM_i}\left[\csdot\mid \IF_i \right]$ denotes the conditional expectation given $\IF_i$ under $\QM_i$ and $Y_{-i} \eqdef (Y_j)_{j \in N \setminus \{i\}}$.
\end{definition}

The popular solution concept of Bayes Nash Equilibrium (BNE) typically assumes a setting in which the beliefs are derived from a common prior, i.e., the beliefs are consistent. In our setting, we relax this assumption and do not require the beliefs to be consistent. Our pure strategy Bayesian equilibrium in Definition \ref{def:BNE} is the pure strategy version of a Bayesian equilibrium. (See, e.g., \cite{Maschler/Solan/Zamir:13:GameTheory}, Definition 10.39 on page 408.)

\subsection{Belief-Free Hindsight Regrets}

Following the insights from \cite{Kalai:04} and \cite{Deb/Kalai:15} in economic theory, we consider a hindsight regret approach which replaces the inequality \eqref{eq:icc} by its \textit{ex post} version,
\[
	u_i(Y_i,Y_{-i};\typ_i) - u_i(1-Y_i,Y_{-i};\typ_i) \ge -\lambda_{i,\rho},
\]
where $\lambda_{i,\rho} \ge 0$ is a random variable chosen to ensure that the inequality holds with the subjective probability at least $1 - \rho$ conditionally on $\IF_i$, for small $\rho>0$. With subjective probability at least $1 - \rho$, the compensation $\lambda_{i,\rho}$ leads player $i$ to stay with his action $Y_i$ in equilibrium, once the payoff states of all the players are revealed. Let us introduce the notion of hindsight regret formally as follows.
\begin{definition}
	Given an equilibrium $\Y$ and $\IF_i$-measurable non-negative random variable $\lambda_{i,\rho}$ and $\rho \in (0,1)$, we say that $\lambda_{i,\rho}$ is a $\rho$-\textit{hindsight regret} for player $i\in N$ if
	\begin{equation}
		\label{eq:hr_def0}
		\QR[i]{u_i^{\Delta}(Y_i,Y_{-i};\typ_i) \ge -\lambda_{i,\rho} \mid \IF_i} \ge 1-\rho, \quad(\QM_i\text{-a.s.}),
	\end{equation}
   where
   \begin{align}
		\label{eq:u_i^Delta}
   		u_i^\Delta(y_i,y_{-i};\tau_i) = u_i(y_i,y_{-i};\tau_i) - u_i(1-y_i,y_{-i};\tau_i).
   \end{align}
\end{definition}
In order to use the inequalities \eqref{eq:hr_def0} to derive testable implications, we introduce two assumptions. The first assumption is a conditional independence assumption of unobserved heterogeneities as often used in the literature.
\begin{assumption}[\textit{Conditionally Independent Types}]
	\label{assum:CI}
	The unobserved heterogeneities, $\eta_i$, $i \in N$, are conditionally independent given $\X$ under $\PM$ and each $\QM_i$.
\end{assumption}
While this assumption prevents spillover of information across players in a way unobserved by the researcher, it is weaker than the often-used assumption in the literature that unobserved heterogeneities are independent of covariates. Independence or conditional independence of unobserved payoff states across players has been used in the literature, for example, by \cite{dePaula/Tang:12} and \cite{Xu:18}. Note that \cite{Aradillas-Lopez:10} allows correlation between unobserved payoff states across players flexibly. Through Assumption \ref{assum:CI}, we exclude such correlation between unobserved payoff states.

Using Assumption \ref{assum:CI}, we can determine the value of $\lambda_{i,\rho}$ for each $\rho \in (0,1)$, once the payoff function is specified. To see this, consider the example of the payoff function differential between actions 1 and 0:
\begin{equation}
	\label{eq:sp1}
	u_i^{\Delta}(1,y_{-i} ;\tau_i) = v_1(X_i;\theta_0)+v_2(X_i;\theta_0) \times \frac{1}{n(i)}\sum_{j \in N(i)} y_j + \eta_i,
\end{equation}
for some parametric functions $v_1$ and $v_2$, and a finite dimensional parameter $\theta_0 \in \Theta \subset \mathbf{R}^d$. The component $v_2(X_i;\theta_0)$ captures the strategic interactions between players. Then,
\begin{align}
	\label{eq:hindsight_regret}
	\lambda_{i,\rho}  = \gamma_i \sqrt{-\frac{\ln \rho}{2}} \qtext{with}\quad \gamma_i = \frac{|v_2(X_i;\theta_0)|}{\sqrt{n(i)}}.
\end{align}
Then,
\begin{align}
	\label{eq:mcdiarmid_bound}
	\begin{aligned}
		&\QR[i]{u_i^{\Delta}(Y_i,Y_{-i};\typ_i) < -\lambda_{i,\rho}  \mid \IF_i} \\
		&\qquad\le \QR[i]{u_i^{\Delta}(Y_i,Y_{-i};\typ_i)-\E_{\QM_i}[u_i^{\Delta}(Y_i,Y_{-i};\typ_i)\mid \IF_i] < -\lambda_{i,\rho}  \mid \IF_i}\\
		&\qquad\le \exp\left( -2\lambda_{i,\rho}^2/\gamma_i^2 \right) = \rho, \quad (\QM_i \text{-a.s.}).
	\end{aligned}
\end{align}
The first inequality follows by the equilibrium constraint, and the second inequality follows from a concentration inequality called McDiarmid's inequality combined with the conditional independence assumption. (See Lemma \ref{lemma:aux_mcdiarmid} in the Supplemental Note.) Hence,
\begin{align}
	\label{hindsight regret}
	\QR[i]{u_i^{\Delta}(Y_i,Y_{-i};\typ_i) \ge -\lambda_{i,\rho}  \mid \IF_i} \ge 1 - \rho, \quad (\QM_i \text{-a.s.}),
\end{align}
confirming that $\lambda_{i,\rho}$ is a $\rho$-hindsight regret.

The second assumption relates the subjective probabilities to the objective one, so that we can translate (\ref{hindsight regret}) into testable restrictions. For each $\rho \in (0,1)$ and player $ i \in N$, we define a collection of events as follows:
\begin{align}
	\label{H(rho)}
	\mathcal{H}_i(\rho) \eqdef \left\{H \in \mathcal{I}_{N(i)}: \QR[i]{H \mid \IF_i} \ge 1 - \rho, \quad (\QM_i\text{-a.s.}) \right\},
\end{align}
where $\mathcal{I}_{N(i)} \eqdef \vee_{j \in N(i)} \mathcal{I}_j$, i.e., the smallest $\sigma$-field generated by $\mathcal{I}_j$, $j \in N(i)$. We define
\begin{align}
	\label{delta rho}
	\delta_i(\rho) \eqdef \sup_{H \in \mathcal{H}_i(\rho)} |\QR[i]{H \mid \IF_i} - \PR{H \mid \IF_i}|,
\end{align}
so that $\delta_i(\rho)$ measures the discrepancy between $\QM_i$ and $\PM$ on the events of which player $i$'s belief is at least $1-\rho$, conditional on his information.\footnote{Here we take $\delta_i(\rho)$ to be the minimal measurable majorant of the supremum on the right hand side when the supremum is not measurable.}
\begin{assumption}
	\label{assum:RE}
	For each $i \in N$, the following two conditions hold:\medskip
	
	(i) for each $H \in \mathcal{H}$ such that $\PR{H}>0$, we have $\QR[i]{H}>0$, and 
	
	(ii) there exists $\rho \in (0,1)$ such that $\delta_i(\rho) = 0, \quad (\PM\text{-a.s.})$.
\end{assumption}
Condition (i) says that any event that is possible in terms of the objective probability is believed to be so by every player. Condition (ii) says that each player $i$ has belief $\QM_i$ such that the discrepancy between the objective and subjective probability measures of high $\QM_i$ events conditional on each player's information set is zero. Hence, Assumption \ref{assum:RE} is substantially weaker than the commonly used rational expectations assumption: $\QM_i=\PM$ for all $i\in N$.\footnote{Our framework can be extended to a setting with a weaker variant of Assumption \ref{assum:RE}(ii), where we require that there exist $\rho \in (0,1)$ and $\kappa \in [0,(1-\rho)/\rho)$ such that the fraction of players $i$ with $\delta_i(\rho) > \kappa \rho$, $(\PM\text{-a.s.})$, is asymptotically negligible. Here, the bound $\kappa \rho$ captures the degree of belief heterogeneity on high probability events that is allowed in the model, and $\kappa$ and $\rho$ are constants that the researcher specifies as part of the model specification. See the Appendix for details.}

To see how these assumptions yield testable implications, let $\rho \in (0,1)$ be the constant in Assumption \ref{assum:RE}, and let $\lambda_{i,\rho}$ be a $\rho$-hindsight regret for player $i$. By Assumption \ref{assum:RE},
\begin{align}
	\label{inequality}
	\begin{aligned}
		\PR{u_i^{\Delta}(Y_i,Y_{-i};\typ_i) \ge -\lambda_{i,\rho}  \mid \IF_i} &\ge \QR[i]{u_i^{\Delta}(Y_i,Y_{-i};\typ_i) \ge -\lambda_{i,\rho}  \mid \IF_i} -\delta_i(\rho) \\
		&\ge 1 - \rho, \quad(\PM\text{-a.s.}).
	\end{aligned}
\end{align}
Later, we use this inequality to perform inference on the payoff parameters. For this, we regard the constant $\rho$ in Assumption \ref{assum:RE} to be part of the researcher's specification on the beliefs, which plays a role analogous to the highest level $K$ in econometric models of level-$k$ thinking, $k = 0,1,...,K$. See, e.g., \cite{An:17}. From our simulations, we find that the inference is robust to a wide range of $\rho = 0.01 \sim 0.00001$. For a practical purpose, we propose to choose $\rho = 0.001$.

While Assumption \ref{assum:RE} relaxes the standard rational expectation assumption substantially, it is not entirely innocuous. To see this, let us consider the following example of the payoff differential:
\[
		u_i^{\Delta}(1,y_{-i};(X_i,\eta_i))=X_i\beta_0+\eta_{i,1}\times \frac{\phi_0}{n(i)}\sum_{j\in N(i)} y_j +\eta_{i,2},
\]
where $\beta_0$ and $\phi_0$ are parameters, $\eta_i \eqdef (\eta_{i,1},\eta_{i,2})$ is an unobserved component such that $\PR{\eta_{i,1} = 1} = 1$, and the distribution of $\eta_{i,2}$ is the same under $\PM$ and $\QM_i$. (This payoff specification is used later in Monte Carlo simulations.) Each player $i$ with information $\mathcal{I}_i$ believes that $\eta_{j,1}$'s are independently drawn from Bernoulli distribution with parameters $q_{n,ij} \in (0,1]$. Under $\QM_i$ and $\PM$, we assume that $\eta_{j,1}$'s and $\eta_{j,2}$'s are independent of each other, both independent of $\X$. Suppose that for a given $\rho \in (0,1)$, the beliefs satisfy
\begin{align}
	\label{eq:Q_inequality}
	\prod_{j \in N(i)} q_{n, ij}>\rho.
\end{align}
Then, for $\boldsymbol{a} = (1,...,1) \in \{0,1\}^{n(i)}$,
\begin{align*}
	\QR[i]{\eta_{N(i),1} \ne \boldsymbol{a}} = 1-\prod_{j \in N(i)} q_{n, ij} < 1 - \rho,
\end{align*} 
where $\eta_{N(i),1} = (\eta_{j,1})_{j \in N(i)}$. If \eqref{eq:Q_inequality} holds for all players, Assumption \ref{assum:RE} is satisfied with this $\rho$. Otherwise, Assumption \ref{assum:RE} may fail.

\subsection{Inference on Large Social Interactions}
\label{subsec: econometric inference}

\subsubsection{Moment Inequalities from Belief-free Hindsight Regrets}
Let us present a method of econometric inference when we observe a single large Bayesian game satisfying Assumptions \ref{assum:CI}-\ref{assum:RE}. First, we focus on a model of large social interactions with the payoff function specified as in \eqref{eq:sp1}. Later in Section \ref{sec:hindsight_regret}, we extend the approach to models with general payoff functions. Throughout the paper, we assume that the researcher knows the strategic neighborhoods $N(i)$, $i\in N$.

The existing approaches in the literature derive moment inequalities from the constraints \eqref{eq:icc} to perform inference on the payoff parameters. We assume that the researcher observes a realization of $\arr{(Y_i,X_i)}_{i\in N}$, where $Y_i$ is the binary action taken by player $i\in N$. The following assumption relates the observed action profile, $\mathcal{Y} = (Y_i)_{i \in N}$, to the underlying game.

\begin{assumption}[\textit{Observed One Equilibrium}]
	\label{assum:A4}
	The observed action profile, $\mathcal{Y}$, is generated from a pure strategy Bayesian equilibrium.
\end{assumption}

To appreciate this assumption in the presence of multiple equilibria, first let $\mathcal{E}$ be the collection of pure strategy Bayesian equilibria. Since each member $e = (e_1,...,e_n) \in \mathcal{E}$ is a pure strategy equilibrium, for each $i \in N$, $e_i$ is $\mathcal{I}_i$-measurable, and we can write $e_i = f_{e,i}(\eta_i, \mathcal{X})$ for some measurable map $f_{e,i}$ taking values in $\{0,1\}$.  (The dependence of this map on the chosen equilibrium $e \in \mathcal{E}$ is made explicit through the subscript $e$ in $f_{e,i}$.) For each $e \in \mathcal{E}$, we let $f_e = (f_{e,i})_{i \in N}$. Then the action profile $\mathcal{Y} = (Y_i)_{i \in N}$ that the researcher observes is generated as follows:
\begin{align}
	\label{Y}
	\mathcal{Y} =  f_{e_0}(\tau),
\end{align}
where $e_0$ is a point in $\mathcal{E}$, and $\tau = (\eta_i, \mathcal{X})_{i \in N}$. Note that for each $i \in N$, we have $Y_i = f_{e_0,i}(\eta_i,\mathcal{X})$. Hence, Assumptions \ref{assum:CI} and \ref{assum:A4} imply that for the given equilibrium $e_0$, $Y_i$'s are conditionally independent across $i$'s given $\mathcal{X}$. We allow that the researcher does not  know which equilibrium in the game the observed outcomes are generated from, that is, the researcher does not know $e_0$. 

The conditional independence of $Y_i$'s is not testable in our framework. The main reason is that $Y_i$'s can be heterogeneously distributed, due to the heterogeneity of beliefs. Hence, there is no way we can recover even the marginal distribution of $Y_i$ from only one sample from the large game.\footnote{For this reason, although we may obtain a sharp identified set, for example, by following the proposal in \cite{Chesher/Rosen:17:Eca}, we cannot consistently estimate this identified set, because we observe only one sample from the joint distribution of $\{(Y_i,X_i): i \in N\}$ in our large game setting, and the distributions of $(Y_i,X_i)$'s are all potentially different across $i$'s due to the heterogenous beliefs.}

\begin{assumption}[\textit{Parametric Specification}]
	\label{assum:A5}
	For each $i\in N$, and $x$ in the support of $X_i$,
	\begin{align}
		\label{eq:piL_piU}
		\PR{\eta_i\le \csdot\mid X_i=x}&=F_{\theta_0}(\csdot\mid x),
	\end{align}
	for some parameter $\theta_0 \in \Theta$, where $F_{\theta}$ is a parametric distribution function with a quasi-concave density function.\footnote{Without loss of generality, we use the same parameter $\theta_0$ and the parameter space $\Theta$ for all parametrized quantities throughout the paper. This loses no generality, because we can expand the parameter space to unify idiosyncratic incidences in which each parameter appears in a parametrization.} 
\end{assumption}

Assumption \ref{assum:A5} states that the conditional CDF of $\eta_i$ given $X_i$ and the payoff function are parameterized by $\theta_0 \in \Theta$. The assumption of the quasi-concavity of the density function is made only to facilitate the explicit expression of certain quantities used for inference later. This assumption is satisfied by many parametric distributions including normal or logistic distributions.

Let us derive moment inequalities from \eqref{eq:hr_def0}. First, we let $\lambda_{i,\rho}$ be the $\rho$-hindsight regret defined in \eqref{eq:hindsight_regret}, where $\rho$ is the constant satisfying Assumption \ref{assum:RE}. We introduce the following probabilities: 
\begin{align*}
	\pi_{i,L}(\theta_0) &= F_{\theta_0}\Bigg(-v_{1}(X_i;\theta_0)- v_{2}(X_i;\theta_0) \times \frac{1}{n(i)}\sum_{j\in N(i)}Y_j +\lambda_{i,\rho}  \mid X_i \Bigg) \qtext{and} \\
	\pi_{i,U}(\theta_0) &=1 - F_{\theta_0}\Bigg(-v_{1}(X_i;\theta_0)- v_{2}(X_i;\theta_0) \times \frac{1}{n(i)}\sum_{j\in N(i)}Y_j - \lambda_{i,\rho}  \mid X_i\Bigg),
\end{align*}
and define pseudo residuals:
\begin{align}
	\label{eq:sample_e}
	\begin{aligned}
		e_{i,L}(\theta_0) \eqdef Y_i - \mleft(1-\frac{\pi_{i,L}(\theta_0)}{1-r_i(\theta_0) }\mright) \qtext{and}\quad e_{i,U}(\theta_0) \eqdef Y_i - \frac{\pi_{i,U}(\theta_0)}{1-r_i(\theta_0)},
	\end{aligned}
\end{align}
where $r_i(\theta_0) \eqdef \rho \cdot \ind\{\lambda_{i,\rho}  >0\}$.\footnote{Note that we allow for the possibility that the true parameter value $\theta_0$ is such that $v_2(X_i;\theta_0)=0$. In such a case, there is no strategic interaction between players, and hence, $\lambda_{i,\rho} = 0$.} Then, we obtain the following moment inequalities. (The result follows from a general result (Proposition \ref{prop:pA1}) which is found and proved in the Appendix.)
\begin{theorem}
	\label{thm: test implications}
	Suppose that Assumptions \ref{assum:CI}-\ref{assum:A5} are satisfied. Then, for each $i \in N$,
	\begin{align*}
		\E_{\PM}\left[e_{i,L}(\theta_0) \mid \X \right] \ge 0 \qtext{ and }\quad \E_{\PM}\left[e_{i,U}(\theta_0) \mid \X\right] \le 0, \quad (\PM\text{-a.s.}),
	\end{align*}
	where $\E_{\PM}[\csdot \mid \X]$ denotes the conditional expectation given $\X$ under $\PM$.
\end{theorem}
In general, the inequality restrictions in Theorem \ref{thm: test implications} tend to become tighter when $\lambda_{i,\rho}$ becomes smaller, i.e., the strategic relevance of the players among each other is weaker. This is a cost to the researcher for not being able to observe the beliefs of individual players in the presence of strong strategic interactions among them. Note that when there is no strategic interaction, we have $\lambda_{i,\rho}=0$.

\subsubsection{Bootstrap Inference}

Let us consider how we can use the moment inequalities in Theorem \ref{thm: test implications} to develop an inference procedure on $\theta_0$.\medskip

\textit{A. Constructing Sample Moments:.} To construct sample moments, we choose a vector of non-negative measurable functions $\vec{g}_i\eqdef [g_{i,1},\ldots, g_{i,m}]^{\top}:\R^{d_X} \to [0,\infty)^m$, with $d_X$ denoting the dimension of $X_i$, and construct the following sample moments in a spirit similar to \cite{Andrews/Shi:13}:
\begin{align}
	\label{eq:sample_moments}
	\begin{aligned}
		\vecg{\hat{\mu}}_L(\theta_0) &\eqdef \frac{1}{n}\sum_{i \in N} e_{i,L}(\theta_0)\,\vec{g}_{i}(X_i) \qtext{and} \\
		\vecg{\hat{\mu}}_U(\theta_0) &\eqdef \frac{1}{n}\sum_{i \in N} e_{i,U}(\theta_0)\,\vec{g}_{i}(X_i).
	\end{aligned}
\end{align}

Although the sample moments in \eqref{eq:sample_moments} are similar to those employed in the literature of moment inequalities \cite[see, e.g.,][]{Rosen:08,Andrews/Soares:10,Andrews/Shi:13}, they are not necessarily sums of independent or conditionally independent random variables. The summands $e_{i,L}(\theta_0) \vec{g}_i(X_{i})$ and $e_{i,U}(\theta_0) \vec{g}_i(X_{i})$ involve $Y_{-i}$ so that they are dependent across $i$'s in a complicated manner. On the other hand, the moments, $\vecg{\tilde{\mu}}_L(\theta_0)$ and $\vecg{\tilde{\mu}}_U(\theta_0)$, which are defined as $\vecg{\hat{\mu}}_L(\theta_0)$ and $\vecg{\hat{\mu}}_U(\theta_0)$ in \eqref{eq:sample_moments} except that $\pi_{i,L}(\theta_0)$ and $\pi_{i,U}(\theta_0)$ are replaced by their conditional expectations given $\X$, are sums of conditionally independent random variables but infeasible to construct using data. In other words, the moments $\vecg{\hat{\mu}}_L$ and $\vecg{\hat{\mu}}_U$ are feasible, yet hard to derive their limiting distribution, while the moments $\vecg{\tilde{\mu}}_L$ and $\vecg{\tilde{\mu}}_U$ facilitate asymptotic analysis, yet are infeasible. Thus, we modify the sample moments as we explain next.\medskip

\textit{B. Constructing a Test Statistic:} We construct our test statistic as follows. We first introduce notation. For a vector $x = [x_j] \in \R^d$, we denote $[x]_{+}\eqdef [x_j\vee 0]_{j=1}^d$, $[x]_{-}\eqdef -[x_j\wedge 0]_{j=1}^d$, and $\|x\|_1 = \sum_j |x_j|$. We take our test statistic to be of the following form:
\begin{align}
	\label{eq:T}
	T(\theta_0) \eqdef \left\|\left[\sqrt{n}(\vecg{\hat{\mu}}_L(\theta_0)+\vec{w}_L(\theta_0)) \right]_- + \left[\sqrt{n}(\vecg{\hat{\mu}}_U(\theta_0)-\vec{w}_U(\theta_0))\right]_+ \right\|_1,
\end{align}
where $\vec{w}_L(\theta_0),\vec{w}_U(\theta_0)\in [0,\infty)^m$ are some non-negative random vectors chosen to satisfy that
\begin{align}
	\label{bound3}
	T(\theta_0) \le \left\|\left[\sqrt{n}\vecg{\tilde{\mu}}_L(\theta_0) \right]_- + \left[\sqrt{n}\vecg{\tilde{\mu}}_U(\theta_0) \right]_+ \right\|_1
\end{align}
with high probability, say, $1 - \nu$ for a small number $\nu>0$, e.g., $\nu = 0.01$ or $\nu = 0.001$. In many applications, given the parametric payoff functions, we can derive an explicit form of $\vec{w}_L(\theta_0),\vec{w}_U(\theta_0)\in [0,\infty)^m$. The explicit form in this case of a payoff function \eqref{eq:sp1} is provided in the Appendix.\medskip

\textit{C. Finding a Bootstrap Critical Value:} To complete our inference procedure using test statistic $T$ in \eqref{eq:T}, we propose a bootstrap critical value by adapting the idea of \cite{Romano/Shaikh/Wolf:14} to our set-up. First, we draw i.i.d.\ standard normal random variables $\{\varepsilon_1,\ldots,\varepsilon_n\}$ and define
\[
	\vecg{\zeta}^{*}(\theta_0)\eqdef \frac{1}{n}\sum_{i \in N}(Y_i-\mu_i^{*}(\theta_0))\vec{g}_i(X_i)\,\varepsilon_i,
\]
where
\[
\mu_i^{*}(\theta_0)\eqdef \left(\frac{1}{2}\mleft(1-\frac{\pi_{i,L}(\theta_0)-\pi_{i,U}(\theta_0)}{1-r_i(\theta_0)}\mright) \vee 0 \right) \wedge 1.
\]
Since we are unable to consistently estimate the conditional expectation of $Y_i$ given $\X$, the random variable $\mu_i^{*}(\theta_0)$ serves as its proxy. In addition, we let
\begin{align*}
	\vecg{\hat\varphi}_L(\theta_0)&\eqdef \mleft[\vecg{\hat{\mu}}_L(\theta_0)-\vec{w}_L(\theta_0)- n^{-1/2} \ind_m\cdot q_{1-\nu/2}^{*}(\theta_0)  \mright]_{+} \qtext{and} \\
	\vecg{\hat\varphi}_U(\theta_0)&\eqdef \mleft[\vecg{\hat{\mu}}_U(\theta_0)+\vec{w}_U(\theta_0)+ n^{-1/2} \ind_m\cdot q_{1-\nu/2}^{*}(\theta_0) \mright]_{-},
\end{align*}
where $q_{1-\nu/2}^{*}(\theta_0)$ is the $(1-\nu/2)$ quantile of the bootstrap distribution of $\sqrt{n}\|\vecg{\zeta}^{*}(\theta_0)\|_\infty$, i.e., the sup-norm of the vector $\sqrt{n}\vecg{\zeta}^{*}(\theta_0)$, and $ \ind_m$ is the $m$-dimensional vector of ones.

For critical values, we consider the following bootstrap test statistic:
\[
	T^{*}(\theta_0) \eqdef \left\|\left[\sqrt{n}(\vecg{\zeta}^{*}+\vecg{\hat\varphi}_L\wedge \vecg{\hat\varphi}_U)(\theta_0)\right]_- + \left[\sqrt{n}(\vecg{\zeta}^{*}-\vecg{\hat\varphi}_L\wedge \vecg{\hat\varphi}_U)(\theta_0)\right]_+ \right\|_1,
\]
where the minimum between $\vecg{\hat\varphi}_L$ and $\vecg{\hat\varphi}_U$ is taken element-wise. Let $c_\gamma^{*}(\theta)$ is the $\gamma\eqdef (1-\alpha+2\nu)$-quantile of the bootstrap distribution of $T^{*}(\theta)$. The tuning parameter $\nu$ should obviously satisfy $\nu<\alpha/2$ and can be chosen via a Monte Carlo study. The choice of $\nu$ does not affect the asymptotic validity of the bootstrap inference, as long as it is fixed to be independent of $n$. For example, for $\alpha = 0.05$, one may choose $\nu = 0.01$; our Monte Carlo simulation study shows a reasonable finite sample behavior for such a choice.\medskip

\textit{D. Constructing a Confidence Region:} Equipped with the test statistic $T(\theta)$ and the bootstrap critical value $c_\gamma^*(\theta)$ for each $\theta \in \Theta$, we construct the bootstrap-based confidence set for $\theta_0\in\Theta$ at nominal level $1-\alpha$ by test-inversion. That is, the confidence set is constructed as follows:
\[
	CS_\epsilon \eqdef \left\{\theta\in\Theta : T(\theta) \le c_\gamma^{*}(\theta) \vee \epsilon \right\},
\]
where $\epsilon>0$ is a fixed small number. (We introduce $\epsilon>0$ here for bootstrap critical values to ensure uniform validity because the statistic $T$ can take the value of zero with a positive probability.) Below we establish that this confidence region is uniformly valid in the collection of probabilities that satisfy Assumptions \ref{assum:CI}, \ref{assum:RE}, and \ref{assum:A5} and an additional assumption that requires that the sample moments are not multicollinear.

\subsubsection{Uniform Validity of Bootstrap Inference}
Let $\PS_0$ be a family of objective probability measures $\PM$ on $(\Omega,\H)$ satisfying Assumptions \ref{assum:CI}, \ref{assum:RE}, and \ref{assum:A5}. Let 
\begin{align}
	\label{eq:zeta}
	\vecg{\zeta}_i \eqdef (Y_i-\E_{\PM}[Y_i\mid \X]) \mathbf{g}_i(X_i).
\end{align} 
We introduce the following technical assumption on the minimum eigenvalue of $\E_{\PM}\left[\vecg{\zeta}_i \vecg{\zeta}_i^{\top}\mid \X \right]$.

\begin{assumption}
	\label{assump: nondeg}
    There exists a positive sequence $\{r_n\}$ such that $r_n \rightarrow 0$, $n^{1/6} r_n \rightarrow \infty$, and
    \begin{align}
    	\label{eq:nondeg}
    	\lim_{n\to\infty}\sup_{\PM\in \PS_0}\PR{\min_{1 \le i \le n} \lambda_{\min}\left(\Sigma_i \right) < r_n}=0,
    \end{align}
    where $\lambda_{\min}\left(\Sigma_i \right)$ is the smallest eigenvalue of $\Sigma_i \eqdef \E_{\PM}\left[\vecg{\zeta}_i \vecg{\zeta}_i^{\top}\mid \X \right]$.
\end{assumption}
The condition \eqref{eq:nondeg} prevents the conditional variance $\Sigma_i$ from being degenerate fast as $n \rightarrow \infty$. This condition requires that the components of $\mathbf{g}_i(X_i)$ are not multicollinear.

The following theorem establishes the uniform validity of the bootstrap confidence set. (See Theorem \ref{thm:bootstrap_consisteny_mlt} for its version in a more general setting, which is presented and proved in the Appendix.)
\begin{theorem}
	\label{thm:bootstrap_consistency}
	Suppose that Assumptions \ref{assum:CI}-\ref{assump: nondeg} hold, and there exists $C_{\vec{g}}>0$ such that
	\begin{align}
		\label{eq:bound_Cg}
		\max_{i\in N}\max_{1\le \ell \le m} \sup_{x \in \mathbf{R}^{d_X}} |g_{i,\ell}(x)| \le C_{\vec{g}},
	\end{align}
	for all $n \ge 1$. Then,
	\[
		\liminf_{n\to\infty}\inf_{\PM\in \PS_0}\PR{\theta_0 \in CS_\epsilon}\ge 1 - \alpha.
	\]
\end{theorem}
The condition \eqref{eq:bound_Cg} is satisfied by many choices of $g_{i,\ell}$ such as indicator functions. While we can relax this condition, we do not believe it adds much to the value of the contribution of this paper. It is important to note that the uniform validity holds regardless of whether the strategic neighborhoods are large or small.

\subsection{Extension to General Payoff Functions}
\label{sec:hindsight_regret}
\subsubsection{Belief-Free Hindsight Regrets}
In this section, we extend our approach to a general payoff function: $u_i(y_i,y_{-i};\tau_i)$. Let us first introduce a generalized version of the hindsight regret in \eqref{eq:hindsight_regret}. For a real function $f:\{0,1\}^n\to \R$ on action profiles of players, and for each player $j \in N$, define
\begin{equation}
\label{eq:max_var}
	V_j(f)\eqdef \sup_{(y_1,...,y_n)\in \{0,1\}^n,y'\in \{0,1\}}\abs{f(y_1,...,y_n)-f(y_1,\ldots, y_{j-1},y',y_{j+1},\ldots,y_n)}.
\end{equation}
We call $V_j(f)$ the \textit{maximal variation of $f$ due to player $j$}. In order to characterize a belief-free hindsight regret, we let
\begin{equation}
\label{eq:hr0}
	\lambda_i(\tau_i) \eqdef \sqrt{-\frac{\ln \rho}{2}\cdot \Lambda_i(\tau_i)}, \qtext{where}\quad \Lambda_i(\tau_i)\eqdef \sum_{j\in N\setminus \{i\}}V_{j}^2(u_i^{\Delta}(1,\csdot;\tau_i)),
\end{equation}
and $u_i^\Delta$ is given in \eqref{eq:u_i^Delta}.\footnote{Here $u_i^\Delta(1,y_{-i}; \tau_i)$ is viewed as a function of $y_1,...,y_n$ (constant in the first argument) and the maximal variation $V_j(u_i^\Delta(1,\cdot; \tau_i))$ is with respect to $y_j$, i.e., the action of player $j$, \textit{not} the $j$-th entry of $y_{-i}$.} Note that $V_{j}(u_i^{\Delta}(1,\csdot;\tau_i))$ measures the largest variation in the player $i$'s payoff differential $u_i^{\Delta}$ between actions $1$ and $0$ which can be caused by player $j$'s arbitrary choice of action. The function $\Lambda_i(\tau_i)$ in \eqref{eq:hr0} measures the overall strategic relevance of other players to player $i\in N$.

The hindsight regret increases with strategic interdependence among the players. This is intuitive because player $i$'s \textit{ex post} payoff loss due to the inability to fully observe the other players' types is large when actions by those players can have a large impact on player $i$'s payoff. 

\subsubsection{Inference}
\label{sec:econ_obs}
As before, we assume that the researcher observes $\arr{(Y_i,X_i)}_{i\in N}$, where $(Y_i)_{i \in N}$ is a pure strategy Bayesian equilibrium from the game, and $X_i$ is the vector of observable covariates of that player. We also assume that the payoff state $\typ_i$ is specified as $\typ_i = (X_i,\eta_i)$, where $\eta_i$ is a payoff component unobserved by the researcher. As for the payoff function, we consider a general parametric payoff function as follows.

\begin{assumption}[\textit{Parametric Specification}]
	\label{assum:A6}
	For each $i\in N$,
	\begin{align*}
		u_i(\csdot,\csdot; \csdot) = u_{i,\theta_0}(\csdot,\csdot; \csdot),
	\end{align*}
	for some $\theta_0$ in a parameter space $\Theta\subset \R^d$, where $u_{i,\theta}(\csdot,\csdot; \csdot)$ is a parametric function.
\end{assumption}

To construct testable implications, define: for $y_{-i} = (y_j)_{j \ne i}$, $y_j \in \{0,1\}$,
\begin{align}
	\label{eq:pi_LU}
	\begin{aligned}
		\pi_{i,L}(y_{-i},X_i;\theta_0)&\eqdef \int \mathbf{1}\left\{u_{i}^{\Delta}(0,y_{-i};\typ_i,\theta_0) \ge -\lambda_{i,\rho}(X_i, \overline \eta;\theta_0)\right\} dF_{\theta_0}(\overline \eta \mid X_i) \qtext{ and } \\
		\pi_{i,U}(y_{-i},X_i;\theta_0)&\eqdef \int \mathbf{1}\left\{u_{i}^{\Delta}(1,y_{-i};\typ_i,\theta_0) \ge -\lambda_{i,\rho}(X_i, \overline \eta;\theta_0)\right\} dF_{\theta_0}(\overline \eta \mid X_i),
	\end{aligned}
\end{align}
where $u_{i}^{\Delta}(0,y_{-i};\typ_i,\theta_0) $ and $\lambda_{i,\rho}(X_i, \overline \eta;\theta_0)$ are the same as $u_{i}^{\Delta}(0,y_{-i};\typ_i) $ and $\lambda_{i,\rho}(X_i, \overline \eta)$; we make explicit their dependence on $\theta_0$ for later use. These probabilities are explicitly known in many settings (see Example \ref{example: Bayesian Game} below), or at least can be simulated from the parametric distribution of $\eta_i$ in Assumption \ref{assum:A5}. 

Let us choose a vector of non-negative functions $\vec{g}_i\eqdef [g_{i,1},\ldots, g_{i,m}]^{\top}:\R^{d_X}\to [0,\infty)^m$ as before. We let 
\begin{align}
	\label{eq:sample_e2}
	\begin{aligned}
		e_{i,L}(\theta_0) &\eqdef Y_i-\mleft(1-\frac{\pi_{i,L}(Y_{-i},X_i;\theta_0)}{1-r_i(\theta_0)}\mright) \qtext{and} \\
		e_{i,U}(\theta_0) &\eqdef Y_i-\frac{\pi_{i,U}(Y_{-i},X_i;\theta_0)}{1-r_i(\theta_0)},
	\end{aligned}
\end{align}
where
\begin{align}
	\label{ri theta0}
	r_i(\theta_0)\eqdef \rho \cdot \ind\left\{\sup_{\overline \eta}{\lambda_{i,\rho}(X_i,\overline \eta;\theta_0)}>0 \right\}.
\end{align}
Then we can show that for all $i \in N$,
\begin{align*}
	\E_{\PM}\left[e_{i,L}(\theta_0) \mid \X \right] \ge 0 \qtext{and}\quad \E_{\PM}\left[e_{i,U}(\theta_0) \mid \X \right] \le 0, \quad (\PM\text{-a.s.}).
\end{align*}
We construct the sample moments as follows:
\begin{align*}
	\vecg{\hat{\mu}}_L(\theta_0) \eqdef \frac{1}{n}\sum_{i \in N} e_{i,L}(\theta_0) \vec{g}_{i}(X_i) \qtext{and}\quad
	\vecg{\hat{\mu}}_U(\theta_0) \eqdef \frac{1}{n}\sum_{i \in N} e_{i,U}(\theta_0) \vec{g}_{i}(X_i),
\end{align*}
and consider the following as our test statistic:
\begin{align*}
	T(\theta_0)\eqdef \left\|\left[\sqrt{n}(\vecg{\hat{\mu}}_L+\vec{w}_L)(\theta_0)\right]_- + \left[\sqrt{n}(\vecg{\hat{\mu}}_U-\vec{w}_U)(\theta_0)\right]_+ \right\|_1,
\end{align*}
where $\vec{w}_L(\theta_0),\vec{w}_U(\theta_0) \in [0,\infty)^m$ are non-negative random vectors motivated similarly as before. Details on these random vectors are found in the Appendix. Having constructed $\pi_{i,U}$ and $\pi_{i,L}$ and the quantities $\vec{w}_L$ and $\vec{w}_U$, we can proceed precisely as before to perform the bootstrap inference.

\begin{example}[Bayesian Game with Large Intersecting Reference Groups]
	\label{example: Bayesian Game}

Let us illustrate the flexibility of our approach by considering a large private information game with large intersecting reference groups. First, let $N_g \subset N$, $g \in G$, for some finite index set $G$, where each set $N_g$ represents a social or demographic group as a reference group. The groups may intersect, as each player may belong to multiple reference groups simultaneously. For example, $N_1$ may represent a high education group and $N_2$ represent a high income group so that $N_1 \cap N_2$ represents the set of those people with high education and income. The neighborhoods are such that the average of the actions by players in each group affects the payoff of the players in the group. 

More specifically, we consider the payoff differential of the following form:
\begin{align}
	\label{eq:sp3}
	u_i^\Delta(1,y_{-i};X_i,\eta_i,\theta_0) &=v_1(X_i;\theta_0)+\frac{v_2(X_i;\theta_0)}{\abs{G_i}}\sum_{g\in G_i}\mleft(\frac{1}{n_g-1}\sum_{j\in N_g\setminus \{i\}} y_j\mright) + \eta_i,
\end{align}
where the strategic neighborhood of player $i$ is given by $N(i) = \bigcup_{g \in G_i} N_g \setminus\{i\}$, $v_1$ and $v_2$ are parametric functions, $n_g = |N_g|$, and $G_i$ is the set of group indices that player $i$ belongs to.\footnote{This model with large intersecting reference groups can be viewed as a special case of a game on a network, where each strategic neighborhood defines a neighborhood in the network. (See \cite{Jackson/Zenou:15:Handbook} for a review of games on networks in economic theory. See also \cite{Bramoulle/Djebbari/Fortin:20:ARE} and references for a review of applications of games on networks for the study of peer effects.) Here, we have in mind a situation where there are many players in each reference group, and each player belongs to multiple reference groups differently from many other players.}

From the payoff specification \eqref{eq:sp3}, we observe that for $i,j\in N$ such that $i\ne j$,
\[
	V_{j}(u_i^{\Delta}(1,\csdot;X_i, \eta_i,\theta_0)) = \frac{\abs{v_2(X_i;\theta_0)}}{\abs{G_i}}\sum_{g\in G_i}\frac{\ind\{j\in N_g\}}{n_g-1}.
\]
Since only those players who belong to at least one of player $i$'s strategic neighborhoods are strategically relevant, we find from \eqref{eq:hr0} the hindsight regret as follows:
\[
	\lambda_{i,\rho}(X_i,\eta_i;\theta_0)= \abs{v_2(X_i;\theta_0)}\sqrt{-\frac{\ln\rho}{2}\sum_{j\in N\setminus \{i\}}\mleft(\frac{1}{\abs{G_i}}\sum_{g\in G_i}\frac{\ind\{j\in N_g\}}{n_g-1}\mright)^2}.
\]
Hence, players with large strategic neighborhoods tend to have negligible hindsight regrets.

As for the probabilities $\pi_{i,L}$ and $\pi_{i,U}$ in \eqref{eq:pi_LU}, we have the following explicit form:
\begin{align*}
		\pi_{i,L}(y_{-i},X_i;\theta_0)&=F_{\theta_0}\Bigg(-u_i^\Delta(1,y_{-i};X_i,\eta_i,\theta_0) + \lambda_{i,\rho}(X_i,\eta_i;\theta_0)  \mid X_i \Bigg) \qtext{and} \\ \notag
	\pi_{i,U}(y_{-i},X_i;\theta_0)&=1 - F_{\theta_0}\Bigg(-u_i^\Delta(1,y_{-i};X_i,\eta_i,\theta_0) - \lambda_{i,\rho}(X_i,\eta_i;\theta_0) \mid X_i\Bigg).
\end{align*}
With these definitions of $\lambda_{i,\rho}$, $\pi_{i,L}$, and $\pi_{i,U}$, we can proceed to construct bootstrap-based confidence intervals for the parameter $\theta_0$.

\end{example}

\section{Monte Carlo Simulations}
\label{sec:monte_carlo}

\subsection{Data Generating Process}

We consider a private information Bayesian game on a network, where the information group of a player consists of his direct neighbors. The underlying network is an undirected network constructed as follows. For a given sample size $n$, we randomly sample $n$
points, $\{U_1, \ldots , U_n\}$, from the uniform distribution on $[0, 1]^2$. These points represent the nodes of a random graph. Two nodes $i$ and $j$ become connected with probability that is inversely proportional to the Euclidean distance between $U_i$ and $U_j$, that is,
\[
    \PR{i\leftrightarrow j\mid U_i,U_j}=\exp\left(-\|U_i-U_j\|_2\sqrt{2\pi n/\delta}\right),
\]
where $\delta$ is a positive constant that determines the average degree of the resulting graph. The scale $2\pi$ is used to make $\delta$ roughly match to the average degree of the generated networks. The random graphs generated this way are called random geometric graphs \citep[see, e.g.,][]{Penrose:03:RandomGeometricGraphs}. Its many variants have been used as a network formation model in the literature \citep[see, e.g.,][and references therein]{Leung:19:JOE}. The specification above follows the design in the simulation study of \cite{Kojevnikov/Marmer/Song:JOE:2020}. We take $n\in \{1000,3000,5000\}$ and $\delta\in \{50,100,150\}$. Table \ref{tbl:net_stats} shows the average degree of the generated graphs.

\bigskip

\begin{table}[t]
\renewcommand{\arraystretch}{1.2}	
\aboverulesep = 0pt
\belowrulesep = 0pt
\caption{Statistics on the Degree Distribution of the Simulated Networks (Mean, \thsup{25}, \thsup{50}, and \thsup{75} Quantiles).}
\label{tbl:net_stats}
\begin{tabular}{c|cccc|cccc|cccc}
	\toprule
	\multirowthead{2}{$n$} & \multicolumn{4}{c|}{$\delta=50$}  & \multicolumn{4}{c|}{$\delta=100$} & \multicolumn{4}{c}{$\delta=150$} \\
	\cmidrule{2-13}
	& \text{Mean} & \thsup{25} & \thsup{50} & \thsup{75} & \text{Mean} & \thsup{25} & \thsup{50} & \thsup{75} & \text{Mean} & \thsup{25} & \thsup{50} & \thsup{75} \\
	\midrule\midrule
	1000 & 39.6 & 33 & 40 & 46 & 69.7 & 57 & 71 & 83 & \phantom{1}97.0 & \phantom{1}80 & \phantom{1}99 & 115 \\
	3000 & 44.0 & 38 & 45 & 51 & 82.7 & 70 & 86 & 97 & 117.7 & \phantom{1}99 & 122 & 138 \\
	5000 & 45.0 & 39 & 46 & 52 & 85.9 & 76 & 89 & 98 & 125.2  & 107 & 131 & 146 \\
	\bottomrule
\end{tabular}
\bigskip
\parbox{6.2in}{\footnotesize \medskip \medskip
	Notes: This table reports some statistics on the distribution of the simulated networks. The parameter $\delta$ controls the density of the networks used in the simulation study.}
\end{table}

For each $i \in N$, we let $N(i)$ be the set of players $j$ such that $i$ is adjacent to $j$ in the network generated as above. The action space of each player is $\{0,1\}$, and the payoff differential of player $i$ is given by
\begin{equation}
\label{eq:payoff_sim}
	u_i^{\Delta}(1,y_{-i};(X_i,\eta_i))=X_i\beta_0+\eta_{i,1}\times \frac{\phi_0}{n(i)}\sum_{j\in N(i)} y_j +\eta_{i,2},
\end{equation}
where $|\phi_0|<1$ and $\PR{\eta_{i,1}=1}=1$. This payoff specification is often used in the literature of social interactions, where $\phi_0$ measures the magnitude of interactions. Equation \eqref{eq:payoff_sim} implies the following form of belief-free hindsight regret:
\begin{equation}
\label{eq:HB2}
	\lambda_{i,\rho}=\abs{\phi_0}\sqrt{-\frac{\ln\rho}{2n(i)}}.
\end{equation}
(Note that player $i$ observes $\eta_{i,1}$ which realizes as 1 with probability one.) We specify the observed part of the type of player $i$ as follows:
\begin{equation}
\label{eq:XW_sim}
	X_i=Z_i+\frac{1}{n(i)}\sum_{j\in N(i)} Z_j -0.2.
\end{equation}
The random variables $Z_i$ and $\eta_{i,2}$ are drawn independently from $\ND{0}{1}$, and $\arr{(Z_i, \eta_{i,2}):i\in N}$ are independent across the players.

In our simulation design, the beliefs of the players differ from the objective probability. Specifically, for $j\in N(i)$, we set $\QM_i(\eta_{j,2}\le x\mid \IF_i)=\PR{\eta_{j,2}\le x\mid \IF_i}$, $x \in \mathbf{R}$, and
\[
    \QM_i(\eta_{j,1}=1\mid \mathcal{I}_i)=\begin{cases}
        1, & X_i< -1 \text{ or } X_i \wedge X_j \ge -1, \\
        q_{ij}, & X_i\ge -1 \text{ and } X_j < -1,
    \end{cases}
\]
$\QM_i(\eta_{j,1}=0\mid \mathcal{I}_i)=1-\QM_i(\eta_{j,1}=1\mid \mathcal{I}_i)$, where $\{q_{ij}:j\in N(i)\}$ are drawn i.i.d. from $\operatorname{Beta}(8,2)$ at each iteration of Monte Carlo simulations to reduce the dependence of the results on the particular realizations of the beliefs. Let $N'\eqdef\{i\in N: X_i<-1\}$ and $N'(i)\eqdef \{j\in N(i):X_j<-1\}$. To generate equilibrium outcomes, we draw realizations of $\{X_i:i\in N\}$, using \eqref{eq:XW_sim}, and $\{q_{ij}:i\in N\setminus N', j\in N'(i)\}$ from the corresponding distribution, and find a solution to the following system of equations for $\{s_i:i\in N\}$:
\begin{align*}
	s_i&=\Phi\left(X_i\beta_0+\frac{\phi_0}{n(i)}\sum_{j\in N(i)}s_j\right),\quad i\in N', \text{ and} \\
	s_i&=\Phi\left(X_i\beta_0+\frac{\phi_0}{n(i)}\left\{\sum_{j\in N'(i)}\left[q_{ij}s_j+(1-q_{ij})\Phi(X_j\beta_0)\right]+\sum_{j\in N(i)\setminus N'(i)}s_j\right\}\right), \quad i\in N\setminus N',
\end{align*}
where $\Phi(\csdot)$ denote the standard normal CDF. Finally, we set
\[
	Y_i=1\left\{X_i\beta_0+\frac{\phi_0}{n(i)}\sum_{j\in N(i)}s_j+\eta_{i,2}\ge 0\right\}.
\]

For the construction of the moment inequalities, we use the following functions:
\begin{align*}
	&g_1(x)= 1, \quad g_2(x)=\sqrt{2}\,\ind\{x\ge 0\}, \\
	&g_3(x)=2\arctan(\abs{x}), \quad g_4=g_2\times g_3.
\end{align*}%
Throughout the simulation studies, we choose $\nu =0.01$ and $\rho=0.001$. The Monte Carlo simulations number is set to $5000$. Note that our framework permits a fraction of players to have beliefs such that $\delta_i(\rho) > 0$. The average fractions for different $\delta$'s are shown in Table \ref{tbl:frac_out}. As we shall see later, our Monte Carlo study shows that the finite sample validity of our bootstrap inference is not affected by that. (Additional simulation results for different values of $\rho$ are found in the Supplemental Note.)

\begin{table}[t]
\renewcommand{\arraystretch}{1.2}	
\aboverulesep = 0pt
\belowrulesep = 0pt
\caption{Average Fraction of Players Such That $\delta_i(\rho) > 0$ (in Percentages).}
\label{tbl:frac_out}
\begin{tabular}{c|SSS}
\toprule
$n$ & {$\delta=50$} & {$\delta=100$} & {$\delta=150$} \\
\midrule\midrule
1000 & 1.5 & 8.6 & 17.5 \\
3000 & 2.2 & 12.7 & 24.2 \\
5000 & 2.3 & 13.7 & 26.6 \\
\bottomrule
\end{tabular}
\bigskip
\parbox{6.2in}{\footnotesize \medskip \medskip
	Notes: This table shows the fraction of players such that $\delta_i(\rho) > 0$ in percentages. The fraction is nonnegligible when $\delta = 150$, i.e., the average neighborhood size is larger. However, as shown in Table \ref{tbl:sim1} below, the coverage probabilities do not show any deterioration.}
\end{table}

\subsection{Finite Sample Coverage Probabilities of the Bootstrap Test}

We first investigate the finite sample validity of the confidence sets. For this study, we choose $\phi_{0}$ from $\{0,1/4\}$ and $\beta_0 = 0.5$. Since the belief-free hindsight regret in \eqref{eq:HB2} is increasing in $\phi _{0}$, we expect that as $\phi _{0}$ moves away from zero, the hindsight regret increases, sending the moment inequalities away from being binding, and the confidence set becomes more conservative. The main interest here is to investigate how conservative the confidence set becomes in finite samples.

\begin{table}[t]
\renewcommand{\arraystretch}{1.2}	
\aboverulesep = 0pt
\belowrulesep = 0pt
\caption{Finite Sample Coverage Probabilities at the Nominal Level of $95\%$.}
\label{tbl:sim1}
\begin{tabular}{cc|ccc}
\toprule
& & $\phi_0=0$ & $\phi_0=1/4$ \\
\midrule\midrule
$n=1000$ & $\delta=50\phantom{0}$ & 0.956 & 1.000 \\
 & $\delta=100$ & 0.957 & 1.000 \\
 & $\delta=150$ & 0.955 & 1.000 \\
\midrule
$n=3000$ & $\delta=50\phantom{0}$ & 0.956 & 1.000 \\
 & $\delta=100$ & 0.955 & 1.000 \\
 & $\delta=150$ & 0.957 & 1.000 \\
\midrule
$n=5000$ & $\delta=50\phantom{0}$ & 0.958 & 1.000 \\
 & $\delta=100$ & 0.952 & 1.000 \\
 & $\delta=150$ & 0.956 & 1.000 \\
\bottomrule
\end{tabular}
\bigskip
\parbox{6.2in}{\footnotesize \medskip \medskip
	Notes: This table reports the finite sample coverage probabilities of the bootstrap inference at the nominal level of $95\%$ with $\beta_0 = 0.5$ and $\phi_0 \in \{0,1/4\}$. As expected, when $\phi_0$ is larger, the inference becomes conservative. However, as we see below from false coverage probabilities (Figure \ref{figure:sim_power2}), this does not mean that the inference will mostly be uninformative.}
\end{table}

Table \ref{tbl:sim1} reports the finite sample coverage probabilities for different sample sizes and values of $\delta$. As we see, when $\phi_0=0$ the results are close to the nominal size of $95\%$, and they are not much affected by the average degree of the underlying network. However, as we expected, the test becomes conservative when $\phi_0$ increases. As we shall see later, this does not necessarily mean that the inference will mostly be uninformative.

\subsection{Finite Sample Power of the Bootstrap Test}

\begin{figure}[ht]
	\caption[False Coverage Probability of the Confidence Sets at $95\%$]{
		False Coverage Probability of the Confidence Sets for $\phi_{0}=0$ and $\beta_{0}=0.5$ at the Nominal Level of $95\%$.}
	\label{figure:sim_power1}
	\medskip
	\begin{center}
		\includegraphics[scale = 0.4]{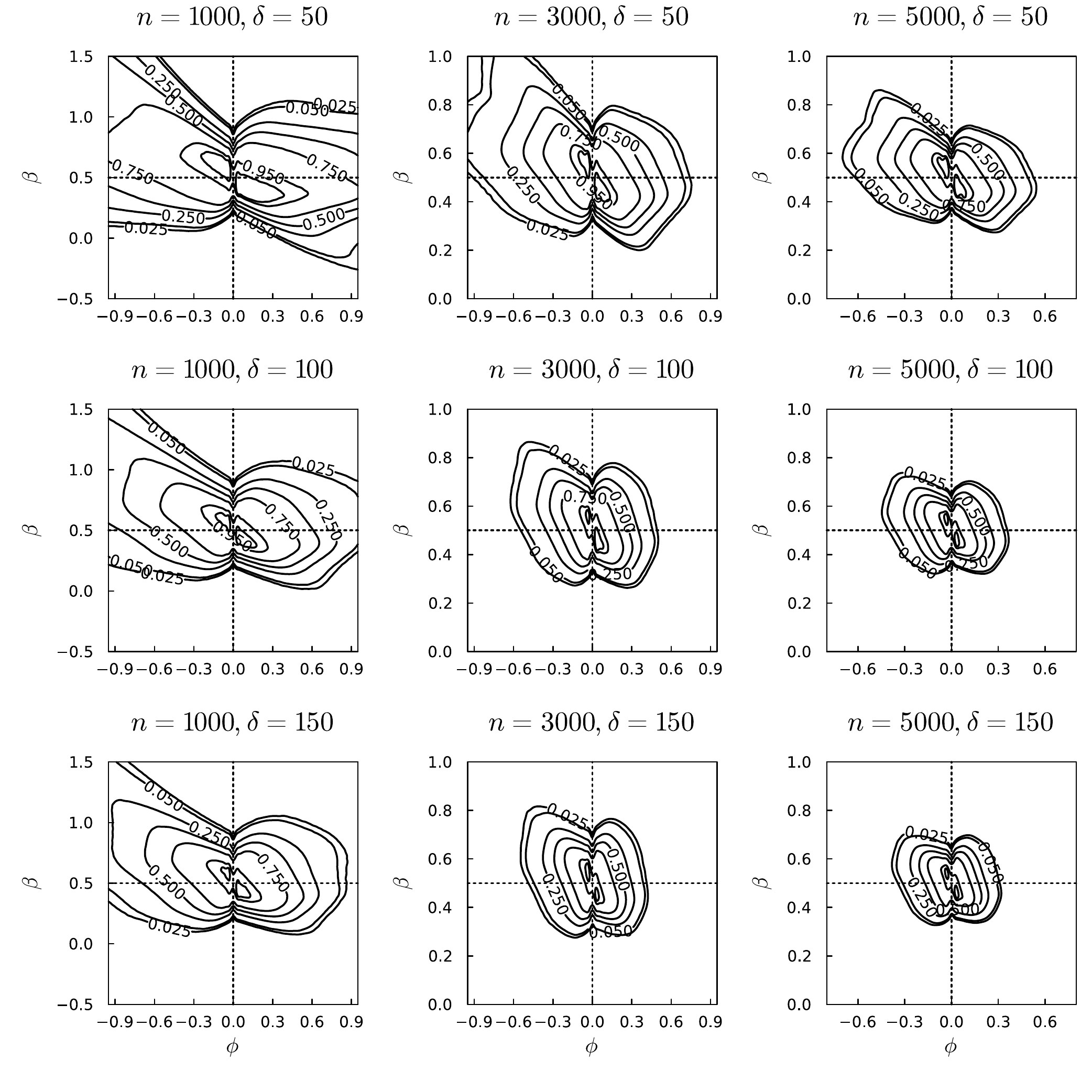}
	\end{center}
   \parbox{6.2in}{\footnotesize \medskip
	Notes: The intersecting point between two dotted lines in each panel indicates the true parameter $(\phi_0,\beta_0)$. The horizontal axis represents the hypothesized value of $\phi$ and the vertical axis that of $\beta$. In each panel, the area surrounded by the innermost contour line consists of the parameter values $(\phi,\beta)$ that are included in the confidence set at least 97.5\% of the times in the Monte Carlo loops, and the area surrounded by the outermost contour line consists of the parameter values $(\phi,\beta)$ that are included in the confidence set at least 2.5\% of the times in the Monte Carlo loops. As expected, since $\phi_0 = 0$, i.e., there is no strategic interaction, the inference shows a strong power property regardless of the average degrees used in the simulation design.}
\bigskip
\end{figure}

\begin{figure}[t]
	\caption[False Coverage Probability of the Confidence Sets at $95\%$]{False Coverage Probability of the Confidence Sets for $\phi_{0}=0.25$ and $\beta_{0}=0.5$ at the Nominal Level of $95\%$}
	\label{figure:sim_power2}
	\medskip
	\begin{center}
		\includegraphics[scale = 0.4]{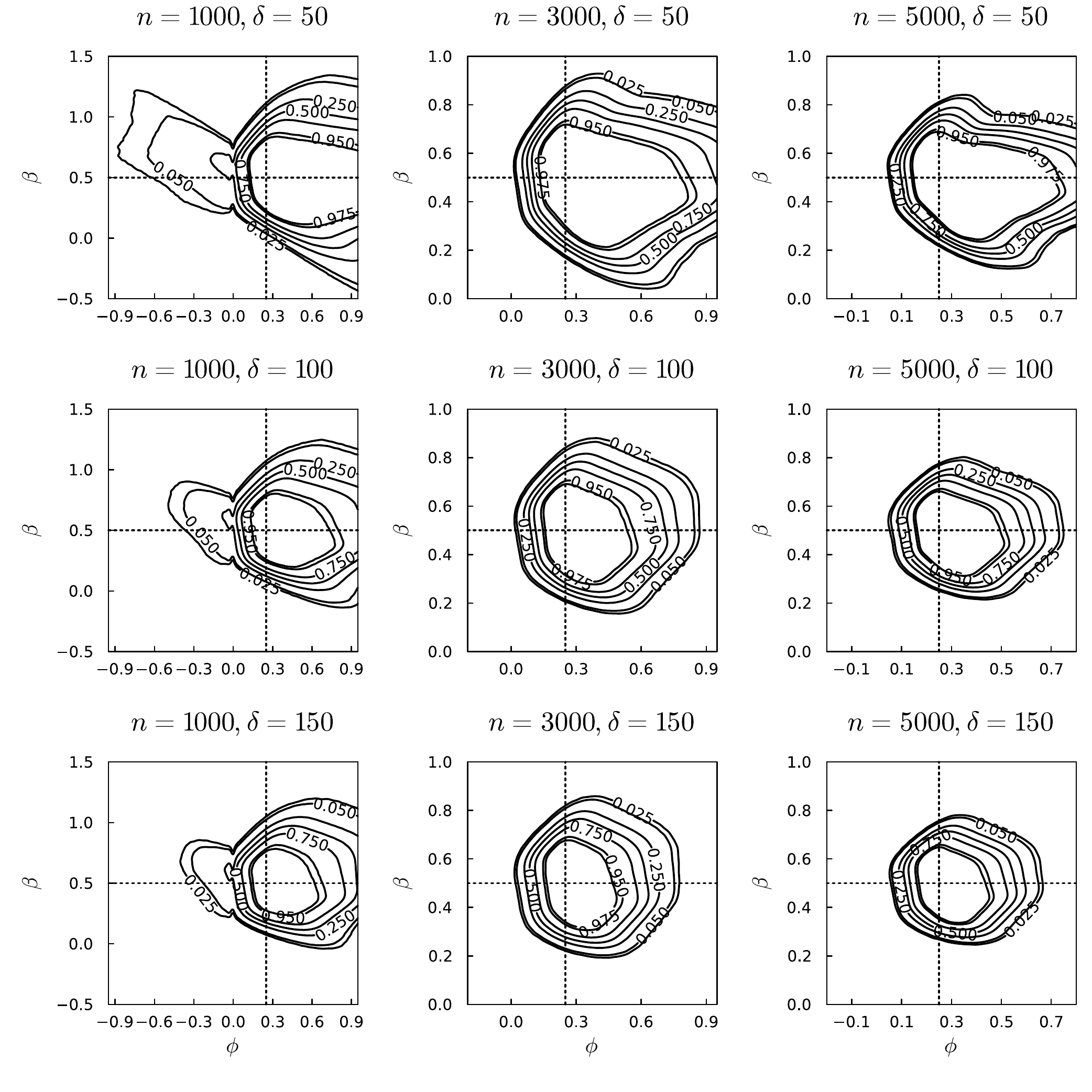}
	\end{center}
\parbox{6.2in}{\footnotesize \medskip
	Notes: The intersecting point between two dotted lines in each panel indicates the true parameter $(\phi_0,\beta_0)$. The horizontal axis represents the hypothesized value of $\phi$ and the vertical axis that of $\beta$. In each panel, the area surrounded by the innermost contour line consists of the parameter values $(\phi,\beta)$ that are included in the confidence set at least 97.5\% of the times in the Monte Carlo loops, and the area surrounded by the outermost contour line consists of the parameter values $(\phi,\beta)$ that are included in the confidence set at least 2.5\% of the times in the Monte Carlo loops. Note that the false coverage probability shows good performance when the average degree becomes larger despite the fact that the coverage probabilities were $1$ in Table \ref{tbl:sim1}.}
\bigskip
\end{figure}

We present results showing the finite sample power properties for different sample sizes and values of $\delta$. The nominal coverage probability is set at 95\%. The results are shown in Figures \ref{figure:sim_power1} and \ref{figure:sim_power2}. The horizontal axis represents the hypothesized value of $\phi$ under the null hypothesis while the vertical axis that of $\beta$. The intersecting point between two dotted lines indicates the true parameter $(\beta_0,\phi_0)$. We set $\beta_0=0.5$ and choose $\phi_0$ from $\{0, 1/4\}$. The bootstrap results show higher false coverage probability rates as $\phi_{0}$ moves away from zero, and substantial improvement as the sample size increases.

There are two notable features. First, when $\phi_{0}=1/4$ and $\delta=150$, we saw that the coverage probabilities were equal to one in Table \ref{tbl:sim1}, suggesting extreme conservativeness of the procedure. However, Figure \ref{figure:sim_power2} shows that even in this case, the false coverage probabilities are reasonably small. This shows that the conservative coverage probabilities (or conservative size properties) do not necessarily imply trivial or weak power properties in finite samples.

Second, when $\phi_0=1/4$ and $n>1000$, the false coverage probability at value $0$ is almost zero. This means that when $\phi_0$ is away from zero, the confidence set has almost zero probability of covering $0$. As $\phi_0$ is away from zero, the power of the bootstrap test (testing the null hypothesis of $\phi_0=0$) naturally increases, while the moment inequalities become more conservative. Despite this conservativeness, the bootstrap test does not lose power to detect the deviation from the null hypothesis of $\phi_0=0$. This has a significant implication in empirical applications because often we are interested in testing for the presence of strategic interactions among the players, and $\phi_0=0$ in this context means the absence of such interactions.

The hindsight regret also affects the estimation of $\beta_0$. In the Supplemental Note, we report the simulation results for the case with $\beta_0 = 1$. In this case, the confidence intervals on $\phi_0$ tend to become slightly larger. This is expected because, with larger $\beta_0$, the variations in the average actions of other players become relatively insignificant as compared to the variations of $X_i$, which leads to a low power for the inference on $\phi_0$.

We also experimented with various values of $\rho$. The results are found in the Supplemental Note. It turns out that the results are stable in the range of $\rho = 0.01 \sim 0.00001$. However, when $\rho$ is too small or too large, the confidence sets become substantially larger. This is expected from our construction of the test. When $\rho$ is extremely small, the hindsight regret becomes larger, due to the $\ln \rho$ term. When $\rho$ is larger, the bound in (\ref{inequality}) becomes smaller, yielding a larger denominator in (\ref{eq:sample_e2}). In either case, the test becomes more conservative, leading to a larger confidence set.

\section{Conclusion}

This paper focuses on a large Bayesian game perspective for social interactions models and develops an inference method that is robust to heterogeneous beliefs among the players. Utilizing the strategic interdependence among the players and the assumption of conditionally independent types, we derive testable implications from the equilibrium constraints.

The framework proposed in this paper may have limitations in some applications for several reasons. First, the framework assumes that the information groups are exogenously given in the beginning of the game. This does not cause any problem, if the current game's types satisfy the conditional independence assumption given any information used by the agents in the endogenous group formation that occurs prior to the game. However, this conditional independence assumption is violated when the agents observe the groups formed before they decide to enter the current game. Second, the framework assumes that the idiosyncratic component of the types is not shared between two different players. This assumption excludes a large network model where the information flows along connected neighborhoods. Third, we restrict our attention to the solution concept of pure strategy Bayes equilibria, and hence in a sense, our robustness to the beliefs is somewhat restricted in the light of weaker solution concepts such as iterated dominance or rationalizability as mentioned in the introduction. Explorations on these fronts require further research beyond this paper.

\section{Acknowledgments}

We thank Andres Aradillas-Lopez, Aureo de Paula, Hiro Kasahara, Jinwoo Kim, Sokbae Lee, Wei Li, Vadim Marmer and Mike Peters for useful conversations and comments. We are grateful to Yoram Halevy, Wei Li, Qingmin Liu, and Mike Peters for their kind and patient answers to our numerous elementary questions on Bayesian games. We thank Co-Editor and two anonymous referees for extremely valuable comments which led to a substantial improvement over earlier versions of this paper. All errors are ours. Song acknowledge that this research was supported by Social Sciences and Humanities Research Council of Canada. This research was enabled in part by support provided by Compute Canada (\href{http://www.computecanada.ca}{www.computecanada.ca}).

\putbib[large_games]

\newpage
\appendix
\section*{Appendices}
\renewcommand{\thesubsection}{\Alph{subsection}}
\numberwithin{equation}{subsection}
\numberwithin{theorem}{subsection}
\numberwithin{prop}{subsection}
\numberwithin{lemma}{subsection}
\numberwithin{claim}{subsection}
\numberwithin{definition}{subsection}
\numberwithin{assumption}{subsection}

\subsection{The Explicit Forms of $\mathbf{w}_U(\theta_0)$ and $\mathbf{w}_L(\theta_0)$}
\label{app:vg's}

\subsubsection{Payoff Function of the Form \eqref{eq:sp1}}

Given the parametric specifications of the utility functions and the distribution of unobserved heterogeneities, we can compute an explicit form of $\mathbf{w}_U(\theta_0)$ and $\mathbf{w}_L(\theta_0)$ which satisfy \eqref{bound3} with probability $1 - \nu$. The $\ell$-th entries of $\vec{w}_L(\theta_0),\vec{w}_U(\theta_0)\in [0,\infty)^m$ are given by
\begin{align*}
	w_{\ell,L}(\theta_0) &\eqdef \sqrt{-\frac{1}{2}\ln\mleft(\frac{\nu}{4m} \mright)\sum_{j \in N} c_{j,\ell,L}^2(\theta_0)} \qtext{and}\\
	w_{\ell,U}(\theta_0) &\eqdef \sqrt{-\frac{1}{2}\ln\mleft(\frac{\nu}{4m}\mright)\sum_{j \in N} c_{j,\ell,U}^2(\theta_0)},
\end{align*}
and whenever $j\in N(i)$, we set
\begin{align*}
	c_{j,\ell,L}(\theta_0) &\eqdef \frac{1}{n} \sum_{i\in N(i)\setminus\{j\}} \frac{v_i^-(X_i;\theta_0) g_{i,\ell}(X_i) }{1-r_i(\theta_0)} \qtext{ and }\\
	c_{j,\ell,U}(\theta_0) &\eqdef \frac{1}{n} \sum_{i\in N(i)\setminus\{j\}} \frac{v_i^+(X_i;\theta_0) g_{i,\ell}(X_i) }{1-r_i(\theta_0)},
\end{align*}
and $v_i^-(X_i;\theta_0)$ and $v_i^+(X_i;\theta_0)$ are functions of $X_i$ that are given as follows.

Let us define the following function: for $x,a,b \in \mathbf{R}$,
\[
	\varphi(x;a,b) = 1\left\{-x < a\right\}a + 1\left\{a \le -x < b\right\}\left(-x\right) + 1\left\{b \le -x \right\} b,
\]
and
\[
	\Psi_{\theta_0}(x;X_i) = \left|F_{\theta_0}\left(z_{\theta_0}(x;X_i) + \frac{v_2(X_i;\theta_0)}{n(i)} \mid X_i \right) - F_{\theta_0}\left(z_{\theta_0}(x;X_i) \mid X_i \right)\right|,
\]
where
\begin{equation*}
	z_{\theta_0}(x;X_i) = \begin{cases}
		\varphi\left(\frac{\displaystyle v_2(X_i;\theta_0)}{\displaystyle 2 n(i) } ;x,x + \frac{\displaystyle v_2(X_i;\theta_0)(n(i) - 1)}{\displaystyle n(i)}\right), & \text{ if } v_2(X_i;\theta_0) \ge 0, \\[2ex]
		\varphi\left(\frac{\displaystyle v_2(X_i;\theta_0)}{\displaystyle 2 n(i) } ;x + \frac{\displaystyle v_2(X_i;\theta_0)(n(i) - 1)}{\displaystyle n(i)},x \right), & \text{ if } v_2(X_i;\theta_0) < 0.
	\end{cases}
\end{equation*}
Then, we define
\begin{align*}
	v_i^-(X_i;\theta_0) &= \Psi_{\theta_0}\left(v_1(X_i;\theta_0) - \lambda_{i,\rho};X_i\right) \qtext{and} \\ 
	v_i^+(X_i;\theta_0) &= \Psi_{\theta_0}\left(v_1(X_i;\theta_0) + \lambda_{i,\rho};X_i\right),
\end{align*}
where $\lambda_{i,\rho}$ is as given in \eqref{eq:hindsight_regret}.

\subsubsection{Arbitrary Payoff Functions}

First, for any map $f$, recall the definition of $V_{j}(f)$ in \eqref{eq:max_var}. For $j \in N$ and $1\le \ell \le m$, let
\begin{align*}
		c_{j,\ell,L}(\theta_0) &\eqdef \frac{1}{n} \sum_{i\in N\setminus\{j\}} \frac{V_{j}(\pi_{i,L}(\csdot, X_i;\theta_0))g_{i,\ell}(X_i)}{1-r_i(\theta_0)} \qtext{ and }\\
		c_{j,\ell,U}(\theta_0) &\eqdef \frac{1}{n} \sum_{i\in N\setminus\{j\}} \frac{V_{j}(\pi_{i,U}(\csdot, X_i;\theta_0))g_{i,\ell}(X_i)}{1-r_i(\theta_0)},
\end{align*}
where $r_i(\theta_0)$ is as defined in \eqref{ri theta0}. For a given $\nu \in (0,1)$, we define the $\ell$-th element of $\vec{w}_L(\theta_0)$ to be
\begin{align*}
	\sqrt{-\frac{1}{2}\ln\mleft(\frac{\nu}{4m}\mright)\sum_{j \in N} c_{j,\ell,L}^2(\theta_0)}.
\end{align*}
The elements of $\vec{w}_U(\theta_0)$ are defined similarly, using $c_{j,\ell,U}(\theta_0)$ in place of $c_{j,\ell,L}(\theta_0)$.

\subsection{Extension to Multinomial Action Sets}
\label{app:mult_action_set}

We show how our framework can be extended to the case with a general parametric payoff function and a multinomial action set. We provide formal results and their proofs here. The results in the main text follow from these as corollaries.

For the rest of the Appendix, an inequality between two vectors $x = [x_j]$ and $y = [y_j]$, say, $x \ge y$, represents the corresponding elementwise inequalities, i.e., $x_j \ge y_j$ for all $j$. The proofs of the results here appear in Section \ref{app:proofs} below. From here on, we suppress from notation the dependence of various quantities on $\theta_0$ for simplicity.

\subsubsection{Belief-free Hindsight Regrets and Testable Implications}

Let $A$ be a finite action set, and let $k \eqdef |A|$. For $i\in N$, $a,a'\in A$, and $y_{-i} \in A^{n-1}$, we define
\[
	u_i^{\Delta}(a,a',y_{-i};\tau_i)\eqdef u_i(a,y_{-i};\tau_i)-u_i(a',y_{-i};\tau_i),
\]
which is player $i$'s payoff differential between choosing $a$ and $a'$ when the other players choose $y_{-i}\in A^{n-1}$. Recall that $\mathcal{I}_i$ is the $\sigma$-field generated by $\left(\eta_i, \X\right)$, where $\X = (X_j)_{j \in N}$.

\begin{definition}
Given an equilibrium $\Y = (Y_i)_{i \in N}$ and $\rho \in (0,1)$, an $\IF_i$-measurable random vector $\vecg{\lambda}_{i} \in [0,\infty)^{k-1}$ is a $\rho$-\textit{hindsight regret} for player $i\in N$ if
\begin{equation*}
	\QR[i]{\vec{u}_i^{\Delta}(Y_i,Y_{-i};\typ_i) \ge -\vecg{\lambda}_{i} \mid \IF_i} \ge 1-\rho, \quad (\QM_i\text{-a.s.}),
\end{equation*}
where $\vec{u}_i^{\Delta}(a,y_{-i};\tau_i)\eqdef \left[u_i^{\Delta}(a,a',y_{-i};\tau_i) \right]_{a'\in A\setminus \{a\}}$.
\end{definition}

Let $\vecg{\lambda}_{i,\rho}(a;\tau_i)\eqdef [\lambda_{i,\rho}(a, a';\tau_i)]_{a'\in A\setminus \{a\}}$ be a vector in $\R^{k-1}$ whose elements are given by
\begin{align}
\label{eq:lambda_i_rho}
	\lambda_{i,\rho}(a, a';\tau_i)\eqdef \sqrt{-\frac{1}{2}\ln\left(\frac{\rho}{k-1}\right)\csdot \Lambda_i(a,a';\tau_i)}, \quad a' \in A,
\end{align}
where
\[
	\Lambda_i(a,a';\tau_i)\eqdef \sum_{j\in N\setminus \{i\}} V_j^2(u_i^{\Delta}(a,a',\csdot;\tau_i)),
\]
and $V_j(\csdot)$ denotes the maximal variation due to player $j$ defined in \eqref{eq:max_var}.

For $\kappa \ge 0$ and $\rho \in (0,1)$, let
\begin{align}
	\label{N(kappa)}
	\begin{aligned}
	N_{\PM}(\kappa,\rho) &\eqdef \left\{i \in N: \PM \ll \QM_i,\, \PM\left(\delta_i\left(\frac{\rho}{k-1}\right) \le \frac{\kappa\rho}{k-1}\right) = 1 \right\} \qtext{and} \\ 
	 n_{\PM}(\kappa,\rho) &\eqdef |N_{\PM}(\kappa,\rho)|,
	\end{aligned}
\end{align}
where $\PM \ll \QM_i$ means that $\QM_i$ dominates $\PM$, i.e., any $\QM_i$-null event is also a $\PM$-null event. Later, we require the fraction of the players outside $N_{\PM}(\kappa,\rho)$ to be asymptotically negligible at least at a certain rate. (See Assumption \ref{assum:RE-B} below.)

From here on, we assume that the observed profile of actions $Y = (Y_i)_{i \in N}$ is generated from one equilibrium of the game, as in \eqref{Y}. The following theorem generalizes the observation in \eqref{inequality}.

\begin{theorem}
\label{thm:tA1}
Suppose that Assumptions \ref{assum:CI} holds. Then, for any $\rho \in (0,1)$ and any pure strategy Bayesian equilibrium $\Y = (Y_i)_{i \in N}$, $\vecg{\lambda}_{i,\rho}(Y_i;\typ_i)$ is a $\rho$-hindsight regret for player $i\in N$. Moreover, for any $\kappa \in [0,(1- \rho)/\rho)$ and $i \in N_{\PM}(\kappa,\rho)$, we have
\[
	\PR{\vec{u}_i^{\Delta}(Y_i,Y_{-i};\typ_i) \ge -\vecg{\lambda}_{i,\rho}(Y_i;\typ_i)\mid \IF_i} \ge 1-r_{i,\rho}(Y_i;\typ_i), \quad (\PM\text{-a.s.}),
\]
where
\begin{align}
	\label{ri a}
	r_{i,\rho}(a;\tau_i)\eqdef \frac{(\kappa +1)\rho}{k-1}\sum_{a'\in A\setminus\{a\}}1\{\lambda_{i,\rho}(a,a';\tau_i)>0\}.	
\end{align}
\end{theorem}
\medskip

\subsubsection{Testable Implications}
First, we make assumptions of parametric functions for the conditional distribution of $\eta_i$ given $X_i$ and the payoff function as in Assumptions \ref{assum:A5} and \ref{assum:A6}.

For each $i\in N$, $a\in A$, and $y_{-i} \in A^{n-1}$, we define
\begin{align*}
	\pi_{i,a,L}(y_{-i},X_i)&\eqdef \int \mathbf{1}\left\{\overline \eta \in H_{i,a,L}(y_{-i},X_i) \right\} dF_{\theta_0}(\overline \eta \mid X_i) \qtext{and} \\
	\pi_{i,a,U}(y_{-i},X_i)&\eqdef \int \mathbf{1}\left\{\overline \eta \in H_{i,a,U}(y_{-i},X_i)  \right\} dF_{\theta_0}(\overline \eta \mid X_i),
\end{align*}
where $F_{\theta_0}(\csdot \mid X_i)$ denotes the conditional CDF of $\eta_i$ given $X_i$,
\begin{align*}
	H_{i,a,L}(y_{-i},X_i) &\eqdef \left\{\overline \eta\in \R^w: \exists a'\in A\setminus \{a\} \text{ s.t. } \vec{u}_i^{\Delta}(a',y_{-i};X_i,\overline \eta) \ge -\vecg{\lambda}_{i,\rho}(a';X_i,\overline \eta) \right\} \qtext{and} \\
	H_{i,a,U}(y_{-i},X_i) &\eqdef \left\{\overline \eta\in \R^w: \vec{u}_i^{\Delta}(a,y_{-i};X_i, \overline \eta) \ge -\vecg{\lambda}_{i,\rho}(a;X_i,\overline \eta) \right\},
\end{align*}
where $\mathbf{R}^w$ denotes the space that $\eta_i$ takes values from. Once we parametrize the conditional distribution of $\eta_i$ given $X_i$ as in Assumption \ref{assum:A5}, we can simulate $\pi_{i,a,L}(Y_{-i},X_i)$ and $\pi_{i,a,U}(Y_{-i},X_i)$ by drawing $\eta_i$'s from this conditional distribution.

For each $a\in A$, we define
\begin{align}
	\label{eq:sample_e_mlt}
	\begin{aligned}
		e_{i,a,L}(\kappa,\rho)\eqdef &\ind\{Y_i=a\}-\left(1-\frac{\pi_{i,a,L}(Y_{-i},X_i)}{1-r_{i,a,L}(\kappa,\rho)} \right) \qtext{and} \\
		e_{i,a,U}(\kappa,\rho)\eqdef &\ind\{Y_i=a\}-\frac{\pi_{i,a,U}(Y_{-i},X_i)}{1-r_{i,a,U}(\kappa,\rho)},
	\end{aligned}
\end{align}
where
\begin{align*}
	r_{i,a,L}(\kappa,\rho)&\eqdef \frac{(\kappa + 1)\rho}{k-1}\max_{c \in A\setminus\{a\}} \sum_{a' \in A \setminus \{a\}} \ind\left\{ \sup_{\overline \eta}\lambda_{i}\left(c,a';X_i,\overline \eta \right)>0\right\} \qtext{and} \\
	r_{i,a,U}(\kappa,\rho)&\eqdef \frac{(\kappa + 1)\rho}{k-1}\sum_{a'\in A\setminus\{a\}} \ind\left\{\sup_{\overline \eta}\lambda_{i}\left(a,a';X_i,\overline \eta \right)>0\right\}.
\end{align*}
Let $\vec{e}_{i,L}(\kappa,\rho)$ and $\vec{e}_{i,U}(\kappa,\rho)$ be vectors of dimension $k-1$ whose elements are $e_{i,a,L}(\kappa,\rho)$ and $e_{i,a,U}(\kappa,\rho)$ with $a$ running in $A\setminus \{a_1\}$. The following result establishes moment inequalities as testable implications.
\begin{prop}
	\label{prop:pA1}
	Suppose that Assumptions \ref{assum:CI}, \ref{assum:A4}, \ref{assum:A5}, and \ref{assum:A6} hold. Then, for $\rho \in (0,1)$, $\kappa \in [0,(1- \rho)/\rho)$, and $i \in N_{\PM}(\kappa,\rho)$,
	\begin{align*}
		\E_{\PM}\left[\vec{e}_{i,L}(\kappa,\rho) \mid \X \right] \ge 0 \qtext{and}\quad \E_{\PM}\left[\vec{e}_{i,U}(\kappa,\rho) \mid \X \right] \le 0, \quad (\PM\text{-a.s.}).
	\end{align*}
\end{prop}
\medskip

\subsubsection{Bootstrap Inference}

Choosing a vector of non-negative measurable functions $\vec{g}_i\eqdef [g_{i,1},\ldots, g_{i,m}]^{\top}:\R^{d_X} \to [0,\infty)^m$, we define sample moments as follows
\begin{align}
	\label{eq:sample_moments_mlt}
	\vecg{\hat{\mu}}_L(\kappa,\rho) \eqdef \frac{1}{n}\sum_{i \in N} \vec{e}_{i,L}(\kappa,\rho) \otimes \vec{g}_{i}(X_i) \qtext{and}\quad \vecg{\hat{\mu}}_U(\kappa,\rho) \eqdef \frac{1}{n}\sum_{i \in N} \vec{e}_{i,U}(\kappa,\rho) \otimes \vec{g}_{i}(X_i),
\end{align}
where $\otimes$ denotes the Kronecker product. Using the sample moments, we consider the following test statistic:
\begin{align}
\label{eq:def_T}
	T(\kappa,\rho) \eqdef \T\left(\sqrt{n}(\vecg{\hat{\mu}}_L+\vec{w}_L)(\kappa,\rho),\sqrt{n}(\vecg{\hat{\mu}}_U-\vec{w}_U)(\kappa,\rho)\right),
\end{align}
where $\T:\R^{(k-1)m}\times\R^{(k-1)m}\to \R$ is a function defined by $\T(x,y)\eqdef\norm{[x]_{-}+[y]_{+}}_1$, and $\vec{w}_L$ and $\vec{w}_U$ are constructed as follows. For any map $f$ from a vector of actions $(y_1,...,y_n)$ of the $n$ -players to a number, we define its maximal variation with respect to player $j$ as
\begin{equation}
	\label{eq:max_var2}
	V_j(f)\eqdef \sup_{(y_1,...,y_n)\in A^n,y'\in A}\abs{f(y_1,...,y_n)-f(y_1,\ldots, y_{j-1},y',y_{j+1},\ldots,y_n)}.
\end{equation}
For $j\ne i$, $1\le \ell\le m$, and $a\in A$, let\footnote{Similarly as in (\ref{eq:hr0}), we view $\pi_{i,a,L}(\cdot,X_i)$ as a function of $y_1,...,y_n$, so that $V_j(\pi_{i,a,L}(\csdot, X_i))$ denotes the maximal variation of $\pi_{i,a,L}(\cdot,X_i)$ with respect to the action of player $j$, not with respect to the $j$-th argument of $y_{-i}$.}
\begin{align}
	\label{eq:diL_diU}
	\begin{aligned}
		c_{j,\ell,L}(a)&\eqdef \frac{1}{n}\sum_{i\in N\setminus\{j\}} \frac{V_j(\pi_{i,a,L}(\csdot, X_i))g_{i,\ell}(X_i)}{1-r_{i,a,L}(\kappa,\rho)} \qtext{and} \\
		c_{j,\ell,U}(a)&\eqdef \frac{1}{n}\sum_{i\in N\setminus\{j\}} \frac{V_j(\pi_{i,a,U}(\csdot, X_i))g_{i,\ell}(X_i)}{1-r_{i,a,U}(\kappa,\rho)}.
	\end{aligned}
\end{align}
Then, for a given $\nu\in (0,1)$, the elements of $\vec{w}_{L}(\kappa,\rho)$ are defined to be
\begin{align}
	\label{wL(a)}
	w_{\ell,L}(a) \eqdef \sqrt{-\frac{1}{2}\ln\left(\frac{\nu}{4(k-1)m}\right) \sum_{j \in N} c_{j,\ell,L}^2(a)}
\end{align}
with $a$ running in $A\setminus \{a_1\}$ and $\ell$ running in $\{1,\ldots,m\}$, keeping the same order of elements as in the sample moments $\vecg{\hat{\mu}}_L(\kappa,\rho)$ and $\vecg{\hat{\mu}}_U(\kappa,\rho)$. The elements of $\vec{w}_U(\kappa,\rho)$ are defined similarly, with $c_{j,\ell,L}(a)$ replaced by $c_{j,\ell,U}(a)$.

For a bootstrap statistic, we draw i.i.d.\ random variables, $\varepsilon_1,\ldots,\varepsilon_n$, from $\ND{0}{1}$ and set
\begin{align}
\label{eq:zeta*}
	\vecg{\zeta}^{*}(\kappa,\rho) \eqdef \frac{1}{n}\sum_{i \in N}\left((\vec{Y}_i-\vecg{\mu}_i^{*}(\kappa,\rho))\otimes \vec{g}_i(X_i)\right)\,\varepsilon_i,
\end{align}
where $\vec{Y}_i$ and $\vecg{\mu}_i^{*}(\kappa,\rho)$ are column vectors formed by $\ind\{Y_i=a\}$ and
\[
	\left(\frac{1}{2}\left(1-\frac{\pi_{i,a,L}(Y_{-i},X_i)}{1-r_{i,a,L}(\kappa,\rho)} +\frac{\pi_{i,a,U}(Y_{-i},X_i)}{1-r_{i,a,U}(\kappa,\rho)}\right) \vee 0 \right) \wedge 1,
\]
respectively, with $a$ running in $A\setminus \{a_1\}$. In addition, given fixed $\nu \in (0,1)$, we define
\begin{align*}
	\vecg{\hat\varphi}_L(\kappa,\rho) &\eqdef \left[(\vecg{\hat{\mu}}_L-\vec{w}_L)(\kappa,\rho)-\ind_d\csdot q_{1-\nu/2}^{*}(\kappa,\rho)/\sqrt{n}\right]_{+} \qtext{and} \\
	\vecg{\hat\varphi}_U(\kappa,\rho) &\eqdef \left[(\vecg{\hat{\mu}}_U+\vec{w}_U)(\kappa,\rho)+\ind_d\csdot q_{1-\nu/2}^{*}(\kappa,\rho)/\sqrt{n}\right]_{-},
\end{align*}
where $q_{1-\nu/2}^{*}(\kappa,\rho)$ is the $(1 - \nu/2)$ quantile of the bootstrap distribution of $\sqrt{n}\|\vecg{\zeta}^{*}(\kappa,\rho)\|_\infty$.

We consider the following bootstrap test statistic:
\begin{align}
\label{eq:def_T*}
	T^{*}(\kappa,\rho) \eqdef \T\left(\sqrt{n}(\vecg{\zeta}^{*}+\vecg{\hat\varphi}_L\wedge \vecg{\hat\varphi}_U)(\kappa,\rho),\sqrt{n}(\vecg{\zeta}^{*}-\vecg{\hat\varphi}_L\wedge \vecg{\hat\varphi}_U)(\kappa,\rho)\right),
\end{align}
(the minimum between $\vecg{\hat\varphi}_L$ and $\vecg{\hat\varphi}_U$ is taken element-wise). The confidence set for $\theta_0\in\Theta$ at nominal level $1-\alpha$ is given by
\[
	CS_\epsilon(\kappa,\rho) \eqdef \{\theta\in\Theta: T(\kappa,\rho) \le c_\gamma^{*}(\kappa,\rho)\vee \epsilon\},
\]
where $\epsilon>0$ is a fixed small number and $c_\gamma^{*}(\kappa,\rho)$ is the $\gamma\eqdef (1-\alpha+2\nu)$-quantile of the bootstrap distribution of $T^{*}$.

We let $\vecg{\zeta}_i \eqdef (\vec{Y}_i -\E_{\PM}[ \vec{Y}_i \mid \X ]) \otimes \mathbf{g}_i(X_i)$ and $\Sigma_i \eqdef \E_{\PM}\left[\vecg{\zeta}_i \vecg{\zeta}_i^{\top}\mid \X \right]$. Let $\PS_0$ denote a family of objective probability measures $\PM$ on $(\Omega,\H)$ satisfying Assumptions \ref{assum:CI} and \ref{assum:A5}. We reintroduce Assumption \ref{assump: nondeg} corresponding to the redefined vectors, $\vecg{\zeta}_i$'s, as above.

\begin{assumption}
	\label{assum: nondeg2}
	There exists a positive sequence $\{r_n\}$ such that $r_n \rightarrow 0$, $n^{1/6} r_n \rightarrow \infty$, and
	\begin{align}
		\label{eq:nondeg2}
		\lim_{n\to\infty}\sup_{\PM\in \PS_0}\PR{\min_{1 \le i \le n} \lambda_{\min}\left(\Sigma_i \right) < r_n}=0,
	\end{align}
	where $\lambda_{\min}\left(\Sigma_i \right)$ is the smallest eigenvalue of $\Sigma_i \eqdef \E_{\PM}\left[\vecg{\zeta}_i \vecg{\zeta}_i^{\top}\mid \X \right]$.
\end{assumption}

The following assumption is a weaker version of Assumption \ref{assum:RE}.

\begin{assumption}
	\label{assum:RE-B}
	There exist constants $\rho \in (0,1)$, $\kappa \in [0, (1-\rho)/\rho)$, and $M_{\kappa,\rho} >0$ such that for all $\PM\in \PS_0$ and $n \ge 1$,
	\begin{align}
		\label{nkappa}
		 \frac{n_{\PM}(\kappa,\rho)}{n} \ge 1 - \frac{M_{\kappa,\rho} r_n}{\sqrt{n}},
	\end{align}
	where $r_n$ is the vanishing positive sequence of numbers in Assumption \ref{assum: nondeg2}.
\end{assumption}

This assumption requires that the fraction of the players outside $N_{\PM}(\kappa,\rho)$ is asymptotically negligible at the rate of $r_n / \sqrt{n}$ uniformly over $\mathsf{P} \in \mathcal{P}_0$. The following theorem establishes the asymptotic validity of the confidence set $CS_{\epsilon}(\kappa,\rho)$.

\begin{theorem}
\label{thm:bootstrap_consisteny_mlt}
Suppose that Assumptions \ref{assum:CI}, \ref{assum:A4}, \ref{assum:A5}, \ref{assum:A6}, \ref{assum: nondeg2}, and \ref{assum:RE-B} hold, and there exists $C_{\vec{g}}>0$ such that for all $n\ge 1$, $\max_{i\in N}\max_{1\le \ell\le m} \sup_{x \in \mathbf{R}^{d_X}} |g_{i, \ell}(x)| \le C_{\vec{g}}$.

Then, for a positive sequence $\{\epsilon_n\}$ such that $\epsilon_n^{-1}=o(n^{1/8})$,
\[
	\limsup_{n\to\infty}\sup_{\PM\in \PS_0}\PR{\theta_0\notin CS_{\epsilon_n}(\kappa,\rho) }\le \alpha,
\]
where $\rho \in (0,1)$ and $\kappa \in [0, (1-\rho)/\rho)$ are the constants that appear in Assumption \ref{assum:RE-B}.
\end{theorem}
\medskip

\subsection{Proofs of the Results in Section \ref{app:mult_action_set}}
\label{app:proofs}

First, for $a\in A$, let
\begin{align}
	\begin{aligned}
	\label{eq:infeasible_e}
		\tilde{e}_{i,a,L}(\kappa,\rho) &\eqdef \ind\{Y_i=a\}-\left(1-\frac{\E_{\PM}\left[\pi_{i,a,L}(Y_{-i},X_i)\mid \X \right]}{1-r_{i,a,L}(\kappa,\rho)}\right) \qtext{and} \\
		\tilde{e}_{i,a,U}(\kappa,\rho) &\eqdef \ind\{Y_i=a\}-\frac{\E_{\PM}\left[\pi_{i,a,U}(Y_{-i},X_i)\mid \X \right]}{1-r_{i,a,U}(\kappa,\rho)},
	\end{aligned}
\end{align}
and let $\vec{\tilde{e}}_{i,L}(\kappa,\rho)$ and $\vec{\tilde{e}}_{i,U}(\kappa,\rho)$ be column vectors whose elements are $\tilde{e}_{i,a,L}(\kappa,\rho)$ and $\tilde{e}_{i,a,U}(\kappa,\rho)$ with $a$ running in $A\setminus \{a_1\}$. The infeasible moments are given by

\begin{align}
\label{eq:infeasible_moments}
	\begin{aligned}
		\vecg{\tilde{\mu}}_L(\kappa,\rho) &\eqdef \frac{1}{n}\sum_{i \in N} \vec{\tilde{e}}_{i,L}(\kappa,\rho) \otimes \vec{g}_i(X_i) \qtext{and}\quad \\
		\vecg{\tilde{\mu}}_U(\kappa,\rho) &\eqdef \frac{1}{n}\sum_{i \in N} \vec{\tilde{e}}_{i,U}(\kappa,\rho) \otimes \vec{g}_i(X_i).
	\end{aligned}
\end{align}
For any $A \subset N$, let $\mathcal{I}_A$ be the smallest $\sigma$-field containing $\mathcal{I}_i$, $i \in A$, $X_A \eqdef (X_j)_{j \in A}$, and $Y_A \eqdef (Y_j)_{j \in A}$.

Define the event
\[
	\mathcal{M}(\kappa,\rho)\eqdef\left\{ |\vecg{\hat{\mu}}_L-\vecg{\tilde{\mu}}_L|(\kappa,\rho) \le \vec{w}_L(\kappa,\rho), |\vecg{\hat{\mu}}_U-\vecg{\tilde{\mu}}_U|(\kappa,\rho) \le \vec{w}_U(\kappa,\rho) \right\},
\]
where, for a vector $\mathbf{x} = [x_j]$, we write $\abs{\mathbf{x}} = [|x_j|]$, the inequalities above are elementwise, and $\vec{w}_L(\kappa,\rho)$ and $\vec{w}_U(\kappa,\rho)$ are constructed using \eqref{eq:diL_diU} and \eqref{wL(a)}. The following proposition shows that the event $\mathcal{M}(\kappa,\rho)$ occurs with a large probability.
\begin{prop}
\label{prop:pA2}
	Suppose that Assumptions \ref{assum:CI}, \ref{assum:A4}, and \ref{assum:A5} hold. Then, for any $\rho \in (0,1)$ and $\kappa \in [0, (1- \rho)/\rho)$,
	\[
	\PR{\mathcal{M}(\kappa,\rho)}\ge 1-\nu.
	\]
\end{prop}
\begin{proof}
	By Assumption \ref{assum:CI}, $\vec{Y}_i$, $i \in N$, are conditionally independent given $\X$. For brevity, we suppress from notation the dependence of various quantities on $(\kappa,\rho)$. Note that
	\begin{align*}
		\vecg{\hat{\mu}}_L - \vecg{\tilde{\mu}}_L = \frac{1}{n}\sum_{i \in N} (\vec{\hat{e}}_{i,L} - \vec{\tilde{e}}_{i,L})\otimes \vec{g}_i(X_i),
	\end{align*}
	and similarly with $\vecg{\hat{\mu}}_U - \vecg{\tilde{\mu}}_U$. The entries of the left hand side vector are given by
	\begin{align*}
		\frac{1}{n} \sum_{i \in N} \varphi_{i,\ell,a}(Y_{-i},X_i), \quad \ell = 1,...,m, \quad a \in A \setminus\{a_1\},
	\end{align*}
	where
	\begin{align*}
		\varphi_{i,\ell,a}(y_{-i},X_i) = - \frac{(\pi_{i,a,L}(y_{-i},X_i) - \E_{\PM}[\pi_{i,a,L}(Y_{-i},X_i) \mid \X])g_{i,\ell}(X_i)}{1 - r_{i,a,L}}.
	\end{align*}
	Therefore, for each $j \in N$,
	\begin{align*}
		V_j(\varphi_{i,\ell,a}(\csdot, X_i))
		\le \frac{V_j(\pi_{i,a,L}(\csdot,X_i))g_{i,\ell}(X_i)}{1 - r_{i,a,L}}.
	\end{align*}
	Hence, by McDiarmid's inequality (see Lemma \ref{lemma:aux_mcdiarmid} in the Supplemental Note), for $a \in A$,
	\begin{align*}
		\PR{\left|\frac{1}{n} \sum_{i \in N} \varphi_{i,\ell,a}(Y_{-i},X_i) \right| > w_{\ell,L}(a) \mid \X} \le 2 \exp\left( - \frac{2 w_{\ell,L}^2(a)}{\sum_{j \in N} c_{j,\ell,L}^2(a)}\right) = \frac{\nu}{2(k-1)m},
	\end{align*}
	where $w_{\ell,L}(a)$ is defined in \eqref{wL(a)}. This yields the following bound:
	\begin{align*}
		1 - \PR{ |\vecg{\hat{\mu}}_L-\vecg{\tilde{\mu}}_L| \le  \vec{w}_L} &\le \sum_{\ell =1}^m \sum_{a \in A \setminus \{a_1\}} \PR{\left|\frac{1}{n} \sum_{i \in N} \varphi_{i,\ell,a}(Y_{-i},X_i) \right| > w_{\ell,L}(a)}
		\le \frac{\nu}{2}.
	\end{align*}
	Arguing similarly for $\vec{w}_U$, we find that
	\begin{align*}
		1 - \PR{ \mathcal{M}(\kappa, \rho)} &\le 2 - \PR{ |\vecg{\hat{\mu}}_L-\vecg{\tilde{\mu}}_L| \le  \vec{w}_L} - \PR{ |\vecg{\hat{\mu}}_U-\vecg{\tilde{\mu}}_U| \le  \vec{w}_U} \le \nu. \qedhere
	\end{align*}
\end{proof}

\begin{proof}[\textbf{Proof of Theorem \ref{thm:tA1}}] For each $i \in N$, the elements of $Y_{-i}$ are conditionally independent given $\IF_i$ under $\QM_i$ and $\PM$ by Assumption \ref{assum:CI}. Now, by Lemma \ref{lemma:aux_mcdiarmid} in the Supplemental Note, for all $a' \in A$,
\begin{align}
	\label{bound32}
	\begin{aligned}
		&\QR[i]{u_i^\Delta(Y_i,a',Y_{-i}) \mid \IF_i] < - \lambda_{i,\rho}(Y_i,a';\tau_i) \mid \IF_i} \\
		&\qquad\le \QR[i]{u_i^\Delta(Y_i,a',Y_{-i}) -\E_{\QM_i}[u_i^\Delta(Y_i,a',Y_{-i})\mid \IF_i] < - \lambda_{i,\rho}(Y_i,a';\tau_i) \mid \IF_i} \\
		&\qquad=\QR[i]{-u_i^\Delta(Y_i,a',Y_{-i}) +\E_{\QM_i}[u_i^\Delta(Y_i,a',Y_{-i})\mid \IF_i] >  \lambda_{i,\rho}(Y_i,a';\tau_i) \mid \IF_i} \\
		&\qquad \le \rho/(k-1), \quad (\QM_i \text{-a.s.})
	\end{aligned}
\end{align}
on $\{\lambda_{i,\rho}(Y_i,a';\tau_i) > 0\}$, where the second inequality holds because $\E_{\QM_i}[\mathbf{u}_i^\Delta(Y_i,Y_{-i})\mid \IF_i] \ge 0$ due to $\Y = (Y_i)_{i \in N}$ being a pure strategy Bayesian equilibrium.\footnote{
	Note that in applying Lemma \ref{lemma:aux_mcdiarmid}, we replace $f$ by $-u_i^\Delta$, $X$ by $(Y_j)_{j \in N\setminus \{i\}}$, $Y$ by $(Y_i,\tau_i)$, and $\mathcal{F}$ by $\IF_i$.
}
On the event $\{\lambda_{i,\rho}(Y_i,a';\tau_i) = 0\}$, $u_i^\Delta(Y_i,a',Y_{-i})$ does not vary with $Y_{-i}$, and hence, $u_i^\Delta(Y_i,a',Y_{-i}) -\E_{\QM_i}[u_i^\Delta(Y_i,a',Y_{-i})\mid \IF_i] = 0$, ($\QM_{i}$-a.s.). Consequently,
\begin{align*}
	&1- \frac{\rho}{k-1} \sum_{a' \in A \setminus \{Y_i\}} 1\{\lambda_{i,\rho}(Y_i,a';\tau_i)>0\} \\
	&\qquad\le 1-\sum_{a'\in A\setminus \{Y_i\}}\QR[i]{u_i^\Delta(Y_i,a',Y_{-i}) \mid \IF_i] < - \lambda_{i,\rho}(Y_i,a';\tau_i) \mid \IF_i}\\
	&\qquad\le \QR[i]{\mathbf{u}_i^\Delta(Y_i,Y_{-i}) \ge - \vecg{\lambda}_{i,\rho}(Y_i;\tau_i) \mid \IF_i}, \quad (\QM_i \text{-a.s.}),
\end{align*}
implying that $\vecg{\lambda}_{i,\rho}(Y_i;\typ_i)$ is a $\rho$-hindsight regret for player $i\in N$.

Finally, let $i\in N_{\PM}(\kappa,\rho)$. Since $\PM \ll \QM_{i}$, for all $a' \in A$, we have $u_i^\Delta(Y_i,a',Y_{-i})\ge 0$, ($\PM$-a.s.) on $\{\lambda_{i,\rho}(Y_i,a';\tau_i) = 0\}$, and thus,
\begin{align*}
	&\PR{\left(\mathbf{u}_i^\Delta(Y_i,Y_{-i}) \ge - \vecg{\lambda}_{i,\rho}(Y_i;\tau_i)\right)^c \mid \IF_i} \\
	&\qquad \le \sum_{a' \in A\setminus \{Y_i\}:\lambda_{i,\rho}(Y_i,a';\tau_i) > 0} \PR{u_i^\Delta(Y_i, a', Y_{-i}) <  - \lambda_{i,\rho}(Y_i,a';\tau_i) \mid \IF_i} \\
	&\qquad \le \sum_{a' \in A\setminus \{Y_i\}:\lambda_{i,\rho}(Y_i,a';\tau_i) > 0} \left\{\QR[i]{u_i^\Delta(Y_i, a', Y_{-i}) <  - \lambda_{i,\rho}(Y_i,a';\tau_i) \mid \IF_i}+\delta_i\left(\frac{\rho}{k-1}\right)\right\} \\
	&\qquad = r_i(Y_i;\tau_i), \quad (\PM \text{-a.s.}). \qedhere
\end{align*}
	
\end{proof}

\begin{proof}[\textbf{Proof of Proposition \ref{prop:pA1}}] We take $\rho \in (0,1)$, $\kappa \in [0, (1-\rho)/\rho)$, and $i \in N_{\PM}(\kappa,\rho)$. For $a\in A$, we define the events
\[
	S_{i,U}(a)\eqdef\{\vec{u}_i^{\Delta }(a,Y_{-i};\typ_i) \ge -\vecg{\lambda}_{i,\rho}(a;\typ_i)\} \qtext{and}\quad S_{i,L}(a)\eqdef \bigcup_{a' \in A\setminus \{a\}} S_{i,U}(a').
\]
By the definition of $\vecg{\lambda}_{i,\rho}$ and Theorem \ref{thm:tA1}, we have
\begin{align*}
	\sum_{a\in A}\PR{S_{i,U}(a)\mid \IF_i}\ind\{Y_i=a\} \ge 1-\sum_{a\in A}r_{i,\rho}(a;\typ_i)\ind\{Y_i=a\}, \quad (\PM\text{-a.s.}),
\end{align*}
where $r_{i,\rho}(a;\typ_i)$ is as defined in \eqref{ri a}. Therefore, noticing that $r_{i,\rho}(a;\typ_i)\le r_{i,a,U}(\kappa, \rho)$,
\begin{align}
\label{eq:ineqU}
		\ind\{Y_i=a\} \le \ind \left\{\PR{S_{i,U}(a)\mid \IF_i}\ge 1-r_{i,a,U}(\kappa, \rho) \right\}, \quad (\PM\text{-a.s.}).
\end{align}
Since $S_{i,U}(a)\in\sigma(Y_{-i},\tau_i)$, taking the conditional expectation given $\X$ on both sides of the inequality in \eqref{eq:ineqU} and using Markov's inequality, we find that
\begin{align*}
	\PR{Y_i=a \mid \X}&\le \frac{\PR{S_{i,U}(a)\mid \X}}{1-r_{i,a,U}(\kappa, \rho)}, \quad (\PM\text{-a.s.}).
\end{align*}

On the other hand, again by the definition of $\vecg{\lambda}_{i,\rho}$ and Theorem \ref{thm:tA1},
\begin{align*}
	\sum_{a' \in A \setminus \{a\}} \PR{S_{i,U}(a') \mid \IF_i} \ind\{Y_i = a'\} &\ge 1 - \sum_{a' \in A \setminus\{a\}}r_{i,\rho}(a';\tau_i)\ind\{Y_i = a'\}\\
	&\ge 1 - \max_{a' \in A \setminus \{a\}} r_{i,\rho}(a';\typ_i).
\end{align*}
Note that the events $S_{i,U}(a)$ and $S_{i,U}(a')$ with $a\ne a'$ are disjoint because
\begin{align*}
	\lambda_{i,\rho}(a,a';\tau_i) = \lambda_{i,\rho}(a',a;\tau_i) \ge 0.
\end{align*}
(Recall the definition in \eqref{eq:lambda_i_rho}.) Hence,
\begin{align*}
	\ind\{Y_i\ne a\} &\le \ind \left\{\PR{S_{i,L}(a)\mid \IF_i}\ge 1-r_{i,a,L}(\kappa, \rho) \right\}, \quad (\PM\text{-a.s.}),
\end{align*}
and following the same argument as before yields
\begin{align*}
	\PR{Y_i\ne a \mid \X}&\le \frac{\PR{S_{i,L}(a)\mid \X}}{1-r_{i,a,L}(\kappa, \rho)}, \quad (\PM\text{-a.s.}).
\end{align*}
These inequalities give the desired result.
\end{proof}

\begin{proof}[\textbf{Proof of Theorem \ref{thm:bootstrap_consisteny_mlt}}] Throughout the proof, we let $d \eqdef (k-1)m$ for simplicity. Also, for simplicity, we suppress the notation $(\kappa,\rho)$ in the arguments below. Define
\begin{align}
	\begin{aligned}
		\label{eq:mu}
		\vecg{\mu}_L \eqdef \E_{\PM}[\vecg{\hat{\mu}}_L\mid \X] \qtext{and}\quad
		\vecg{\mu}_U \eqdef \E_{\PM}[\vecg{\hat{\mu}}_U\mid \X].
	\end{aligned}
\end{align}
Let $\G\eqdef \sigma(Y_1,\ldots,Y_n,\X)$ and let
\begin{align}
\label{eq:zeta_mlt}
	\vecg{\zeta}\eqdef \frac{1}{n}\sum_{i \in N} (\vec{Y}_i-\E_{\PM}[\vec{Y}_i\mid \X])\otimes \mathbf{g}_i(X_i),
\end{align}
so that we have
\begin{align}
\label{eq:zeta_rpt}
	\vecg{\zeta}=\vecg{\tilde{\mu}}_L-\vecg\mu_L=\vecg{\tilde{\mu}}_U-\vecg\mu_U, \quad (\PM\text{-a.s.}).
\end{align}
Since $Y_1,\ldots, Y_n$ are conditionally independent given $\X$, $\vecg{\zeta}$ is the sum of conditionally independent random vectors given $\X$.

Let $Z$ be a standard normal random vector in $\R^d$ independent of $\G$. Define
\begin{align}
\label{eq:bound_T}
	\begin{aligned}
		\widetilde{T}&\eqdef \T\left(\sqrt{n}\vecg{\tilde{\mu}}_L,\sqrt{n}\vecg{\tilde{\mu}}_U\right) \qtext{and} \\
		\widetilde{T}'&\eqdef \T\left(\sqrt{n}(\vecg{\zeta}+\vecg{\mu}_L\wedge (-\vecg{\mu}_U)),\sqrt{n}(\vecg{\zeta}-\vecg{\mu}_L\wedge (-\vecg{\mu}_U))\right).
	\end{aligned}
\end{align}
We also introduce the following functionals of the random vector $Z$:
\begin{align*}
	S'&\eqdef \T\left(V^{1/2}Z+\sqrt{n}(\vecg{\mu}_L\wedge (-\vecg{\mu}_U)),V^{1/2}Z-\sqrt{n}(\vecg{\mu}_L\wedge (-\vecg{\mu}_U))\right) \qtext{and} \\
	S^{*}&\eqdef \T\left(W^{1/2}Z+\sqrt{n}(\vecg{\mu}_L\wedge (-\vecg{\mu}_U)),W^{1/2}Z-\sqrt{n}(\vecg{\mu}_L\wedge (-\vecg{\mu}_U))\right),
\end{align*}
where
\begin{align}
	\label{eq:V&W}
	\begin{aligned}
		V &\eqdef n\E_{\PM}\left[\vecg{\zeta}\vecg{\zeta}^{\top}\mid \X\right] \qtext{and} \\
		W &\eqdef \frac{1}{n}\sum_{i \in N} \E_{\PM}\left[\left((\vec{Y}_i-\vecg{\mu}_i^{*})\otimes \vec{g}_i(X_i)\right)\left((\vec{Y}_i-\vecg{\mu}_i^{*})\otimes \vec{g}_i(X_i)\right)^{\top}\mid \X\right],
	\end{aligned}
\end{align}
respectively. (Recall the definition of $\vecg{\mu}_i^{*}$ after \eqref{eq:zeta*}.) Let $c_\gamma$ denote the $\gamma$-quantile of the conditional distribution of $S'$ given $\X$, and let $q_\gamma$ denote the $\gamma$-quantile of the conditional distribution of $\|W^{1/2} Z\|_\infty$ given $\X$.

For random variables $X'$ and $X''$ and sub-$\sigma$-fields $\F',\F'' \subset\H$, we define
\begin{equation}
\label{eq:aux_dk}
	\DK{z}{X',X''\mid \F',\F''}\eqdef \sup_{t\ge z}\,\abs{F_{X'}(t\mid\F')-F_{X''}(t\mid\F'')},
\end{equation}
where $F_{X'}(\csdot\mid\F')$ and $F_{X''}(\csdot\mid\F'')$ are the conditional cdfs of $X'$ and $X''$ given $\F'$ and $\F''$, respectively (when $\F'=\F''$ we denote this measure by $\DK{z}{X',X''\mid\F'}$; also we drop the superscript $z$ when the supremum is taken over $\R$). Recall the definition $\vecg{\zeta}^*$ in \eqref{eq:zeta*}. Define
\begin{equation}
\label{eq:bound_T*}
	\widetilde{T}^{*}\eqdef\T\left(\sqrt{n}(\vecg{\zeta}^{*}+\vecg{\mu}_L\wedge (-\vecg{\mu}_U)),\sqrt{n}(\vecg{\zeta}^{*}-\vecg{\mu}_L\wedge (-\vecg{\mu}_U))\right)
\end{equation}
and let $\tilde{c}_\gamma^{*}$ denote the $\gamma$-quantile of the bootstrap distribution of $\widetilde{T}^{*}$. We let
\begin{align*}
	\Delta&\eqdef \DK{\epsilon}{\widetilde{T}^{*},S^{*}\mid \G} \qtext{and}\\
	\tilde \Delta&\eqdef \DK{\epsilon}{\sqrt{n}\|\vecg{\zeta}^*\|_\infty, \|W^{1/2} Z\|_\infty \mid \G}.
\end{align*}

\begin{claim}
\label{claim:B1}
$W - V$ is positive semidefinite, $(\PM\text{-a.s.})$.
\end{claim}

\begin{subproof}
Since $Y_i$'s are conditionally independent given $\X$, and $\vecg{\mu}_i^{*}$'s are $\sigma(\X)$-measurable, we can write
\[
	W = V + \frac{1}{n}\sum_{i \in N} \E_{\PM}\left[\left((\E_{\PM}[\vec{Y}_i\mid \X] - \vecg{\mu}_i^*)\otimes \mathbf{g}_i(X_i)\right)
	 \left((\E_{\PM}[\vec{Y}_i\mid \X] - \vecg{\mu}_i^*)\otimes \mathbf{g}_i(X_i)\right)^\top \mid \X \right], \quad (\PM\text{-a.s.}).
\]
This gives the desired result.
\end{subproof}

\begin{claim}
\label{claim:B2}
For any $\PM\in \PS_0$ and $\upsilon\in (0,\gamma \wedge (1 - \nu/2))$,
\begin{align}
\label{eq:bounds}
	\begin{aligned}
		\PR{c_{\gamma-\upsilon}>\tilde{c}_\gamma^{*} \vee \epsilon} \le C b_n \qtext{and}\quad 
		\PR{q_{1-\nu/2-\upsilon}> q_{1-\nu/2}^{*}} &\le C b_n,
	\end{aligned}
\end{align}
where $C>0$ is a constant that does not depend on $n$ or $\PM$, and
\[
	b_n\eqdef \frac{1}{\upsilon(r_n^2 n)^{1/6}}+\frac{1}{\upsilon \epsilon^3 \sqrt{n}}+\frac{\PR{\min_{1 \le i \le n} \lambda_{\min}(\Sigma_i)<r_n}}{\upsilon}.
\]
\end{claim}

\begin{subproof}
We first prove the first inequality in \eqref{eq:bounds}. Since $W-V$ is positive semidefinite by Claim \ref{claim:B1}, and sets of the form $\{x\in \R^d:\T(x+a,x-a)\le t\}$ with $a\in [0,\infty)^d$ and $t\ge 0$ are convex and symmetric under reflection, Theorem 1 in \cite{Jensen:84} implies that for all $t\in \R$,
\begin{align}
\label{eq:ineq3}
	\PR{S'\le t\mid \X}\ge \PR{S^{*}\le t\mid \X}, \quad (\PM\text{-a.s.}).
\end{align}
On the event $\{\Delta\le \upsilon\}\cap \{c_{\gamma-\upsilon}>\epsilon\}$,
\begin{align*}
	\PR{S'\le \tilde{c}^{*}_{\gamma} \vee \epsilon\mid \G} &\ge \PR{S^{*}\le \tilde{c}_{\gamma}^{*}\vee \epsilon\mid \G} \\
	&\ge \PR{\widetilde{T}^{*}\le \tilde{c}_\gamma^{*} \vee \epsilon\mid \G}-\upsilon \\
	&\ge \gamma-\upsilon=\PR{S'\le c_{\gamma-\upsilon} \mid \G}, \quad (\PM\text{-a.s.}),
\end{align*}
which implies that $\tilde{c}_\gamma^{*}\vee \epsilon \ge c_{\gamma-\upsilon}$. Hence, we obtain that
\[
	\PR{c_{\gamma-\upsilon}>\tilde{c}_{\gamma}^{*} \vee \epsilon\mid \X} \le \PR{\Delta > \upsilon\mid \X}, \quad (\PM\text{-a.s.}).
\]

Since $\vec{Y}_i,\ldots,\vec{Y}_n$ are conditionally independent given $\X$,\footnote{\label{footnote: norms} Here, for a given matrix $A = [a_{ij}]$, $\|A\|_{e,1}$ denotes the elementwise $\ell_1$ norm, i.e., $\|A\|_{e,1} = \sum_{i,j} |a_{ij}|$, and $\|A\|_{e,\infty}$ denotes the elementwise sup-norm, i.e., $\|A\|_{e,\infty} = \max_{i,j}|a_{ij}|$.}
\begin{align*}
	H \eqdef\E_{\PM}\left[\norm{W- n\E_{\PM}\left[\vecg{\zeta}^{*} \vecg{\zeta}^{* \top}\mid \G\right]}_{e,\infty}\mid \X \right]&\le \E_{\PM}\left[\norm{W- n\E_{\PM}\left[\vecg{\zeta}^{*} \vecg{\zeta}^{* \top}\mid \G\right]}_{e,1}\mid \X \right] \\
	&\le \frac{d^2 C_g^2}{\sqrt{n}}, \quad (\PM\text{-a.s.}),
\end{align*}
for the constant $C_{\vec{g}}>0$ in Theorem \ref{thm:bootstrap_consistency}. Applying Lemmas \ref{lemma:aux_psi_bnd} and \ref{lemma:aux_normal_approx} in the Supplemental Note, the fact that $W - V$ is positive semidefinite, and noticing that $\lambda_{\min} (V)  \ge \min_{1 \le i \le n} \lambda_{\min}(\Sigma_i)$, we deduce that
\begin{align}
\label{eq:upsilon_ineq}
	\begin{aligned}
		\upsilon\PR{\Delta>\upsilon}&\le \E_{\PM}\Delta\le \frac{C d^{2/3}}{r_n^{1/3}}\E_{\PM}H^{1/3}+\frac{C \sqrt{d}}{\epsilon^3}\E_{\PM}H+\PR{\min_{1 \le i \le n} \lambda_{\min}(\Sigma_i)<r_n} \\
		&\le \frac{C'}{(r_n^2 n)^{1/6}}+\frac{C'}{\epsilon^3 \sqrt{n}}+\PR{\min_{1 \le i \le n} \lambda_{\min}(\Sigma_i)<r_n},
	\end{aligned}
\end{align}
where $C,C'>0$ are constants that do not depend on $n$ or $\PM$.

Let us turn to the second statement. Similarly as before, we obtain
\[
	\PR{q_{1-\nu/2-\upsilon} > q_{1-\nu/2}^{*} \mid \X}\le \PR{\tilde \Delta > \upsilon\mid \X}.
\]
Using the same arguments as in \eqref{eq:upsilon_ineq}, and noting that $\lambda_{\min}(V) \le \lambda_{\min}(W)$ by Claim \ref{claim:B1}, we obtain the desired result.
\end{subproof}

\begin{claim}
\label{claim:B3}
For any $\PM\in \PS_0$ and $\upsilon \in (0,\gamma)$ and $\epsilon_n'>0$,
\begin{align*}
	\PR{\widetilde{T}> (\tilde{c}_\gamma^{*} - \epsilon_n' ) \vee \epsilon }-(1-\gamma)
	\le C\left(h_{1,n}+\frac{h_{2,n}}{\upsilon}+\left(2+\frac{1}{\upsilon}\right)\PR{\min_{1 \le i \le n} \lambda_{\min}(\Sigma_i)<r_n}+\upsilon\right),
\end{align*}
where $C>0$ is a constant that does not depend on $n$ or $\PM$, and
\begin{align*}
	h_{1,n}&\eqdef \frac{1}{(r_n^3 n)^{1/8}}+\frac{\epsilon_n'}{r_n^{1/2}}+\frac{1}{\epsilon^{4}\sqrt{n}} \qtext{and}\quad h_{2,n}\eqdef \frac{1}{(r_n^2 n)^{1/6}}+\frac{1}{\epsilon^{3}\sqrt{n}}.
\end{align*}
\end{claim}

\begin{subproof}
Let us define the event:
\begin{align*}
	A_n = \left\{ c_{\gamma - \upsilon} \vee \epsilon \le \tilde{c}_\gamma^{*} \vee \epsilon  \right\}.
\end{align*}
Using \eqref{eq:ineq3}, and noting that $\widetilde T' \ge \widetilde T$ from \eqref{eq:zeta_rpt}, we find that 
\begin{align*}
	&\PR{\widetilde{T}> (\tilde{c}_\gamma^{*} - \epsilon_n') \vee \epsilon \mid \X}\\
	&\qquad\le \PR{ \{\widetilde{T}'> (\tilde{c}_\gamma^{*} - \epsilon_n') \vee \epsilon \} \cap A_n \mid \X} + \PR{A_n^c \mid \X}\\
	&\qquad\le \PR{ \widetilde{T}'> (c_{\gamma - \upsilon} - \epsilon_n' ) \vee \epsilon \mid \X} + \PR{c_{\gamma - \upsilon} \vee \epsilon > \tilde{c}_\gamma^{*} \vee \epsilon \mid \X} \\
	&\qquad\le \DK{\epsilon}{\widetilde{T}',S'\mid \X} + \PR{S' > (c_{\gamma - \upsilon} - \epsilon_n' ) \vee \epsilon  \mid \X} + \PR{c_{\gamma - \upsilon} \vee \epsilon > \tilde{c}_\gamma^{*} \vee \epsilon \mid \X},
\end{align*}
where $\DK{\epsilon}{\csdot,\csdot\mid \X}$ is defined in \eqref{eq:aux_dk}. Hence,
\begin{align}
\label{eq:dev}
	\begin{aligned}
		&\PR{\widetilde{T}> (\tilde{c}_\gamma^{*} - \epsilon_n' ) \vee \epsilon \mid \X}-(1-\gamma) \\
		&\qquad\le \PR{\widetilde{T}> (\tilde{c}_\gamma^{*} - \epsilon_n' ) \vee \epsilon \mid \X}-\PR{S'> c_\gamma \vee \epsilon \mid \X} \\ 
		&\qquad\le \DK{\epsilon}{\widetilde{T}',S'\mid \X} +  \PR{c_\gamma \vee \epsilon > \tilde{c}_\gamma^{*} \vee \epsilon \mid \X},\\ 
		&\qquad \qquad + \PR{(c_{\gamma - \upsilon} - \epsilon_n' ) \vee \epsilon  < S' \le c_{\gamma-\upsilon} \vee \epsilon  \mid \X} + \PR{c_{\gamma - \upsilon} \vee \epsilon < S' \le c_\gamma \vee \epsilon  \mid \X}.
	\end{aligned}
\end{align}
We can bound the last probability in \eqref{eq:dev} by $\upsilon$ by the definition of $c_{\gamma}$. The second probability in \eqref{eq:dev} is bounded by $Cb_n$ using Claim \ref{claim:B2}. As for the third probability, we use Lemmas \ref{lemma:aux_psi_bnd} and  \ref{lemma:normal_anticoncentration} in the Supplemental Note to bound it by
\begin{align*}
   \PR{\lambda_{\min}(\Sigma) \le r_n \mid \X} + \frac{d \epsilon_n'}{r_n^{1/2}}.
\end{align*}
Finally, as for the term $\DK{\epsilon}{\widetilde{T}',S'\mid \X}$, the largest eigenvalue $\lambda_{\max}(V)$ of $V$ is bounded, i.e.,
\[
	\lambda_{\max}(V)\le n\E_{\PM}\left[\normin{\vecg{\zeta}}^2 \mid \X \right]\le C_{\vec{g}}^2d, \quad (\PM\text{-a.s.}).
\]
Therefore, using Lemmas \ref{lemma:aux_psi_bnd} and \ref{lemma:aux_CLT} in the Supplemental Note, and setting
\[
	\Gamma\eqdef n^{-3/2}\sum_{i \in N} \E_{\PM}\left[\norm{(\vec{Y}_i-\E[\vec{Y}_i\mid \X])\otimes \mathbf{g}_i(X_i)}_3^3\mid \X \right],
\]
we find that since $\lambda_{\min} (V)  \ge \min_{1 \le i \le n} \lambda_{\min}(\Sigma_i)$,
\begin{align*}
	\E_{\PM}\DK{\epsilon}{\widetilde{T}',S'\mid \X}&\le \frac{Cd^{3/4}}{r_n^{3/8}}\E_{\PM}\Gamma^{1/4}+\frac{C\sqrt{d}}{\epsilon^4}\E_{\PM}\Gamma+\PR{\min_{1 \le i \le n} \lambda_{\min}(\Sigma_i)<r_n} \\
	&\le\frac{C'}{(r_n^3 n)^{1/8}}+\frac{C'}{\epsilon^4 \sqrt{n}}+\PR{\min_{1 \le i \le n} \lambda_{\min}(\Sigma_i)<r_n},
\end{align*}
where $C,C'>0$ are constants that do not depend on $n$ or $\PM$. The desired result follows by combining this bound with that in Claim \ref{claim:B2}.
\end{subproof}

Let
\begin{align*}
	R_L&\eqdef \left\{[\vecg\mu_j] \in \R^d : \min_{1 \le j \le d}\sqrt{n}\left(\vecg\mu_j-\vecg{\tilde\mu}_{L,j}\right)\ge q_{1-\nu/2}^{*}\right\} \qtext{and} \\
	R_U&\eqdef \left\{[\vecg\mu_j] \in \R^d : \max_{1 \le j \le d}\sqrt{n}\left(\vecg\mu_j-\vecg{\tilde\mu}_{U,j}\right)\le -q_{1-\nu/2}^{*}\right\}.
\end{align*}

\begin{claim}
\label{claim:B4}
For any $\PM\in \PS_0$,
\[
	\PR{\vecg{\mu}_L\notin R_L}+\PR{\vecg{\mu}_U\notin R_U}-\nu \le C \left(b_n + n^{-1/4} r_n^{-3/2}\right) + \PR{\min_{1 \le i \le n} \lambda_{\min}(\Sigma_i)<r_n},
\]
where $C>0$ is a constant that does not depend on $n$ or $\PM$, and $b_n$ is defined in Claim \ref{claim:B2}.
\end{claim}

\begin{subproof}
We reuse the notation from the proof of Claim \ref{claim:B3}. In addition, for $x\in \R^d$, let $M(x)\eqdef \max_{1\le j\le d}\{x_j\}$. For any $\upsilon \in (0,\gamma)$,
\begin{align*}
	\PR{\vecg{\mu}_L\notin R_L\mid \X} &= \PR{ M(\sqrt{n} \vecg{\zeta}) > q_{1-\nu/2}^* \mid \X} \\
	&\le \PR{q_{1-\nu/2-\upsilon}> q_{1-\nu/2}^{*}\mid \X}  +\PR{ M(\sqrt{n} \vecg{\zeta}) > q_{1-\nu/2 - \upsilon} \mid \X}.
\end{align*}
We bound the last probability by
\begin{align*}
	&\PR{ M(V^{1/2}Z) > q_{1-\nu/2-\upsilon} \mid \X} + \DK{}{M(\sqrt{n}\vecg{\zeta}),M(V^{1/2}Z)\mid \X} \\
	&\qquad\le \PR{ \|V^{1/2}Z\|_\infty > q_{1-\nu/2-\upsilon} \mid \X} + \DK{}{M(\sqrt{n}\vecg{\zeta}),M(V^{1/2}Z)\mid \X} \\
	&\qquad\le \PR{ \|W^{1/2}Z\|_\infty > q_{1-\nu/2-\upsilon} \mid \X} + \DK{}{M(\sqrt{n}\vecg{\zeta}),M(V^{1/2}Z)\mid \X} \\
	&\qquad\le \frac{\nu}{2} + \upsilon + \DK{}{M(\sqrt{n}\vecg{\zeta}),M(V^{1/2}Z)\mid \X}, \quad (\PM\text{-a.s.}),
\end{align*}
where the second inequality uses Theorem 1 of \cite{Jensen:84}. By Theorem 3.1 of \cite{Kojevnikov/Song:22},
\[
	\DK{}{M(\sqrt{n}\vecg{\zeta}),M(V^{1/2}Z)\mid \X}\le C(1 \vee \ln(d))^{5/4}(\gamma_1 + \gamma_3)^{1/2} n^{-1/4}, \quad (\PM\text{-a.s.}),
\]
where $C$ is a constant that does not depend on $n$ or $\PM$, and for $s=1,3$,
\[
	\gamma_s \eqdef \max_{1 \le i \le n} \left( \E_{\PM}\left[\|\vecg{\zeta}_i\|_\infty^s \mid \X\right] + \bar \sigma_i^s \left(1 \vee \ln(d)\right)^{s/2} \right) / \underline{\lambda}^s,
\]
with
\begin{align*}
	\bar \sigma_i^2 = \frac{1}{n} \max_{1 \le j \le d} [\Sigma_i]_{jj} \qtext{and}\quad \underline{\lambda}^2 = \frac{1}{n} \min_{1 \le i \le n} \lambda_{\min}(\Sigma_i).
\end{align*}
Note that $\bar \sigma_i^2 \le C_g^2 m^2/n$, and hence, on the event $\left\{\min_{1 \le i \le n} \lambda_{\min}(\Sigma_i) \ge r_n\right\}$, we have
\begin{align*}
	\DK{}{M(\sqrt{n}\vecg{\zeta}),M(V^{1/2}Z)\mid \X} &\le Cn^{-1/4}(1 \vee \ln(d))^{5/4}\sum_{s = 1,3} r_n^{-s/2} \left( 1 + n^{-s/2} (1 \vee \ln(d)\right) \\
	&\le Cn^{-1/4} r_n^{-3/2}, \quad (\PM\text{-a.s.}),
\end{align*}
because $r_n$ is a positive, bounded sequence. Therefore, using Claim \ref{claim:B2},
\[
	\PR{\vecg{\mu}_L\notin R_L}\le \frac{\nu}{2}+ \upsilon + C \left(b_n + n^{-1/4} r_n^{-3/2}\right),
\]
for a constant $C>0$ that does not depend on $n$ or $\PM$. Finally, the same bound holds for $\PR{\vecg{\mu}_U\notin R_U}$. Making $\upsilon>0$ arbitrarily small, we obtain the desired bound in Claim \ref{claim:B4}.
\end{subproof}

Let us turn to completing the proof of the theorem. On the event $\mathcal{M}(\vec{w}_L,\vec{w}_U)$, $T\le \widetilde{T}$,
\begin{align*}
	\vecg{\hat\varphi}_L&\le \left[\vecg{\tilde{\mu}}_L-\ind_d\csdot q_{1-\nu/2}^{*}/\sqrt{n}\right]_{+} \qtext{and} \\
	\vecg{\hat\varphi}_U&\le \left[\vecg{\tilde{\mu}}_U+\ind_d\csdot q_{1-\nu/2}^{*}/\sqrt{n}\right]_{-}.
\end{align*}
The latter inequalities imply that under $\PM\in \PS_0$, on the event $\mathcal{M}(\vec{w}_L,\vec{w}_U)\cap \{\vecg{\mu}_L\in R_L\}\cap\{\vecg{\mu}_U\in R_U\}$, we have
\begin{align}
	\label{ineq32}
	\vecg{\hat\varphi}_L \wedge \vecg{\hat\varphi}_U \le [\boldsymbol{\mu}_L]_{+} \wedge [\boldsymbol{\mu}_U]_{-}.
\end{align}
Let us define
\begin{equation}
	\label{eq:bar_T*}
	\overline{T}^{*}\eqdef\T\left(\sqrt{n}\left(\vecg{\zeta}^{*}+[\vecg{\mu}_L]_{+}\wedge [\vecg{\mu}_U]_{-} \right),\sqrt{n}\left(\vecg{\zeta}^{*}-[\vecg{\mu}_L]_{+}\wedge [\vecg{\mu}_U]_{-}\right)\right).
\end{equation}
Observe that
\begin{align*}
	|\overline{T}^{*} - \widetilde{T}^{*}| &\le 2 \sqrt{n}\left\|[\vecg{\mu}_L]_{+}\wedge [\vecg{\mu}_U]_{-} - \vecg{\mu}_L \wedge (-\vecg{\mu}_U) \right\|_1 \\
	&\le 2 \sqrt{n}\left(\left\|[\vecg{\mu}_L]_{+} - \vecg{\mu}_L \right\|_1 + \left\|[\vecg{\mu}_U]_{-} - (-\vecg{\mu}_U) \right\|_1\right),
\end{align*}
where the last inequality uses the fact that $a \wedge b = (a + b + |a-b|)/2$ for $a,b \in \mathbf{R}$. By Proposition \ref{prop:pA1}, we have
\begin{align*}
	\sqrt{n}\left\|[\vecg{\mu}_L]_{+} - \vecg{\mu}_L \right\|_1 &= \frac{1}{\sqrt{n}}\sum_{i \in N \setminus N(\kappa,\rho)} 
	\left\| \E_{\PM}\left[ \mathbf{e}_{i,L} \otimes \mathbf{g}_i(X_i) \mid \X \right]\right\|_1 \le \xi_n,
\end{align*}
where
\begin{align*}
	\xi_n \eqdef \sqrt{n} \left(1 - \frac{n_{\PM}(\kappa,\rho)}{n}\right) \cdot m (k-1) C_{\mathbf{g}}.
\end{align*}
Arguing similarly for $\sqrt{n}\left\|[\vecg{\mu}_U]_{-} - (-\vecg{\mu}_U) \right\|_1$, we find that
\begin{align*}
	|\overline{T}^{*} - \widetilde{T}^{*}| \le 4 \xi_n.
\end{align*}
Hence, under $\PM\in \PS_0$, on the event $\mathcal{M}(\vec{w}_L,\vec{w}_U)\cap \{\vecg{\mu}_L\in R_L\}\cap\{\vecg{\mu}_U\in R_U\}$, from \eqref{ineq32}, we have
\begin{align*}
	T^{*}\ge \widetilde{T}^{*} - 4 \xi_n.
\end{align*}
(Recall the definition of $T^*$ in \eqref{eq:def_T*}.) Consequently,
\begin{align}
\label{eq:tail_prob}
	\begin{aligned}
		\PR{T>c_\gamma^{*}\vee \epsilon}&\le \PR{\widetilde{T}>(\tilde{c}_\gamma^{*} - 4\xi_n) \vee \epsilon  }+\PR{\mathcal{M}(\vec{w}_L,\vec{w}_U)^c} \\
		&\quad+\PR{\vecg{\mu}_L\notin R_L}+\PR{\vecg{\mu}_U\notin R_U}.
	\end{aligned}
\end{align}
We set $\epsilon_n' = 4 m (k-1) C_{\mathbf{g}} M_{\kappa,\rho} r_n$, where $M_{\kappa,\rho}>0$ is a constant that appears in Assumption \ref{assum:RE-B}. By this assumption, we have $4 \xi_n \ge \epsilon_n'$ for all $\mathsf{P} \in \mathcal{P}_0$ and $n \ge 1$. Hence,
\begin{align*}
	\limsup_{n\to\infty}\sup_{\PM\in \PS_0} \PR{\widetilde{T}>(\tilde{c}_\gamma^{*} - 4\xi_n) \vee \epsilon  } \le \limsup_{n\to\infty}\sup_{\PM\in \PS_0} \PR{\widetilde{T}> (\tilde{c}_\gamma^{*} - \epsilon_n') \vee \epsilon  }.
\end{align*}
The last term on the right hand side vanishes by \eqref{nkappa}. We find from Claim \ref{claim:B3} that the first limsup on the right hand side is bounded by $1- \gamma$. Combining Claim \ref{claim:B4}, Assumption \ref{assum: nondeg2}, and Proposition \ref{prop:pA2}, in view of \eqref{eq:tail_prob}, we find that for any $\upsilon\in (0,\gamma)$,
\[
	\limsup_{n\to\infty}\sup_{\PM\in \PS_0}\PR{T>c_\gamma^{*} \vee \epsilon}\le \alpha+\upsilon.
\]
Since $\upsilon$ is arbitrary, the result follows.
\end{proof}
\subsection{Notation List}

For brevity, we suppress the argument notations, $(\theta_0,\rho)$ and $(\kappa,\rho)$, in many places below.

{%
	\footnotesize
	\setlength\LTleft{-3.6mm}
	\begin{longtable}{lll}
		\hline\hline
		Notation & Description & Place of Definition \\
		\hline
		& & \\
		\endhead
		& & \\
		\hline\hline
		\endfoot
		
		$\delta_i$ &: the distance between $\QR[i]{\csdot \mid \mathcal{I}_i}$ and $\PR{\csdot \mid \mathcal{I}_i}$ on high probability events & \eqref{delta rho}\\

		$e_{i,L}$ &: $Y_i- (1-\pi_{i,L}/(1-r_i))$ & \eqref{eq:sample_e} \\

		$e_{i,U}$ &: $Y_i- \pi_{i,U}/(1-r_i)$ & \eqref{eq:sample_e} \\
		
		$e_{i,a,L}$ &: $\ind\{Y_i=a\}-(1-\pi_{i,a,L}(Y_{-i},X_i)/(1-r_{i,a,L}))$ & \eqref{eq:sample_e_mlt} \\
		
		$e_{i,a,U}$ &: $\ind\{Y_i=a\}-\pi_{i,a,U}(Y_{-i},X_i)/(1-r_{i,a,U})$ & \eqref{eq:sample_e_mlt} \\

		$\tilde e_{i,a,L}$ &: $\ind\{Y_i=a\}-(1-\E_{\PM}[\pi_{i,a,L}\mid \X]/(1-r_{i,a,L}))$ & \eqref{eq:infeasible_e} \\

		$\tilde e_{i,a,U}$ &: $\ind\{Y_i=a\}- \E_{\PM}[\pi_{i,a,U}\mid \X]/(1-r_{i,a,U})$ & \eqref{eq:infeasible_e} \\
		
		$\vec{e}_{i,L}$, $\vec{e}_{i,U}$ &: vectors whose elements are $e_{i,a,L}$ and $e_{i,a,U}$ with $a \in A\setminus \{a_1\}$ & Below \eqref{eq:sample_e_mlt}\\

		$\vec{\tilde e}_{i,L}$, $\vec{e}_{i,U}$ &: vectors whose elements are $\tilde e_{i,a,L}$ and $\tilde e_{i,a,U}$ with $a \in A\setminus \{a_1\}$ & Below \eqref{eq:infeasible_e}\\

		$g_{i,\ell}$ &: nonnegative functions of $X_i$ & Above \eqref{eq:sample_moments_mlt} \\

		$\vec{g}_i$ &: $[g_{i,1},\ldots, g_{i,m}]^{\top}$ & Above \eqref{eq:sample_moments_mlt} \\

		$\G$ &: $\sigma(Y_1,\ldots,Y_n,\X)$ & Above \eqref{eq:zeta_mlt} \\

		$\IF_i$ &: The information set of player $i$, $\sigma(\eta_i, \X)$ & \eqref{eq:info} \\

		$\vecg{\mu}_i^{*}$ &: the vector with entries $\left(\frac{1}{2}[1-\pi_{i,a,L}/(1-r_{i,a,L}) + \pi_{i,a,U}/(1-r_{i,a,U})]_+\right) \wedge 1$ & Below \eqref{eq:zeta*}\\

		$\vecg{\mu}_L, \vecg{\mu}_U$ &:  $\E_{\PM}[\vecg{\hat \mu}_L\mid \X], \E_{\PM}[\vecg{\hat \mu}_U\mid \X]$ & \eqref{eq:mu} \\

		$\vecg{\hat{\mu}}_L, \vecg{\hat{\mu}}_U$ &: $n^{-1}\sum_{i \in N} e_{i,L}\,\vec{g}_{i}(X_i)$, $n^{-1}\sum_{i \in N} e_{i,U}\,\vec{g}_{i}(X_i)$ & \eqref{eq:sample_moments}, \eqref{eq:sample_moments_mlt} \\

		$\vecg{\tilde{\mu}}_L, \vecg{\tilde{\mu}}_U$ &: $n^{-1}\sum_{i \in N} \vec{\tilde{e}}_{i,L}\otimes \vec{g}_i(X_i)$, $n^{-1}\sum_{i \in N} \vec{\tilde{e}}_{i,U}\otimes \vec{g}_i(X_i)$ & \eqref{eq:infeasible_moments} \\

		$\lambda_{i,\rho}$ &: a $\rho$-hindsight regret & \eqref{eq:hr0} \\

		$\Lambda_i(\tau_i)$ &: $\sum_{j\in N\setminus \{i\}}V_{j}^2(u_i^{\Delta}(1,\cdot;\tau_i))$ & \eqref{eq:hr0} \\

		$N$ &: the total set of players, $\{1,\ldots,n\}$ & \\

		$N(i)$ &: the strategic neighborhood of player $i$ & \\
		
		$n_{\PM}(\kappa,\rho)$ &: the number of players such that $\delta_i(\rho/(k-1)) \le \kappa \rho/(k-1)$ & \eqref{N(kappa)} \\
		
		$N_{\PM}(\kappa,\rho)$ &: the set of players such that $\delta_i(\rho/(k-1)) \le \kappa \rho/(k-1)$ & \eqref{N(kappa)} \\

		$\pi_{i,L}$ &: $\PR{u_{i}^{\Delta}(0,Y_{-i};\tau_i)\ge -\lambda_{i,\rho}\mid \X, Y_{-i}}$ & \eqref{eq:pi_LU} \\

		$\pi_{i,U}$ &: $\PR{u_{i}^{\Delta}(1,Y_{-i};\tau_i)\ge -\lambda_{i,\rho}\mid \X, Y_{-i}}$ & \eqref{eq:pi_LU} \\

		$\PM$ &: the objective probability &  See Section \ref{subsec:setup} \\

		$\QM_i$ &: the probability as the subjective belief of player $i$ & See Section \ref{subsec:setup} \\

		$\typ_i$ &: the payoff state of player $i$, $(X_i,\eta_i)$ & \eqref{eq:Typ_i} \\

		$T$ &: $\T\left(\sqrt{n}(\vecg{\hat{\mu}}_L+\vec{w}_L),\sqrt{n}(\vecg{\hat{\mu}}_U-\vec{w}_U)\right)$ & \eqref{eq:def_T} \\

		$T^*$ &: $\T\left(\sqrt{n}(\vecg{\zeta}^{*}+\vecg{\hat\varphi}_L\wedge  \vecg{\hat\varphi}_U),\sqrt{n}(\vecg{\zeta}^{*}-\vecg{\hat\varphi}_L\wedge \vecg{\hat\varphi}_U)\right)$ & \eqref{eq:def_T*} \\

		$\tilde T$ &: $\T\left(\sqrt{n}\vecg{\tilde{\mu}}_L,\sqrt{n}\vecg{\tilde{\mu}}_U\right)$ & \eqref{eq:bound_T} \\

		$\widetilde{T}'$ &: $\T(\sqrt{n}(\vecg{\zeta}+\vecg{\mu}_L\wedge (-\vecg{\mu}_U)),\sqrt{n}(\vecg{\zeta}-\vecg{\mu}_L\wedge (-\vecg{\mu}_U)))$ & \eqref{eq:bound_T} \\

		$\widetilde{T}^{*}$ &: $\T\left(\sqrt{n}(\vecg{\zeta}^{*}+\vecg{\mu}_L\wedge (-\vecg{\mu}_U)),\sqrt{n}(\vecg{\zeta}^{*}-\vecg{\mu}_L\wedge (-\vecg{\mu}_U))\right)$ & \eqref{eq:bound_T*} \\

		$\T(x,y)$ &: $\norm{[x]_{-}+[y]_{+}}_1$ & Below \eqref{eq:def_T} \\

		$u_i^{\Delta}(a,b;t)$ &: $u_i(a,b;t)-u_i(1-a,b;t)$ & \eqref{eq:u_i^Delta} \\

		$V$ &: $n\E_{\PM}[\vecg{\zeta} \vecg{\zeta}^\top \mid \X]$ & \eqref{eq:V&W} \\

		$V_j(f)$ &: the maximal variation of $f$ at the player $j$ or at the $j$-th coordinate & \eqref{eq:max_var} \\

		$\vec{w}_L$, $\vec{w}_U$ &: the sample-dependent vectors & See \eqref{eq:diL_diU} and below \\

		$W$ &: $n^{-1} \sum_{i \in N} \E_{\PM}\left[\left((\vec{Y}_i-\vecg{\mu}_i^{*})\otimes \vec{g}_i(X_i)\right)\left((\vec{Y}_i-\vecg{\mu}_i^{*})\otimes \vec{g}_i(X_i)\right)^{\top}\mid \X\right]$ & \eqref{eq:V&W} \\

		$\vec{Y}_i$ &: a vector of $\ind\{Y_i = a\}$ with $a$ running in $A \setminus \{a_1\}$ & \\

		$\vecg{\zeta}$ &: $n^{-1}\sum_{i \in N} (\vec{Y}_i-\E_{\PM}[\vec{Y}_i\mid X])\otimes \mathbf{g}_i(X_i)$ & \eqref{eq:zeta}, \eqref{eq:zeta_mlt}\\

		$\vecg{\zeta}^{*}$ &: $n^{-1}\sum_{i \in N}\left((\vec{Y}_i-\vecg{\mu}_i^{*})\otimes \vec{g}_i(X_i)\right)\, \varepsilon_i$ & \eqref{eq:zeta*} \\

		$\| x  \|$ &: the Euclidean norm of a vector $x$, i.e., $\sqrt{x^\top x}$ &  \\

		$\| x  \|_\infty$ &: the sup-norm of a vector $x = [x_j]$, i.e., $\max_{j} |x_j|$  & \\
		
		$\|x\|_1$ &: the $\ell_1$ norm of a vector $x$, i.e., $\sum_j |x_j|$ & \\

		$\| A\|_{e,1}$ &: the elementwise $\ell_1$ norm of matrix $A = [a_{ij}]$, i.e., $\| A\|_{e,1} = \sum_{i,j} |a_{ij}|$ & See footnote \ref{footnote: norms} \\

		$\| A\|_{e,\infty}$ &: the elementwise sup-norm of matrix $A = [a_{ij}]$, i.e., $\| A\|_{e,\infty} = \max_{i,j} |a_{ij}|$ & See footnote \ref{footnote: norms}
	\end{longtable}
}

\end{bibunit}

\clearpage
\setcounter{page}{1}

\vspace*{5ex minus 1ex}
\begin{center}
\Large \textsc{Supplemental Note to ``Econometric Inference on a Large Bayesian Game with Heterogeneous Beliefs''}
\bigskip
\end{center}
\thispagestyle{plain}
\date{\today}

\vspace*{3ex minus 1ex}
\begin{center}
\textsc{Denis Kojevnikov} and \textsc{Kyungchul Song} \\
\textit{Tilburg University and University of British Columbia} \\
\bigskip\bigskip
\end{center}

\begin{bibunit}[elsart-harv]
The supplemental note consists of two sections. In Section \ref{app:aux_results}, we collect the auxiliary results used to prove the main results of our paper, ``Econometric Inference on a Large Bayesian Game with Heterogeneous Beliefs''. Section \ref{app:sim_results} provides additional simulation results for the choices of $\rho = 0.000001, 0.00001, 0.0001, 0.001, 0.01$. In all cases, the results show stable finite sample coverage probabilities, although the size of the confidence sets become large when $\rho$ is extremely small or large, which is expected from the theory. Hence, we propose using $\rho = 0.001$ in the paper.

\appendix
\renewcommand{\thesubsection}{\Alph{subsection}}
\numberwithin{equation}{subsection}
\numberwithin{lemma}{subsection}
\numberwithin{claim}{subsection}
\numberwithin{definition}{subsection}
\setcounter{subsection}{4}
\setcounter{equation}{0}
\setcounter{lemma}{0}

\subsection{Auxiliary Results}
\label{app:aux_results}

In this section, we collect auxiliary results and their proofs. The notation in this section is self-contained. Let $(\Omega,\H,\PM)$ denote the underlying probability space. First, we present a conditional version of McDiarmid's inequality for a function under the conditional independence assumption. Let us define the maximal variation of a function $f:\X^d\to \R$, $d\ge 1$, at the $i$-th coordinate is given by
\[
	V_i(f)=\sup_{x\in \X^d,x'\in \X}\abs{f(x)-f(x_1,\ldots, x_{i-1},x',x_{i+1},\ldots,x_d)}.
\]

\begin{lemma}[McDiarmid's Inequality]
\label{lemma:aux_mcdiarmid}
Let $X$ be a random vector taking values in $\X^d$ such that $X_1,\ldots,X_d$ are conditionally independent given $\F\subset\H$ and let $Y$ be an $\F$-measurable random element taking values in a measurable space $(E,\mathcal{E})$. Consider a measurable map $f:\X^d \times E\to \R$ such that $\E\abs{f(X,Y)}<\infty$ and let $c_{i}\eqdef V_i(f(\cdot,Y))$. Then for any $\epsilon>0$,
\[
	\PR{f(X,Y)-\E[f(X,Y)\mid \F]\ge \epsilon \mid \F}\le \exp\left(-\frac{2\epsilon^2}{\sum_{i=1}^d c_{i}^{*2}}\right) \qtext{a.s.},
\]
where $c_{i}^{*}$ is the minimal measurable majorant of $c_{i}$.\footnote{
	Note that $c_{i}^*=c_{i}$ if, for example, the set $\X$ is countable.
}
\end{lemma}
The proof can proceed in the same way as the proof of Lemma 1.2 in \cite{McDiarmid:89}.

Next, we establish a number of results regarding Gaussian random vectors in $\R^d$ and their transformation $\T:\R^d\times \R^d\to \R$ given by
\[
	\T(x,y)\eqdef \norm{[x]_{-}+[y]_{+}}_1.
\]

Consider $X\sim \ND{0}{\Sigma}$, where $\Sigma$ is a $d\times d$ positive definite covariance matrix. For $1\le i\le d$, the marginal distribution of $(X_1,\dots,X_i)^{\top}$ is $\ND{0}{\Sigma^{(i)}}$, where $\Sigma^{(i)}$ is a block of $\Sigma$ corresponding to its first $i$ rows and columns, and for $1<i\le d$ the conditional distribution of $X_i$ given $X_1,\dots,X_{i-1}$ is also normal with variance given by the Schur complement $\Sigma^{(i)}/\Sigma^{(i-1)}$. Let $\Pi$ denote the set of permutations of $\{1,\ldots, d\}$. We define
\[
	\psi(\Sigma)\eqdef \min_{\pi\in \Pi}\left\{[\Sigma_{\pi,11}]^{-1/2}+\sum_{i=2}^d[\Sigma_{\pi}^{(i)}/\Sigma_{\pi}^{(i-1)}]^{-1/2}\right\},
\]
where $\Sigma_{\pi}=P_{\pi}\Sigma P_{\pi}$, $\pi\in\Pi$, is the variance of $(X_{\pi(1)},\ldots,X_{\pi(d)})^{\top}$ ($P_{\pi}$ denotes the permutation matrix corresponding to $\pi$). When $d=1$, we set $\psi(\Sigma)=\Sigma^{-1/2}$.

\begin{lemma}
\label{lemma:aux_psi_bnd}
Let $\lambda_{\tmin}(\Sigma)$ and $\lambda_{\tmax}(\Sigma)$ be the smallest and the largest eigenvalues of $\Sigma$. Then
\[
	\frac{1}{\sqrt{\lambda_{\tmax}(\Sigma)}}\left(1+\frac{d-1}{\sqrt{1+\kappa^2}}\right)\le\psi(\Sigma)\le \frac{d}{\sqrt{\lambda_{\tmin}(\Sigma)}},
\]
where $\kappa=\lambda_{\tmax}(\Sigma)/\lambda_{\tmin}(\Sigma)$ is the condition number of $\Sigma$.
\end{lemma}

\begin{proof}
Fix $\pi\in \Pi$ and let
\begin{equation}
\label{eq:aux_psi_bnd0}
	\psi_{\pi}(\Sigma)\eqdef [\Sigma_{\pi,11}]^{-1/2}+\sum_{i=2}^d[\Sigma_{\pi}^{(i)}/\Sigma_{\pi}^{(i-1)}]^{-1/2}.
\end{equation}
In addition, let $\lambda_{\pi,\tmin}^{(i)}$ and $\lambda_{\pi,\tmax}^{(i)}$ denote the smallest and the largest eigenvalues of $\Sigma_{\pi}^{(i)}$. Notice that by the properties of the Rayleigh quotient, $\lambda_{\tmin}(\Sigma) \le \lambda_{\pi,\tmin}^{(i)}$ and $\lambda_{\tmax}(\Sigma)\ge \lambda_{\pi,\tmax}^{(i)}$ \citep[see, e.g.,][Section~6.2]{Serre:10:Matrices}.

For $i>1$, consider the Schur complement $\Sigma_{\pi}^{(i)}/\Sigma_{\pi}^{(i-1)}$, i.e.,
\[
	\sigma_i^2\eqdef \Sigma_{\pi}^{(i)}/\Sigma_{\pi}^{(i-1)}=\Sigma_{\pi,ii}^{(i)}-v_i^{\top}[\Sigma_{\pi}^{(i-1)}]^{-1}v_i,
\]
where $v_i$ is the $i$-th column of $\Sigma_{\pi}^{(i)}$ without the last element, and let
\[
	A^{(i)}\eqdef \begin{bmatrix}
		\Sigma_{\pi}^{(i-1)} & 0 \\
		0 & \sigma_i^2
	\end{bmatrix} \quad\text{and}\quad
	B^{(i)}\eqdef\begin{bmatrix}
		I & -[\Sigma_{\pi}^{(i-1)}]^{-1}v_i \\
		0 & 1
	\end{bmatrix}.
\]
Then
\[
	\sigma_i^2=e_i^{\top}A^{(i)}e_i=(B^{(i)}e_i)^{\top}\Sigma_{\pi}^{(i)}(B^{(i)}e_i)\ge \lambda_{\pi,\tmin}^{(i)}\norm{B^{(i)}e_i}^2\ge \lambda_{\tmin}.
\]
Moreover, $\Sigma_{\pi,11}=e_1^{\top}\Sigma_{\pi}e_1\ge \lambda_{\tmin}$. Combining these inequalities, we get
\begin{equation}
\label{eq:aux_psi_bnd1}
	\psi_{\pi}(\Sigma)\le \frac{d}{\sqrt{\lambda_{\tmin}}}.
\end{equation}

Similarly, $\Sigma_{\pi,11}\le \lambda_{\tmax}$, and since $\normin{[\Sigma_{\pi}^{(i-1)}]^{-1}}\le \lambda_{\tmin}^{-1}$ and $\norm{v_i}\le \lambda_{\tmax}$,
\[
	\sigma_i^2\le \lambda_{\pi,\tmax}^{(i)}\norm{B^{(i)}e_i}^2\le \lambda_{\tmax}\left(1+\norm{[\Sigma_{\pi}^{(i-1)}]^{-1}v_i}^2\right)\le \lambda_{\tmax}(1+\kappa^2).
\]
Therefore,
\begin{equation}
\label{eq:aux_psi_bnd2}
	\psi_{\pi}(\Sigma)\ge \frac{1}{\sqrt{\lambda_{\tmax}}}\left(1+\frac{d-1}{\sqrt{1+\kappa^2}}\right).
\end{equation}
The result follows by noticing that the bounds \eqref{eq:aux_psi_bnd1} and \eqref{eq:aux_psi_bnd2} do not depend on $\pi$.
\end{proof}

\begin{lemma}
\label{lemma:normal_anticoncentration}
For any $\epsilon>0$ and $a,b\in [0,\infty)^d$,
\[
	\sup_{r\ge 0}\PR{r<\T(X+a,X-b)\le r+\epsilon}\le \psi(\Sigma)\epsilon.
\]
\end{lemma}

\begin{proof}
For a given $\pi\in \Pi$ let $Y_i=[X_{\pi(i)}+a_{\pi(i)}]_{-}+[X_{\pi(i)}-b_{\pi(i)}]_{+}$, $W_0= 0$, and $W_i=W_{i-1}+Y_i$, $1\le i\le d$. Then since
\begin{align*}
	\PR{r<W_i\le r+\epsilon}&\le \PR{r<Y_i+W_{i-1}\le r+\epsilon, W_{i-1}\le r} \\
	&\quad+\PR{r<W_{i-1}\le r+\epsilon},
\end{align*}
we find that
\[
	\PR{r<W_d\le r+\epsilon}\le \sum_{i=1}^d\PR{r<Y_i+W_{i-1}\le r+\epsilon, W_{i-1}\le r}.
\]
The conditional distribution of $X_{\pi(i)}$ given $Z_{i-1}\eqdef (X_{\pi(1)},\ldots,X_{\pi(i-1)})^{\top}$ is normal with variance $\sigma_i^2=\Sigma_{\pi}^{(i)}/\Sigma_{\pi}^{(i-1)}$. Also, for any $y\ge 0$, the event $\{y<Y_i\le y+\epsilon\}=\{-y-\epsilon\le X_{\pi(i)}+a_{\pi(i)}<-y\}\cup\{y<X_{\pi(i)}-b_{\pi(i)}\le y+\epsilon\}$. Hence,
\begin{align*}
	\PR{y<Y_i \le y+\epsilon\mid Z_{i-1}=z} &= \PR{-y-\epsilon \le X_{\pi(i)}+a_{\pi(i)} < -y\mid Z_{i-1}=z} \\
	&\quad + \PR{y < X_{\pi(i)} - b_{\pi(i)} \le y+\epsilon\mid Z_{i-1}=z} \\
	&\le 2\sup_{x \in \mathbf{R}} \PR{x < X_{\pi(i)} \le x+\epsilon\mid Z_{i-1}=z} \le \epsilon/\sigma_i.
\end{align*}
Consequently, we find that for $i>1$ and $r\ge 0$,
\begin{align*}
	&\PR{r<Y_i+W_{i-1}\le r+\epsilon, W_{i-1}\le r} \\
	&\qquad=\E[\PR{r<Y_i+W_{i-1}\le r+\epsilon\mid Z_{i-1}}\ind\{W_{i-1}\le r\}]\le \epsilon/\sigma_i.
\end{align*}
In addition, $\PR{r<Y_1\le r+\epsilon}\le [\Var(X_{\pi(1)})]^{-1/2}\epsilon$. Therefore,
\[
	\PR{r<W_d\le r+\epsilon}\le \psi_{\pi}(\Sigma)\epsilon,
\]
where $\psi_{\pi}(\csdot)$ is given in \eqref{eq:aux_psi_bnd0}. Since the probability on the RHS of the last inequality does not depend on $\pi$,
\[
	\sup_{r\ge 0}\PR{r<\T(X+a,X-b)\le r+\epsilon}\le \min_{\pi\in\Pi}\psi_{\pi}(\Sigma)\epsilon. \qedhere
\]
\end{proof}

\begin{remark*}
In the preceding result, the distribution of $\T(X+a,X-b)$ has an atom at $0$ when $(a+b)\in (0,\infty)^d$. Therefore, the uniform bound depending on $\epsilon$ can be established only over the non-negative reals.
\end{remark*}

The next results establish bounds on the conditional Kolmogorov distance between the $\T$-transforms of certain random vectors. Since the function $\T(\csdot,\csdot)$ is not differentiable we use its smooth approximation $\widetilde{\T}_{\kappa}:\R^d\times \R^d\to\R$, $\kappa>0$, defined by
\[
	\widetilde{\T}_{\kappa}(x,y)\eqdef\norm{\vecg{\varphi}_{\kappa}(-x)+\vecg{\varphi}_{\kappa}(y)}_1,
\]
where $\vecg{\varphi}_{\kappa}:\R^d\to\R^d$ is a function of the form $\vecg{\varphi}_{\kappa}(x)=[\varphi_{\kappa}(x_1),\ldots,\varphi_{\kappa}(x_d)]^{\top}$ with $\varphi_{\kappa}(x)=\kappa^{-1}\ln(e^{\kappa x}+1)$. Note that $0\le \varphi_{\kappa}(x)-(x\vee 0)\le \kappa^{-1}\ln(2)$ for all $x\in\R$.

\begin{lemma}
\label{lemma:aux_CLT}
Let $X_1,\ldots,X_n$ be random vectors in $\R^d$ that are conditionally independent given $\F\subset\H$ with $\E[X_i\mid \F]=0$ and $\E[\norm{X_i}_3^3\mid \F]<\infty$ a.s. Let $S\eqdef \sum_{i=1}^n X_i$ and let $N$ be a random vector in $\R^d$ s.t.\ $N\mid \F\sim \ND{0}{V}$, where $V=\E[SS^{\top}\mid \F]$ a.s. Then, assuming that $V$ is a.s.\ positive definite, for any $\epsilon>0$ and $\F$-measurable random vectors $a,b\in [0,\infty)^d$,
\begin{align*}
	&\DK{\epsilon}{\T(S+a,S-b),\T(N+a,N-b)\mid \F} \\
	&\qquad\le C_d\Gamma^{1/4}[\psi(V)]^{3/4} \qtext{a.s.\ on } \{\delta^{*}\le \epsilon^4\},
\end{align*}
where $\Gamma\eqdef \sum_{i=1}^n \E[\norm{X_i}_3^3\mid \F]$, $\delta^{*}\eqdef \Gamma/\psi(V)$, and $C_d>0$ is a constant depending only on $d$.
\end{lemma}

\begin{proof}
Let $f$ be a trice continuously differential function, s.t.\ for a given $\delta>0$, $f(x)=1$ if $x\le 0$, $f=0$ if $x\ge\delta>0$, and $\absin{f^{(j)}(x)}\le D\delta^{-j}\ind_{(0,\delta)}(x)$ for some absolute constant $D>0$ and $1\le j\le 3$. Further, for $\kappa>0$, set
\[
	g_r(s)\eqdef f(\widetilde{\T}_{\kappa}(s+a,s-b)-r).
\]
First, letting $\nu\eqdef 2\ln(2)d\kappa^{-1}$, we find that
\begin{align*}
	&\PR{\T(S+a,S-b)\le r\mid \F}\le \PR{\widetilde{\T}_{\kappa}(S+a,S-b)\le r+\nu\mid \F} \\
	&\qquad\le \E[g_{r+\nu}(S)\mid \F] \\
	&\qquad\le \PR{\T(N+a,N-b)\le r+\delta+\nu\mid \F}+\E[g_{r+\nu}(S)-g_{r+\nu}(N)\mid \F]
\shortintertext{and}
	&\PR{\T(S+a,S-b)> r\mid \F}\le \PR{\widetilde{\T}_{\kappa}(S+a,S-b)> r\mid \F} \\
	&\qquad\le 1-\E[g_{r-\delta}(S)\mid \F] \\
	&\qquad\le \PR{\T(N+a,N-b)> r-\delta-\nu\mid \F}+\E[g_{r-\delta}(S)-g_{r-\delta}(N)\mid \F]
\end{align*}
a.s.\ for all $r\ge 0$. Hence, for $0<\delta+\nu\le \epsilon$ w.p.1,
\begin{align}
\label{eq:aux_CLT1}
	\begin{aligned}
		&\DK{\epsilon}{\T(S+a,S-b),\T(N+a,N-b)\mid \F} \\
		&\qquad\le \sup_{q\in \Q_{\ge 0}}\abs{\E[g_q(S)-g_q(N)\mid \F]} +\sup_{q\in \Q_{\ge 0}}\PR{q<\T(N+a,N-b)\le q+\delta+\nu\mid \F}.
	\end{aligned}
\end{align}

Consider the first term on the RHS of \eqref{eq:aux_CLT1}.

\begin{claim}
There is a constant $B_d>0$ depending only on $d$ such that for any $q\ge 0$,
\begin{align*}
	&\abs{\E[g_q(S)-g_q(N)\mid \F]}\le B_d\left(\frac{1}{\delta^3}+\frac{\kappa}{\delta^2}+\frac{\kappa^2}{\delta}\right)\Gamma \qtext{a.s.}
\end{align*}
\end{claim}
\begin{subproof}
Let $Z_1,\ldots,Z_n$ be i.i.d.\ standard normal random vectors in $\R^d$ independent of $X_1,\ldots,X_n$ and $\F$, and let $Y_i\eqdef V_i^{1/2}Z_i$, where $V_i$ is a version of $\E[X_iX_i^{\top}\mid \F]$. Define
\begin{align*}
	U_i&\eqdef \sum_{k=1}^{i-1} X_k+\sum_{k=i+1}^n Y_k \\
\shortintertext{and}
	W_i&\eqdef g_q(U_i+X_i)-g_q(U_i+Y_i).
\end{align*}
Then $g_q(S)-g_q(N)=\sum_{i=1}^n W_i$ and
\[
	\abs{\E[g_q(S)-g_q(N)\mid \F]}\le \sum_{i=1}^n\abs{\E[W_i\mid \F]} \qtext{a.s.}
\]
Let $h_{i1}(\lambda)\eqdef g_q\left(U_i+\lambda X_i\right)$ and $h_{i2}(\lambda)\eqdef g_q\left(U_i+\lambda Y_i\right)$. Using Taylor expansion up to the third order, we find that
\[
	W_i=\sum_{j=0}^2 \frac{1}{j!}\left(h_{i1}^{(j)}(0)-h_{i2}^{(j)}(0)\right)+\frac{1}{3!}\left(h_{i1}^{(3)}(\lambda_1)-h_{i2}^{(3)}(\lambda_2)\right),
\]
where $\abs{\lambda_1},\abs{\lambda_2}\le 1$. Then, since $U_i$ is $\G_{i}\eqdef \F\vee \sigma(X_1,\ldots,X_{i-1},Z_{i+1},\ldots,Z_n)$-measurable,
\[
	\E[\E[h_{i1}^{(j)}(0)-h_{i2}^{(j)}(0)\mid \G_i]\mid \F]=0 \qtext{a.s.}
\]
for $j\le 2$. Also since $\absin{\varphi_{\kappa}^{(j)}(x)}\le \kappa^{j-1}$, $1\le j\le 3$, we get
\begin{align*}
	\absin{h_{i1}^{(3)}(\lambda_1)-h_{i2}^{(3)}(\lambda_2)}&\le B\Big(\norm{f^{(3)}}_{\infty}\left(\norm{X_i}_1^3+\norm{Y_i}_1^3\right) \\
	&\quad+ \norm{f''}_{\infty}\kappa\left(\norm{X_i}_1\norm{X_i}_2^2+\norm{Y_i}_1\norm{Y_i}_2^2\right) \\
	&\quad+ \norm{f'}_{\infty}\kappa^2\left(\norm{X_i}_3^3+\norm{Y_i}_3^3\right)\Big),
\end{align*}
where $B>0$ is an absolute constant. Finally, since $\E[\norm{Y_i}_3^3\mid \F]\le 2\sqrt{2/\pi}\E[\norm{X_i}_3^3\mid \F]$ a.s.,
\begin{align*}
	&\absin{\E[h_{i1}^{(3)}(\lambda_1)-h_{i2}^{(3)}(\lambda_2)\mid \F]}\le \E[\absin{h_{i1}^{(3)}(\lambda_1)-h_{i2}^{(3)}(\lambda_2)}\mid \F] \\
	&\qquad\le B_d\left(\frac{1}{\delta^3}+\frac{\kappa}{\delta^2}+\frac{\kappa^2}{\delta}\right)\E[\norm{X_i}_3^3\mid \F] \qtext{a.s.} \qedhere
\end{align*}
\end{subproof}

Using Lemma \ref{lemma:normal_anticoncentration}, it follows that
\begin{align}
\label{eq:aux_CLT2}
	\begin{aligned}
		&\DK{\epsilon}{\T(S+a,S-b),\T(N+a,N-b)\mid \F} \\
		&\qquad\le B_d\left(\frac{1}{\delta^3}+\frac{\kappa}{\delta^2}+\frac{\kappa^2}{\delta}\right)\Gamma +\psi(V)(\delta+\nu) \qtext{a.s.}
	\end{aligned}
\end{align}
We set $\nu=\delta$. The since \eqref{eq:aux_CLT2} holds for any $\delta$ a.s., it holds for random $\delta$ on $\{\delta\in(0,\epsilon/2)\}$. Consequently, the result follows by taking $\delta=(\delta^{*})^{1/4}/2$ and noticing that $0<\psi(V)<\infty$ a.s.\ by Lemma \ref{lemma:aux_psi_bnd}.
\end{proof}

\begin{lemma}
\label{lemma:aux_normal_approx}
Suppose that $\G$ and $\F$ are $\sigma$-fields s.t.\ $\F\subset \G\subset\H$, $X$ and $Y$ are random vectors in $\R^d$ s.t.\ $X\mid \G\sim \ND{0}{\Sigma_X}$ and $Y\mid \F\sim \ND{0}{\Sigma_Y}$. Then, assuming that $\Sigma_Y$ is a.s.\ positive definite, for any $\epsilon>0$ and $\F$-measurable random vectors $a,b\in [0,\infty)^d$,
\begin{align}
\label{eq:aux_normal_approx}
	\begin{aligned}
		&\DK{\epsilon}{\T(X+a,X-b),\T(Y+a,Y-b)\mid \G,\F} \\
		&\qquad\le C_d\norm{\Sigma_X-\Sigma_Y}_{e,\infty}^{1/3}\psi(\Sigma_Y)^{2/3} \qtext{a.s.\ on } \{\delta^{*}\le \epsilon ^3\},
	\end{aligned}
\end{align}
where $\delta^{*}\eqdef \norm{\Sigma_X-\Sigma_Y}_{e,\infty}/\psi(\Sigma_Y)$ and $C_d>0$ is a constant depending only on $d$.
\end{lemma}

\begin{proof}
Let $f$ be a twice continuously differential function s.t.\ for a given $\delta>0$, $f(x)=1$ if $x\le 0$, $f(x)=0$ if $x\ge \delta>0$ and $\absin{f^{(j)}}\le D\delta^{-j}\ind_{(0,\delta)}(x)$ for some absolute constant $D>0$ and $1\le j\le 2$. Further, set
\[
	g_r(s)\eqdef f(\T(s+a,s-b)-r).
\]
As in the proof of Lemma \ref{lemma:aux_CLT} for any $0<\delta\le \epsilon$ w.p.1,
\begin{align*}
	&\DK{\epsilon}{\T(X+a,X-b),\T(Y+a,Y-b)\mid \G,\F} \\
	&\qquad\le \sup_{q\in \Q_{\ge 0}}\abs{\E[g_q(X)\mid \G]-\E[g_q(Y)\mid \F]} \\
	&\qquad\quad+\sup_{q\in \Q_{\ge 0}}\PR{q<\T(N+a,N-b)\le q+\delta\mid \F}.
\end{align*}
Let $Z_1$ and $Z_2$ be independent standard normal random vectors in $\R^d$ independent of $\G$. Then
\begin{align*}
	\E[g_q(X)\mid \G]-\E[g_q(Y)\mid \F]&=\E[g_q(\Sigma_X^{1/2}Z_1)\mid \G]-\E[g_q(\Sigma_Y^{1/2}Z_2)\mid \F] \\
	&=h_{q,1}(\Sigma_X)-h_{q,2}(\Sigma_Y) \qtext{a.s.},
\end{align*}
where $h_{q,1}(\sigma)\eqdef\E g_q(\sigma^{1/2}Z_1)$ and $h_{q,2}(\sigma)\eqdef\E g_q(\sigma^{1/2}Z_2)$ (the functions $h_{q,1}$ and $h_{q,2}$ implicitly depend on $a$ and $b$; however, since they are $\F$-measurable we treat them as constants).

\begin{claim}
There exists a constant $B_d$ depending only on $d$ such that for any $q\ge 0$,
\[
	\abs{h_{q,1}(\sigma_X)-h_{q,2}(\sigma_Y)}\le \frac{B_d}{\delta^2}\norm{\sigma_X-\sigma_Y}_{e,\infty}.
\]
\end{claim}
\begin{subproof}
Let $\tilde{g}_q(x)\eqdef f(\widetilde{\T}_{\kappa}(x+a,x-b)-q)$ with $\kappa>0$ and let
\[
	\tilde{h}_{q,1}(\sigma)\eqdef\E\tilde{g}_q(\sigma^{1/2}Z_1) \qtext{and}\quad \tilde{h}_{q,2}(\sigma)\eqdef\E\tilde{g}_q(\sigma^{1/2}Z_2).
\]
For $t\in [0,1]$, define $Z(t)\eqdef\sqrt{t}\sigma_X^{1/2}Z_1+\sqrt{1-t}\sigma_Y^{1/2}Z_2$ and $\phi(t)\eqdef \E\tilde{g}_q(Z(t))$. Then
\[
	\tilde{h}_{q,1}(\sigma_X)-\tilde{h}_{q,2}(\sigma_Y)=\phi(1)-\phi(0)=\int_0^1 \phi'(t)dt.
\]

Using the integration by parts formula \citep[see Equation A.17 in][Section A.6]{Talagrand:11:SpinGlasses} for $t\in (0,1)$,
\begin{align*}
	\phi'(t)&=\frac{1}{2}\E\left[\left(\sigma_X^{1/2}Z_1/\sqrt{t}-\sigma_Y^{1/2}Z_2/\sqrt{1-t}\right)^{\top} \nabla \tilde{g}_q(Z(t))\right] \\
	&=\frac{1}{2}\E\left[\boldsymbol{i}^{\top}(\sigma_X-\sigma_Y)\circ \nabla^2 \tilde{g}_q(Z(t))\boldsymbol{i}\right],
\end{align*}
where $\boldsymbol{i}$ is the vector of ones, and $\circ$ denotes the Hadamard product. Therefore,
\[
	\abs{\int_0^1 \phi'(t)dt}\le \normin{\sigma_X-\sigma_Y}_{e,\infty}\int_0^1 \E\abs{\boldsymbol{i}^{\top}\nabla^2 \tilde{g}_q(Z(t))\boldsymbol{i}}dt.
\]
The $(r,s)$-th element of the Hessian of $\tilde{g}_q$ can be bounded by
\[
	\abs{D_{r,s}(\tilde{g}_q)(z)}\le \norm{f''}_{\infty}+\norm{f'}_{\infty}\frac{\kappa}{2} 1\{r=s\}.
\]
Consequently, the result follows by setting $\kappa=\delta^{-1}$.
\end{subproof}

Using Lemma \ref{lemma:normal_anticoncentration} it follows that
\begin{align}
\label{eq:aux_normal_approx1}
	\begin{aligned}
		&\DK{\epsilon}{\T(X+a,X-b),\T(Y+a,Y-b)\mid \G,\F} \\
		&\qquad\le \frac{B_d}{\delta^2}\norm{\Sigma_X-\Sigma_Y}_{e,\infty} +\psi(\Sigma_Y)\delta \qtext{a.s.}
	\end{aligned}
\end{align}
Finally, since \eqref{eq:aux_normal_approx1} holds for any $0<\delta\le \epsilon$ a.s., it holds for random $\delta$ a.s.\ on $\{\delta\in (0,\epsilon]\}$. Consequently, the result follows by taking $\delta=(\delta^{*})^{1/3}$ and noticing that \eqref{eq:aux_normal_approx} holds trivially on $\{\norm{\Sigma_X-\Sigma_Y}_{e,\infty}=0\}$ and that $0<\psi(\Sigma_Y)<\infty$ a.s.\ by Lemma \ref{lemma:aux_psi_bnd}.
\end{proof}

\renewcommand{\bibsection}{\subsection*{References}}
\putbib[large_games]

\pagebreak

\subsection{Additional Simulation Results}
\label{app:sim_results}

In this section, we present additional simulation results using different values of $\rho$.  As we can see, the coverage probabilities are very similar as we vary $\rho$. However, the size of the confidence sets gets larger when $\rho$ is very small or large. From the simulation results, we propose using $\rho=0.001$ in practice.

\begin{figure}[tbph]
	\caption[False Coverage Probability of the Confidence Intervals at $95\%$]{
		False Coverage Probability of the Confidence Intervals for $(\phi_{0},\beta_{0})$ at $95\%$: $\phi_0 = 0.25, \beta_0 = 0.5, \rho=0.000001$.
	}
	\label{figure:sim_power_11}
	\medskip
	\begin{center}
		\includegraphics[scale = 0.4]{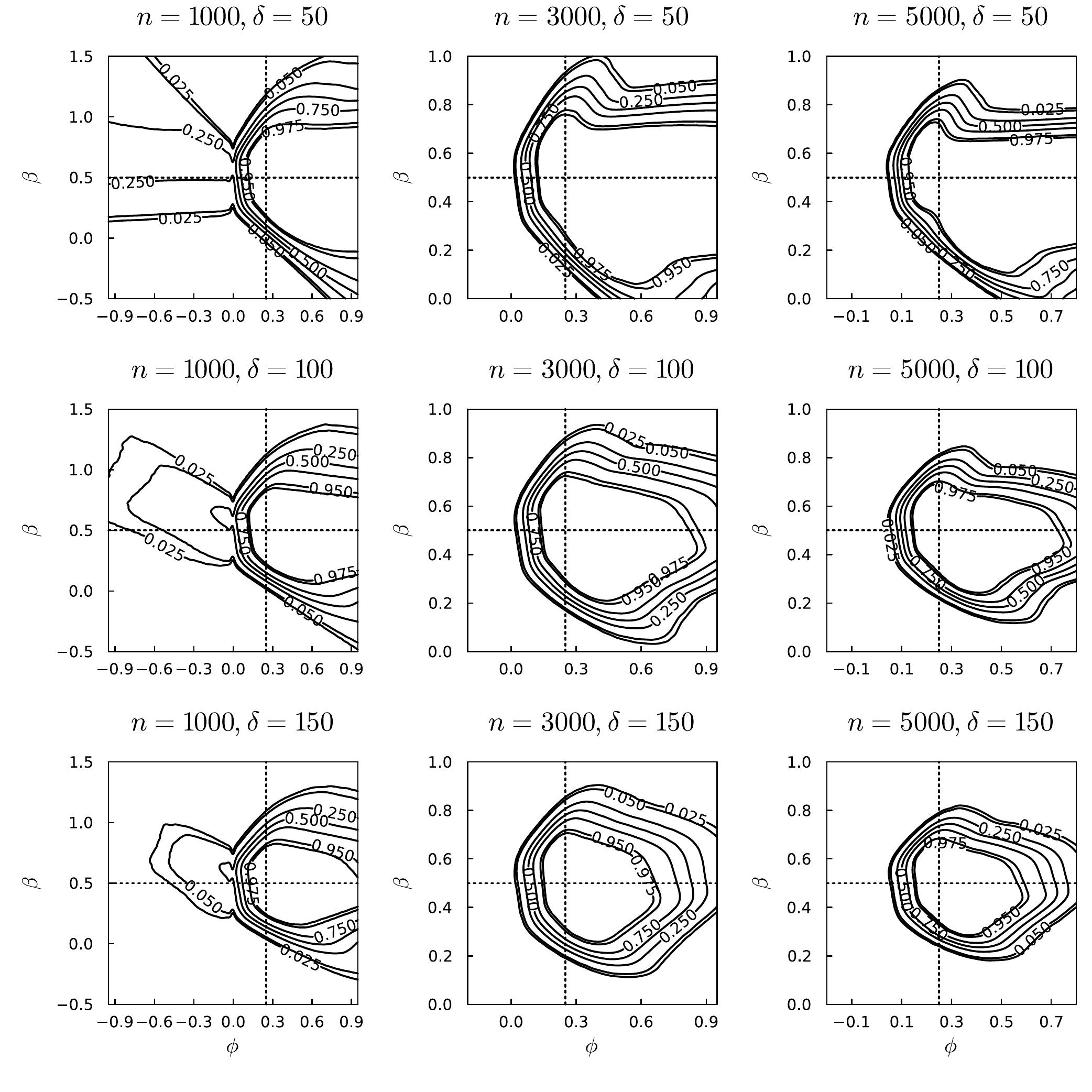}
	\end{center}
\parbox{6.2in}{\footnotesize \medskip
	Notes: The intersecting point between two dotted lines in each panel indicates the true parameter $(\phi_0,\beta_0)$. The horizontal axis represents the hypothesized value of $\phi$ and the vertical axis that of $\beta$. In each panel, the area surrounded by the innermost contour line consists of the parameter values $(\phi,\beta)$ that are included in the confidence set at least 97.5\% of the times in the Monte Carlo loops, and the area surrounded by the outermost contour line consists of the parameter values $(\phi,\beta)$ that are included in the confidence set at least 2.5\% of the times in the Monte Carlo loops.}
\bigskip
\end{figure}

\clearpage

\begin{figure}[tbph]
	\caption[False Coverage Probability of the Confidence Intervals at $95\%$]{
		False Coverage Probability of the Confidence Intervals for $(\phi_{0},\beta_{0})$ at $95\%$: $\phi_0 = 0.25, \beta_0 = 0.5, \rho=0.00001$.
	}
	\label{figure:sim_power_21}
	\medskip
	\begin{center}
		\includegraphics[scale = 0.4]{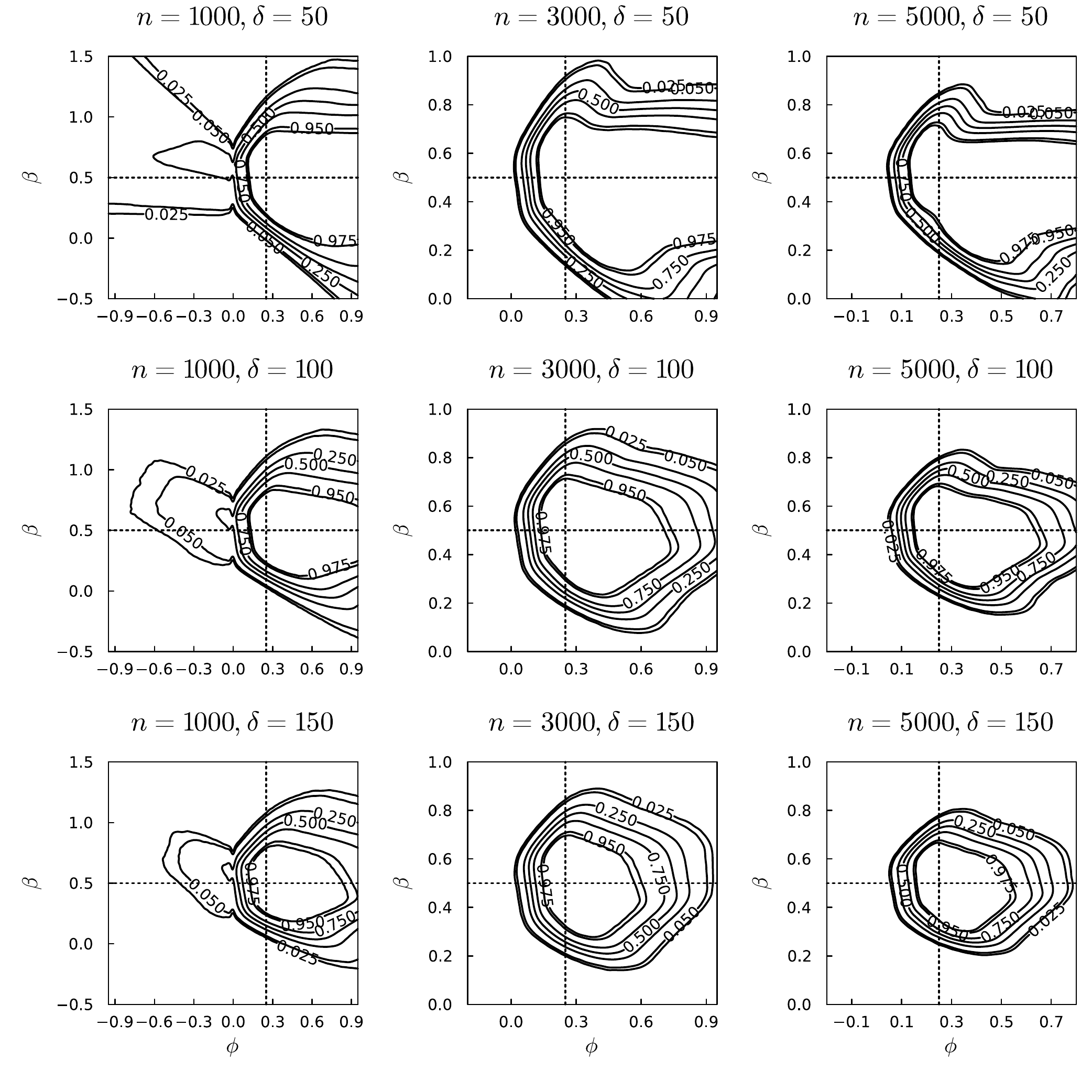}
	\end{center}
\parbox{6.2in}{\footnotesize \medskip
	Notes: The intersecting point between two dotted lines in each panel indicates the true parameter $(\phi_0,\beta_0)$. The horizontal axis represents the hypothesized value of $\phi$ and the vertical axis that of $\beta$. In each panel, the area surrounded by the innermost contour line consists of the parameter values $(\phi,\beta)$ that are included in the confidence set at least 97.5\% of the times in the Monte Carlo loops, and the area surrounded by the outermost contour line consists of the parameter values $(\phi,\beta)$ that are included in the confidence set at least 2.5\% of the times in the Monte Carlo loops.}
\bigskip
\end{figure}

\clearpage

\begin{figure}[tbph]
	\caption[False Coverage Probability of the Confidence Intervals at $95\%$]{
		False Coverage Probability of the Confidence Intervals for $(\phi_{0},\beta_{0})$ at $95\%$: $\phi_0 = 0.25, \beta_0 = 0.5, \rho=0.0001$.
	}
	\label{figure:sim_power_31}
	\medskip
	\begin{center}
		\includegraphics[scale = 0.4]{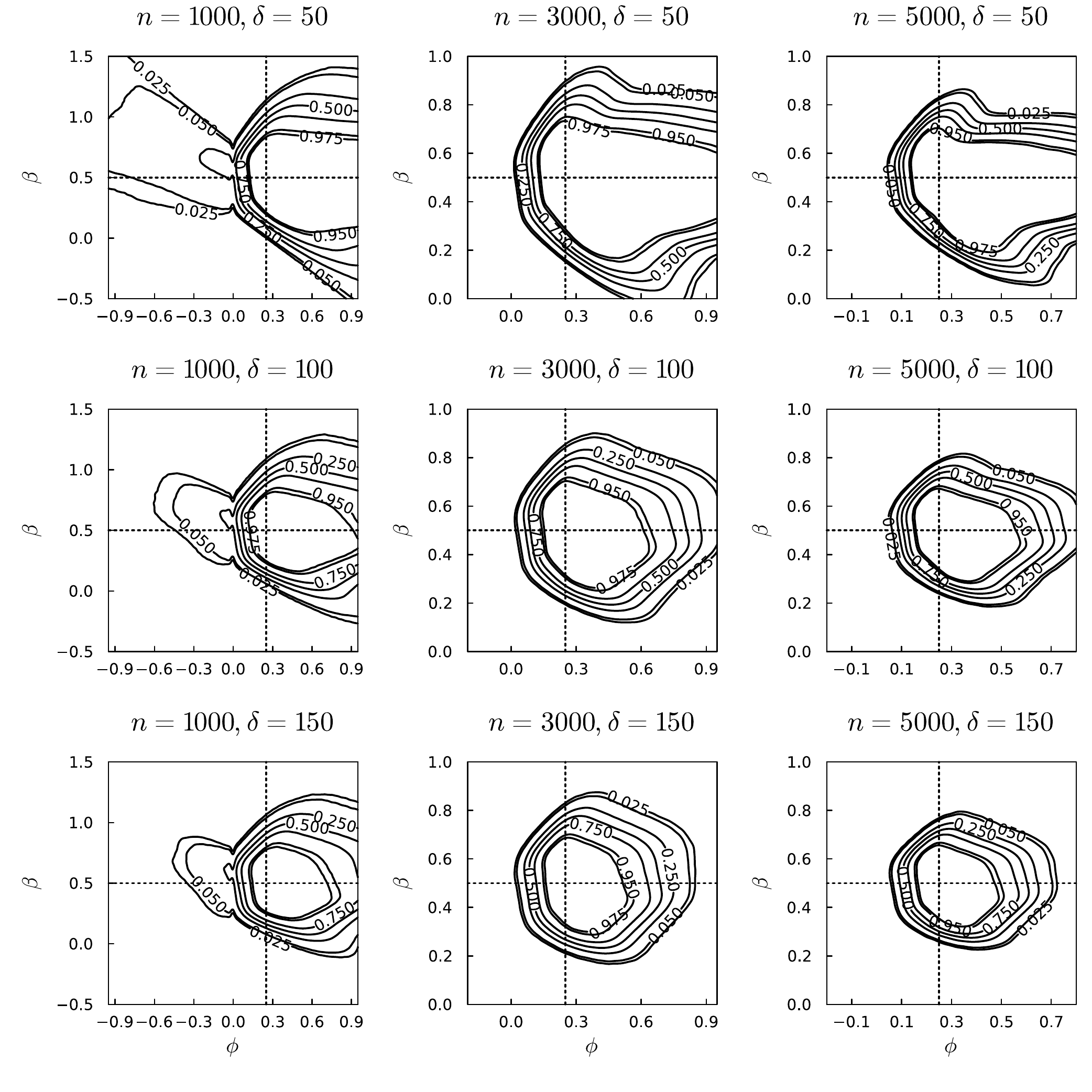}
	\end{center}
\parbox{6.2in}{\footnotesize \medskip
	Notes: The intersecting point between two dotted lines in each panel indicates the true parameter $(\phi_0,\beta_0)$. The horizontal axis represents the hypothesized value of $\phi$ and the vertical axis that of $\beta$. In each panel, the area surrounded by the innermost contour line consists of the parameter values $(\phi,\beta)$ that are included in the confidence set at least 97.5\% of the times in the Monte Carlo loops, and the area surrounded by the outermost contour line consists of the parameter values $(\phi,\beta)$ that are included in the confidence set at least 2.5\% of the times in the Monte Carlo loops.}
\bigskip
\end{figure}

\begin{figure}[tbph]
	\caption[False Coverage Probability of the Confidence Intervals at $95\%$]{
		False Coverage Probability of the Confidence Intervals for $(\phi_{0},\beta_{0})$ at $95\%$: $\phi_0 = 0.25, \beta_0 = 0.5, \rho=0.001$.
	}
	\label{figure:sim_power_41}
	\medskip
	\begin{center}
		\includegraphics[scale = 0.4]{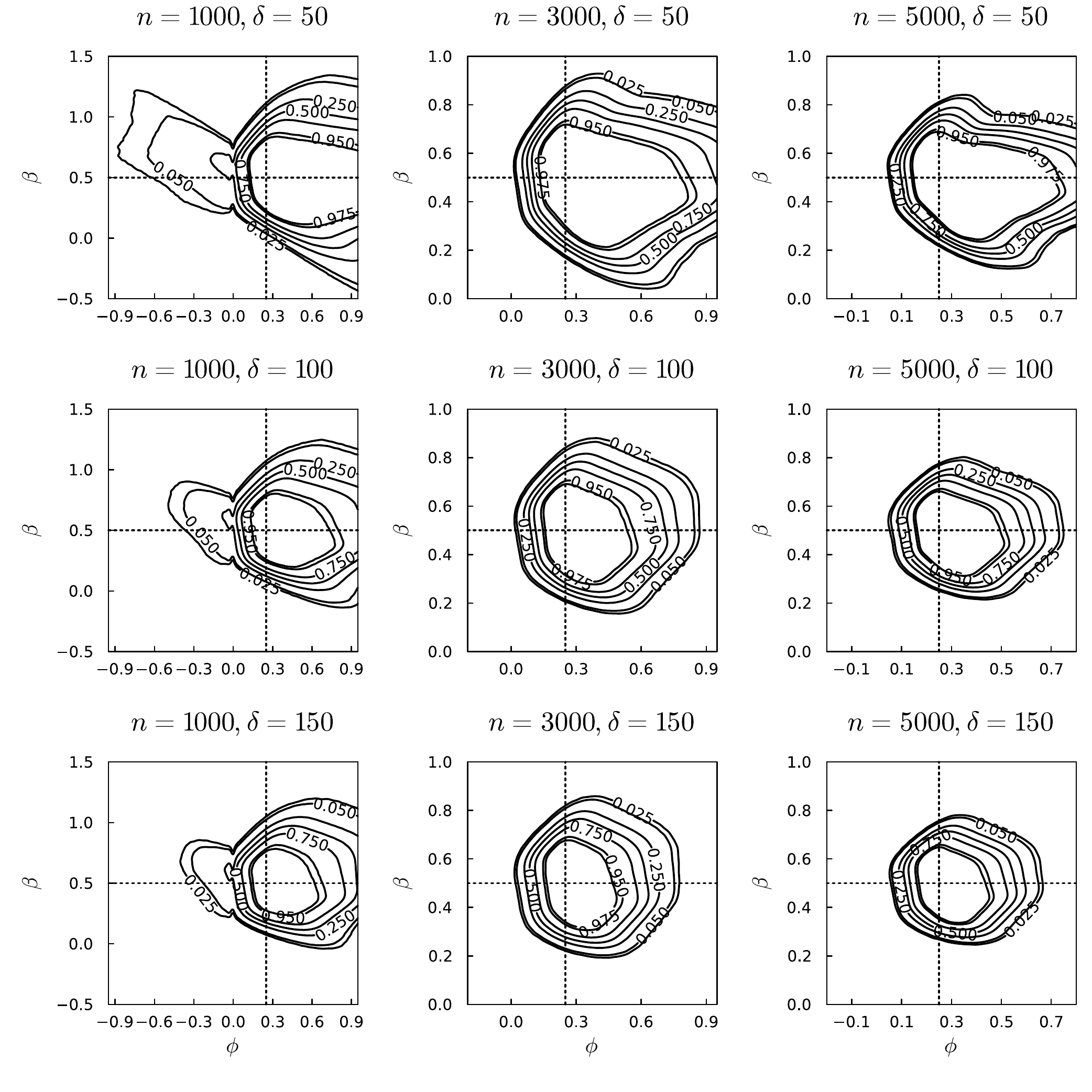}
	\end{center}
\parbox{6.2in}{\footnotesize \medskip
	Notes: The intersecting point between two dotted lines in each panel indicates the true parameter $(\phi_0,\beta_0)$. The horizontal axis represents the hypothesized value of $\phi$ and the vertical axis that of $\beta$. In each panel, the area surrounded by the innermost contour line consists of the parameter values $(\phi,\beta)$ that are included in the confidence set at least 97.5\% of the times in the Monte Carlo loops, and the area surrounded by the outermost contour line consists of the parameter values $(\phi,\beta)$ that are included in the confidence set at least 2.5\% of the times in the Monte Carlo loops.}
\bigskip
\end{figure}

\begin{figure}[tbph]
	\caption[False Coverage Probability of the Confidence Intervals at $95\%$]{
		False Coverage Probability of the Confidence Intervals for $(\phi_{0},\beta_{0})$ at $95\%$: $\phi_0 = 0.25, \beta_0 = 0.5, \rho=0.01$.
	}
	\label{figure:sim_power_51}
	\medskip
	\begin{center}
		\includegraphics[scale = 0.4]{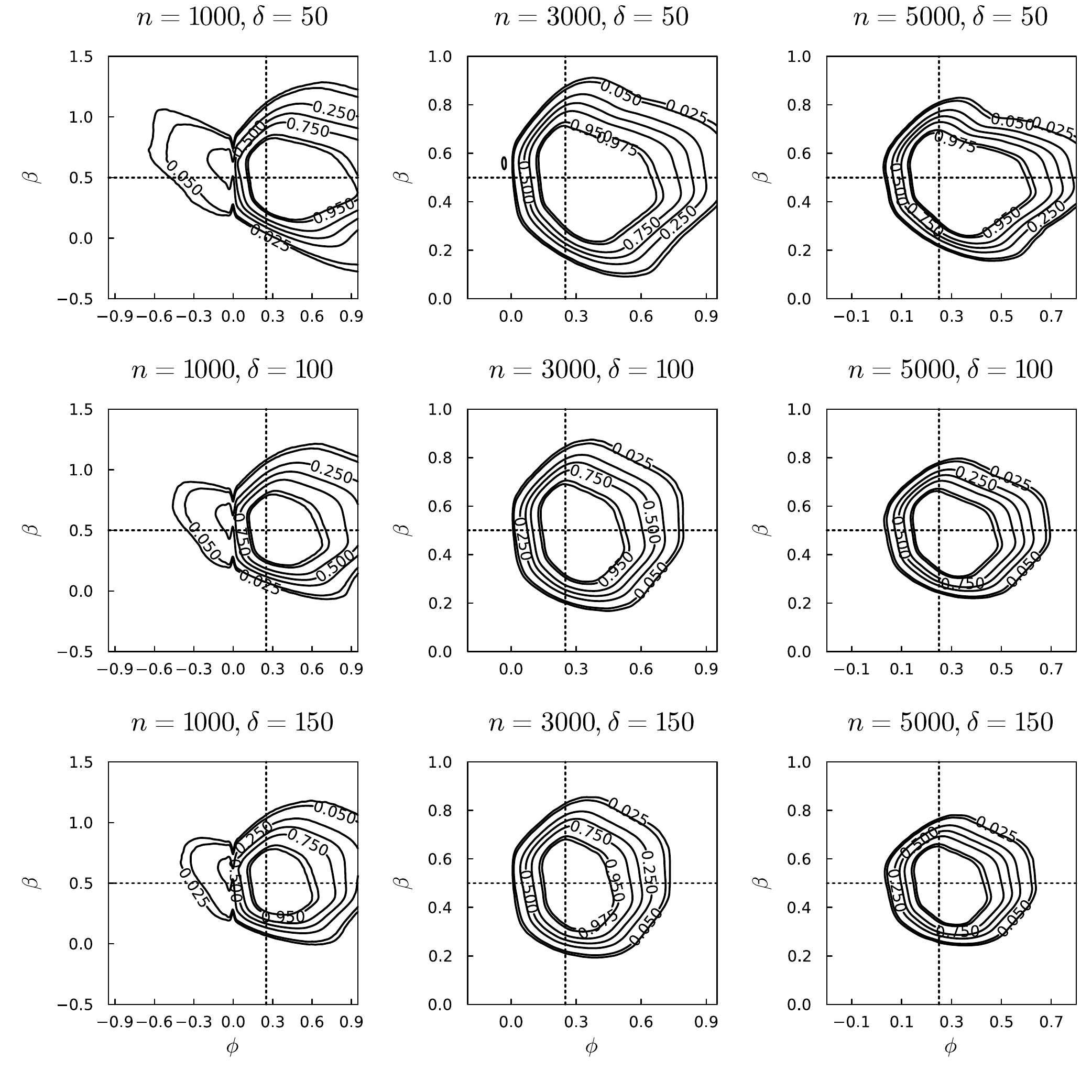}
	\end{center}
\parbox{6.2in}{\footnotesize \medskip
	Notes: The intersecting point between two dotted lines in each panel indicates the true parameter $(\phi_0,\beta_0)$. The horizontal axis represents the hypothesized value of $\phi$ and the vertical axis that of $\beta$. In each panel, the area surrounded by the innermost contour line consists of the parameter values $(\phi,\beta)$ that are included in the confidence set at least 97.5\% of the times in the Monte Carlo loops, and the area surrounded by the outermost contour line consists of the parameter values $(\phi,\beta)$ that are included in the confidence set at least 2.5\% of the times in the Monte Carlo loops.}
\bigskip
\end{figure}

\begin{figure}[tbph]
	\caption[False Coverage Probability of the Confidence Intervals at $95\%$]{
		False Coverage Probability of the Confidence Intervals for $(\phi_{0},\beta_{0})$ at $95\%$: $\phi_0 = 0.25, \beta_0 = 1.0, \rho=0.000001$.
	}
	\label{figure:sim_power_12}
	\medskip
	\begin{center}
		\includegraphics[scale = 0.4]{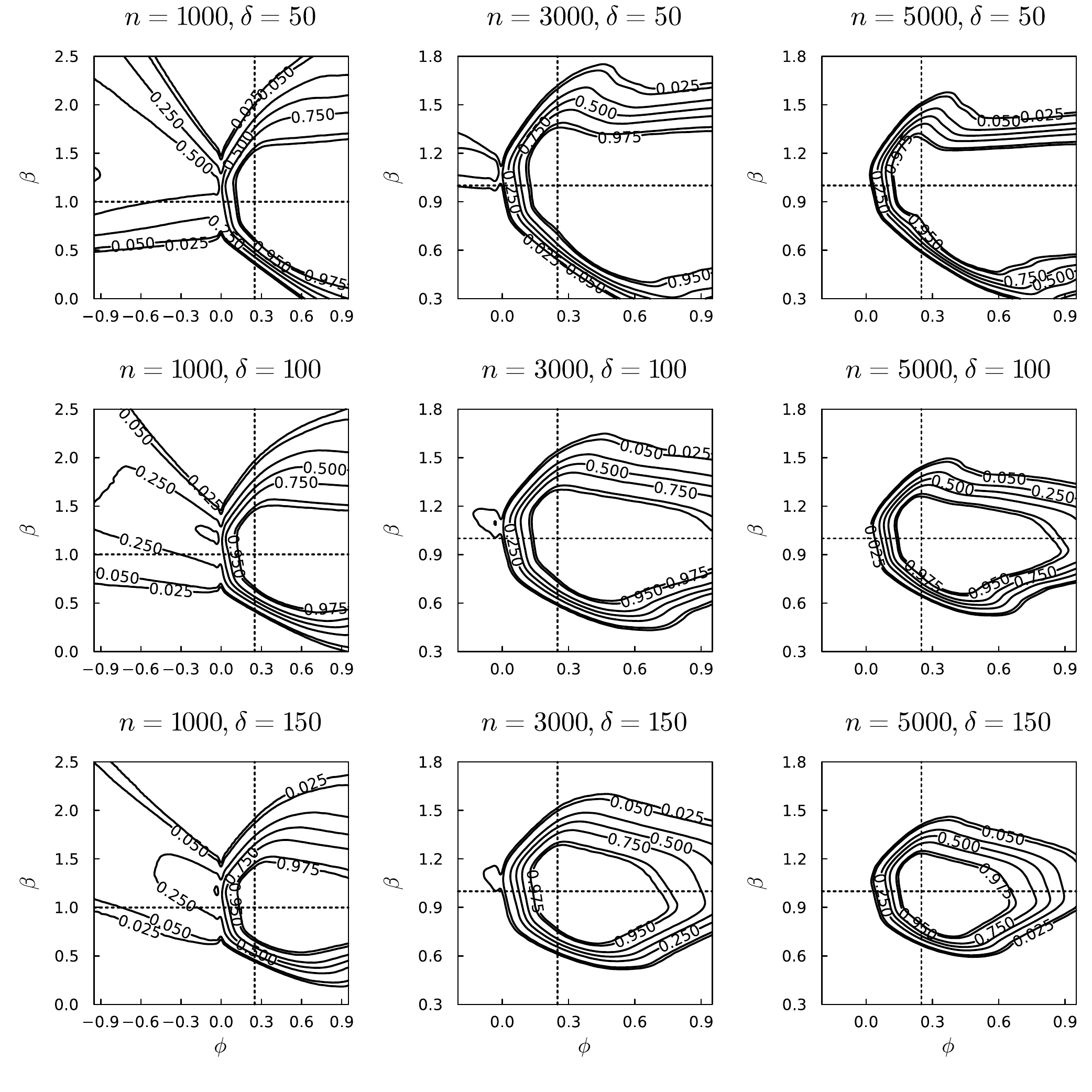}
	\end{center}
\parbox{6.2in}{\footnotesize \medskip
	Notes: The intersecting point between two dotted lines in each panel indicates the true parameter $(\phi_0,\beta_0)$. The horizontal axis represents the hypothesized value of $\phi$ and the vertical axis that of $\beta$. In each panel, the area surrounded by the innermost contour line consists of the parameter values $(\phi,\beta)$ that are included in the confidence set at least 97.5\% of the times in the Monte Carlo loops, and the area surrounded by the outermost contour line consists of the parameter values $(\phi,\beta)$ that are included in the confidence set at least 2.5\% of the times in the Monte Carlo loops.}
\bigskip
\end{figure}

\clearpage

\begin{figure}[tbph]
	\caption[False Coverage Probability of the Confidence Intervals at $95\%$]{
		False Coverage Probability of the Confidence Intervals for $(\phi_{0},\beta_{0})$ at $95\%$: $\phi_0 = 0.25, \beta_0 = 1.0, \rho=0.00001$.
	}
	\label{figure:sim_power_22}
	\medskip
	\begin{center}
		\includegraphics[scale = 0.4]{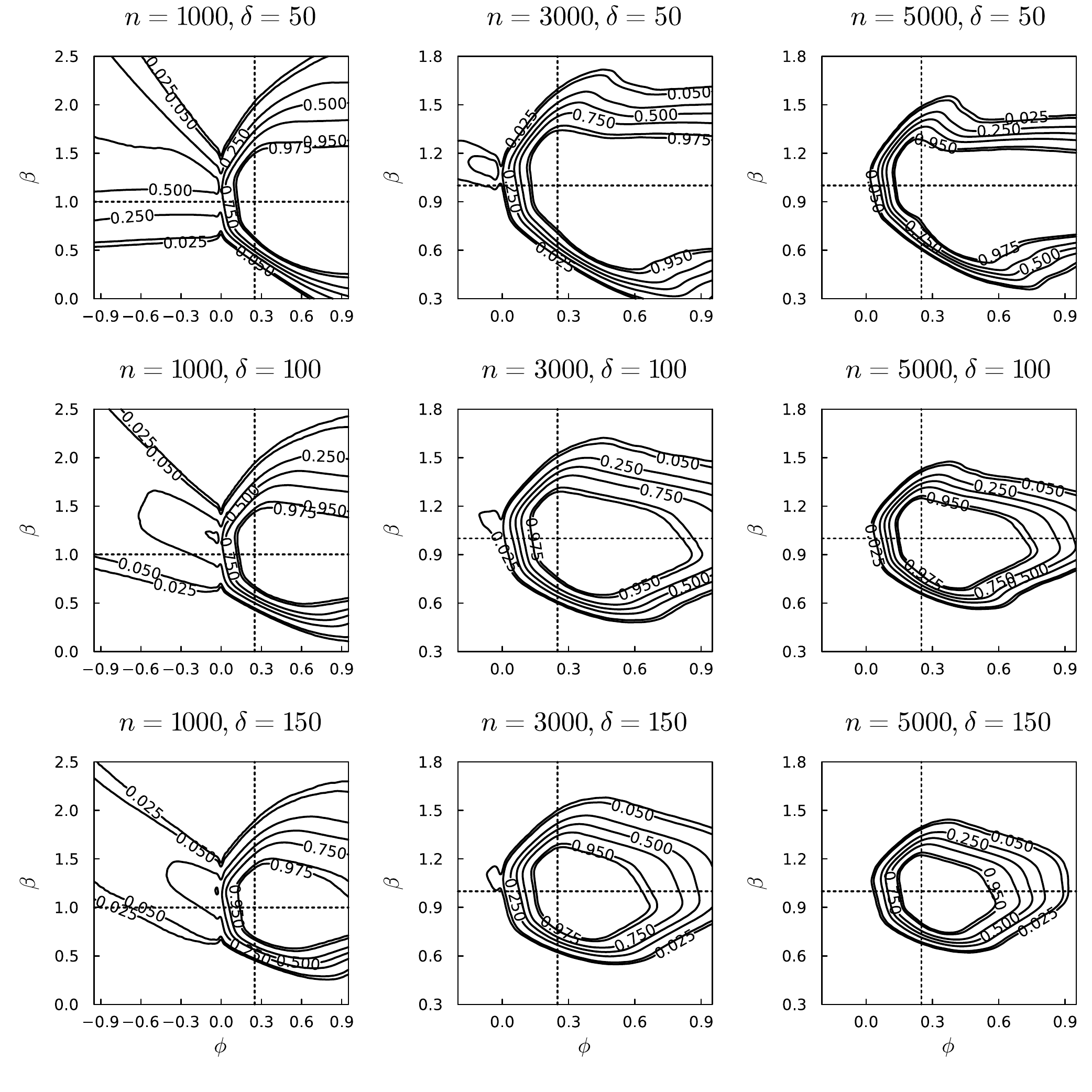}
	\end{center}
\parbox{6.2in}{\footnotesize \medskip
	Notes: The intersecting point between two dotted lines in each panel indicates the true parameter $(\phi_0,\beta_0)$. The horizontal axis represents the hypothesized value of $\phi$ and the vertical axis that of $\beta$. In each panel, the area surrounded by the innermost contour line consists of the parameter values $(\phi,\beta)$ that are included in the confidence set at least 97.5\% of the times in the Monte Carlo loops, and the area surrounded by the outermost contour line consists of the parameter values $(\phi,\beta)$ that are included in the confidence set at least 2.5\% of the times in the Monte Carlo loops.}
\bigskip
\end{figure}

\clearpage

\begin{figure}[tbph]
	\caption[False Coverage Probability of the Confidence Intervals at $95\%$]{
		False Coverage Probability of the Confidence Intervals for $(\phi_{0},\beta_{0})$ at $95\%$: $\phi_0 = 0.25, \beta_0 = 1.0, \rho=0.0001$.
	}
	\label{figure:sim_power_32}
	\medskip
	\begin{center}
		\includegraphics[scale = 0.4]{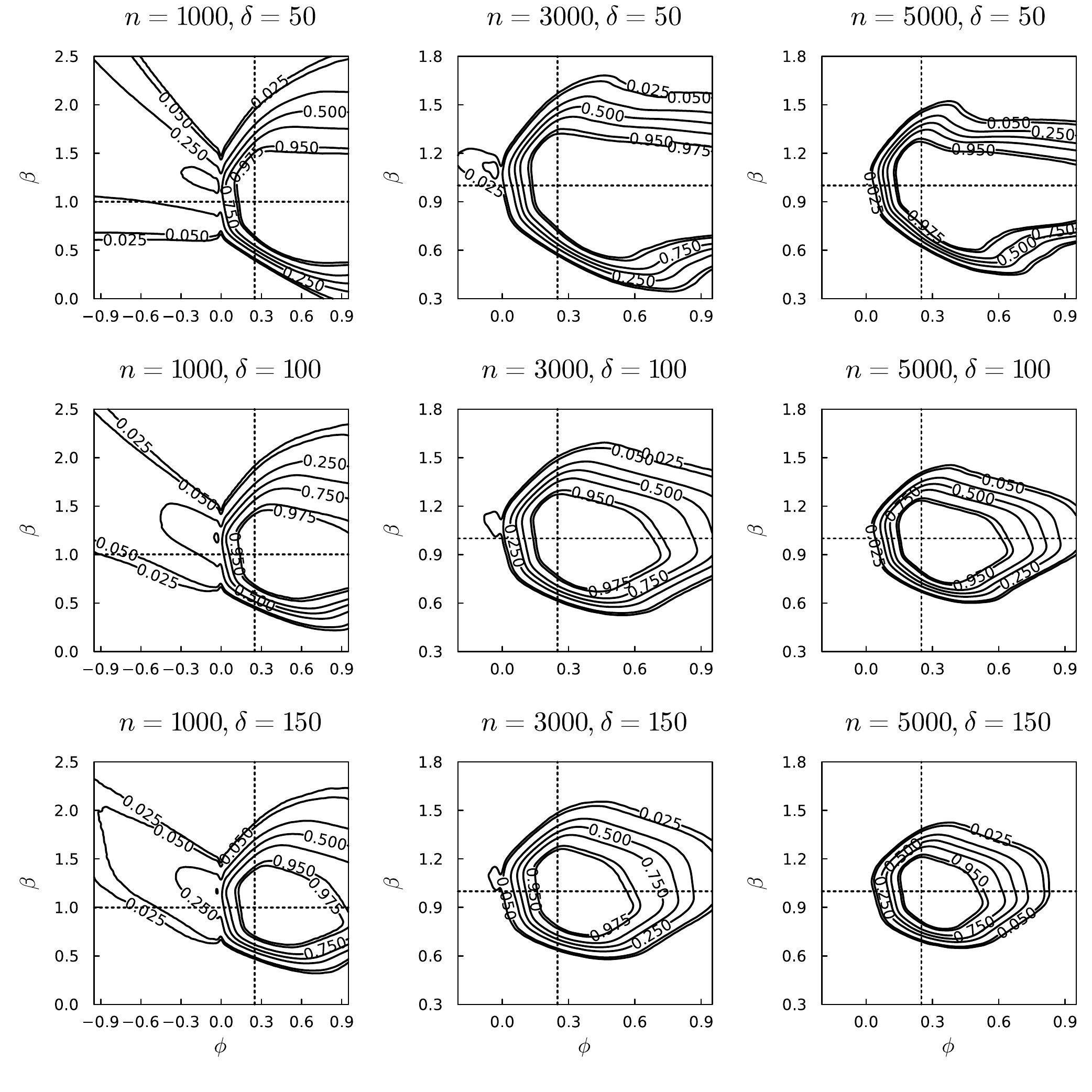}
	\end{center}
\parbox{6.2in}{\footnotesize \medskip
	Notes: The intersecting point between two dotted lines in each panel indicates the true parameter $(\phi_0,\beta_0)$. The horizontal axis represents the hypothesized value of $\phi$ and the vertical axis that of $\beta$. In each panel, the area surrounded by the innermost contour line consists of the parameter values $(\phi,\beta)$ that are included in the confidence set at least 97.5\% of the times in the Monte Carlo loops, and the area surrounded by the outermost contour line consists of the parameter values $(\phi,\beta)$ that are included in the confidence set at least 2.5\% of the times in the Monte Carlo loops.}
\bigskip
\end{figure}

\begin{figure}[tbph]
	\caption[False Coverage Probability of the Confidence Intervals at $95\%$]{
		False Coverage Probability of the Confidence Intervals for $(\phi_{0},\beta_{0})$ at $95\%$: $\phi_0 = 0.25, \beta_0 = 1.0, \rho=0.001$.
	}
	\label{figure:sim_power_42}
	\medskip
	\begin{center}
		\includegraphics[scale = 0.4]{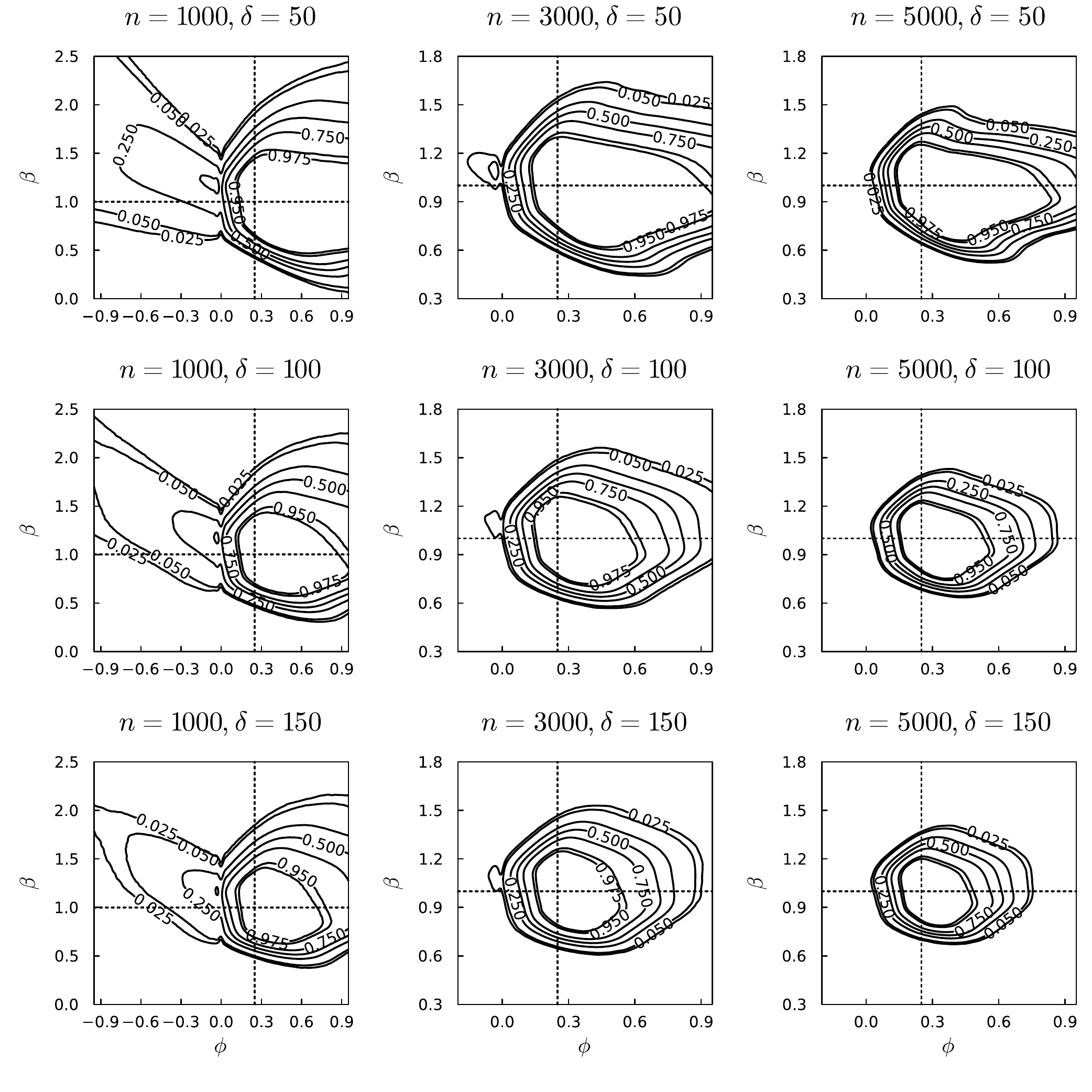}
	\end{center}
\parbox{6.2in}{\footnotesize \medskip
	Notes: The intersecting point between two dotted lines in each panel indicates the true parameter $(\phi_0,\beta_0)$. The horizontal axis represents the hypothesized value of $\phi$ and the vertical axis that of $\beta$. In each panel, the area surrounded by the innermost contour line consists of the parameter values $(\phi,\beta)$ that are included in the confidence set at least 97.5\% of the times in the Monte Carlo loops, and the area surrounded by the outermost contour line consists of the parameter values $(\phi,\beta)$ that are included in the confidence set at least 2.5\% of the times in the Monte Carlo loops.}
\bigskip
\end{figure}

\begin{figure}[tbph]
	\caption[False Coverage Probability of the Confidence Intervals at $95\%$]{
		False Coverage Probability of the Confidence Intervals for $(\phi_{0},\beta_{0})$ at $95\%$: $\phi_0 = 0.25, \beta_0 = 1.0, \rho=0.01$.
	}
	\label{figure:sim_power_52}
	\medskip
	\begin{center}
		\includegraphics[scale = 0.4]{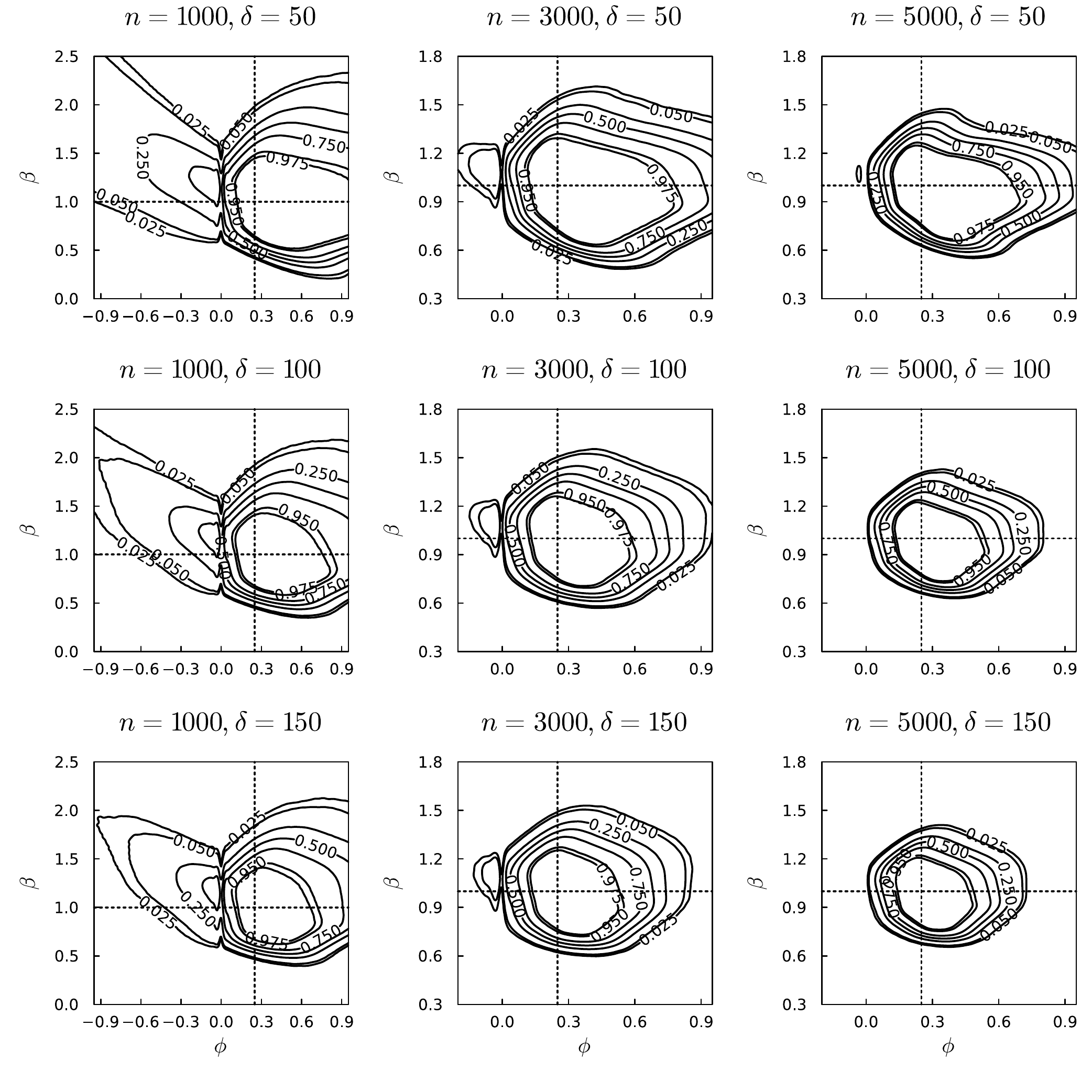}
	\end{center}
\parbox{6.2in}{\footnotesize \medskip
	Notes: The intersecting point between two dotted lines in each panel indicates the true parameter $(\phi_0,\beta_0)$. The horizontal axis represents the hypothesized value of $\phi$ and the vertical axis that of $\beta$. In each panel, the area surrounded by the innermost contour line consists of the parameter values $(\phi,\beta)$ that are included in the confidence set at least 97.5\% of the times in the Monte Carlo loops, and the area surrounded by the outermost contour line consists of the parameter values $(\phi,\beta)$ that are included in the confidence set at least 2.5\% of the times in the Monte Carlo loops.}
\bigskip
\end{figure}

\end{bibunit}

\end{document}